\documentclass[runningheads]{llncs}
\usepackage[english]{babel}
\usepackage{hyphenat}
\usepackage[dvipsnames]{xcolor}
\usepackage{cmll}
\usepackage{xspace}
\usepackage{amsmath}
\usepackage{amssymb}
\usepackage[hidelinks]{hyperref}
\usepackage{ebproof}
\usepackage{multicol}
\usepackage{mdframed}
\usepackage[capitalize]{cleveref}
\usepackage{thmtools}

\usepackage{cancel}

\mdfsetup{innertopmargin=.3cm,innerbottommargin=.6cm}

\usepackage{doc}
\usepackage{ulem} 
\renewcommand{\emph}[1]{{\it #1}}

\newcommand{\lamcc}{\lam_{{\tt gs}}}

\newcommand{\defeq}{:=}

\newcommand{\ctype}{\kappa}
\newcommand{\stype}{\mathcal{S}}

\newcommand{\rdel}{\sig}

\newcommand{\conj}[1]{\{#1\}}

\newcommand{\upd}[3]{\name{upd}_{#1}(#2,#3)}
\newcommand{\get}[3]{\name{get}_{#1}(\lam #2.#3)}
\newcommand{\set}[3]{\name{set}_{#1}(#2,#3)}

\newcommand{\defn}[1]{{\bf #1}}

\newcommand{\final}{final}
\newcommand{\undefined}{\name{undefined}}
\newcommand{\splus}{\Cup}

\newcommand{\eqstate}{\equiv_{\tt c}}
\newcommand{\equivstate}{\equiv}
\newcommand{\estate}{\epsilon}
\newcommand{\gtype}{\mathcal{T}}

\newcommand{\size}[1]{|#1|}

\newcommand{\red}[1][]{\ra_{#1}}
\newcommand{\redn}[1][]{\rra_{#1}}
\newcommand{\redcbv}{\ra_{\beta_v}}

\newcommand{\redbeta}{\red[\beta]}
\newcommand{\redbetav}{\red[\beta_v]}
\newcommand{\redget}{\red[\getname]}
\newcommand{\redset}{\red[\setname]}
\newcommand{\gname}{{\tt r}}
\newcommand{\getname}{{\tt g}}
\newcommand{\setname}{{\tt s}}

\newcommand{\dred}{\ra}
\newcommand{\gsred}{\ra}
\newcommand{\rred}[1][]{\rra_{#1}}
\newcommand{\drred}{\rra}
\newcommand{\gsrred}{\rra}

\newcommand{\subs}[2]{\{#1\!\setminus\!#2\}}
\newcommand{\fv}[1]{{\name{fv}}(#1)}

\newcommand{\dom}[1]{{\tt dom}(#1)}

\newcommand{\seqi}[4]{#1 \vdash^{#4} #2: #3}
\newcommand{\mul}[1]{[ #1 ]}
\newcommand{\emul}{[ \, ]}

\newcommand{\betav}{{\beta_v}}

\newcommand{\M}{\mathcal{M}}
\newcommand{\W}{\mathcal{W}}

\newcommand{\tightp}[1]{\name{tight}(#1)}
\newcommand{\tightt}{\name{tt}}

\newcommand{\tvar}{\name{v}}
\newcommand{\tneutral}{\name{n}}
\newcommand{\tabs}{\name{a}}
\newcommand{\vl}{\name{a}}

\newcommand{\nott}[1]{\overline{#1}}

\newcommand{\ruleUpd}{\name{upd}}
\newcommand{\ruleEmp}{\name{emp}}
\newcommand{\ruleConf}{\name{conf}}

\newcommand{\ruleGet}{\name{get}}
\newcommand{\ruleSet}{\name{set}}
\newcommand{\ruleBeta}{\name{$\beta_\name{v}$}}
\newcommand{\ruleAppL}{\name{appL}}
\newcommand{\ruleAppR}{\name{appR}}

\newcommand{\ruleLift}{\name{$\uparrow$}}
\newcommand{\ruleAx}{\name{ax}}
\newcommand{\ruleLam}{$\lambda$}
\newcommand{\ruleApp}{\name{@}}

\newcommand{\ruleMany}{\name{m}}

\newcommand{\ruleLamP}{\ruleLam$_\name{p}$}

\newcommand{\ruleAppPOne}{\ruleApp$_{\name{p1}}$}
\newcommand{\ruleAppPTwo}{\ruleApp$_{\name{p2}}$}

\newcommand{\isvalue}[1]{\name{val}(#1)}
\newcommand{\isabs}[1]{\name{abs}(#1)}

\newcommand{\neutral}{\name{ne}}
\newcommand{\normal}{\name{no}}

\newcommand{\eset}{\emptyset}
\newcommand{\name}[1]{{\tt #1}}
\newcommand{\sep}{\hspace{.5cm}}
\newcommand{\sm}{\setminus\!\!\!\setminus}
\newcommand{\tr}{\triangleright}

\newcommand{\iI}{i \in I}
\newcommand{\jJ}{j \in J}
\newcommand{\kK}{k \in K}
\newcommand{\ih}{{\it i.h.}\xspace}
\newcommand{\ie}{\textit{i.e.}}

\newcommand{\tim}{\times}
\newcommand{\ta}{\Rightarrow}
\newcommand{\ra}{\rightarrow}
\newcommand{\rra}{\twoheadrightarrow}
\newcommand{\Ra}{\Rightarrow}

\newcommand{\La}{\Leftarrow}

\newcommand{\lam}{\lambda}
\newcommand{\sig}{\sigma}
\newcommand{\del}{\delta}

\newcommand{\kap}{\kappa}
\newcommand{\Gam}{\Gamma}
\newcommand{\Del}{\Delta}

\newcommand{\val}{{\tt Val}}
\renewcommand{\emph}[1]{{\it #1}}
\newcommand{\id}{{\tt I}}
\newcommand{\rel}{\mathcal{R}}

\newcommand{\syscbv}{\mathcal{O}}
\newcommand{\sysgs}{\mathcal{P}}

\crefname{definition}{Def.}{Def.}%
\crefname{section}{Sec.}{Sec.}
\crefname{example}{Ex.}{Ex.}
\crefname{figure}{Fig.}{Fig.}
\crefname{proposition}{Prop.}{Prop.}

\newcommand{\cra}{\gg}

\newcommand{\conftype}[2]{#1 \times #2}  
\newcommand{\comptype}[2]{#1 \cra #2}  
\newcommand{\tcomptype}[3]{#1 \cra (#2 \times #3)}  

\title{Quantitative Global Memory\thanks{Supported by: National Funds through the Portuguese funding agency, FCT - Fundação para a Ciência e a Tecnologia -, within project LA/P/0063/2020, and the project and individual research grant 2021.04731.BD; Base Funding UIDB/00027/2020 of the Artificial Intelligence and Computer Science Laboratory – LIACC - funded by national funds through the FCT/MCTES (PIDDAC); and Cost Action CA20111 EuroProofNet.}}

\author{
  Sandra Alves
  \inst{1}
  \and
  Delia Kesner
  \inst{2,3}
  \and
  Miguel Ramos
  \inst{4,}\thanks{Corresponding author.}
 }

\institute{
  CRACS/INESC-TEC, DCC, Faculdade de Ciências, Universidade do Porto \\
  Rua do Campo Alegre s/n, 4169–007 Porto, Portugal
  \and
  Universit\'e Paris Cit\'e, CNRS, IRIF \\
  \and
  Institut Universitaire de France \\
  \and
  LIACC, DCC, Faculdade de Ciências, Universidade do Porto \\
  Rua do Campo Alegre s/n, 4169–007 Porto, Portugal
}

\authorrunning{Alves, Kesner and Ramos}
\titlerunning{A Quantitative Model for a Language with Global Memory}

\begin{document}

\maketitle

\begin{abstract}
  We show that recent approaches to static analysis based on quantitative typing systems can be extended to programming languages with global state. More precisely, we define a call-by-value language equipped with operations to access a global memory, together with  a semantic model based on a (tight) multi-type system that captures exact measures of time and space related to evaluation of programs. We show that the type system is quantitatively sound and complete with respect to the operational semantics of the language.
\end{abstract}

\setcounter{tocdepth}{2}

%
%


\section{Introduction}

The aim of this paper is to extend \emph{quantitative} techniques of {\it static analysis} based on \emph{multi-types} to programs with {\it effects}.

{\bf Effectful Programs.} Programming languages produce different kinds of \emph{effects} (observable interactions with the environment),  such as handling exceptions, read/write from a global memory outside its own scope, using a database or a file, performing non-deterministic choices, or sampling from probabilistic distributions. The degree to which these side effects are used depends on each programming paradigm~\cite{Jones1993} (imperative programming makes use of them while declarative programming does not). In general, avoiding the use of side effects facilitates the formal verification of programs, thus allowing to (statically) ensure their correctness. For example, the functional language Haskell eliminates side effects by replacing them with {\it monadic} actions, a clean approach that continues to attract growing attention. Indeed, rather than writing a function that returns a raw type, an effectful function returns a raw type inside a useful wrapper -- and that wrapper is a monad~\cite{Wadler1993}. This approach allows programming languages to combine the qualities of both the imperative and declarative worlds: programs produce effects, but these are encoded in such a way that formal verification can be performed very conveniently.

{\bf Quantitative Properties.} We address quantitative properties of programs with effects using {\it multi-types}, which originate in the theory of \emph{intersection} type  systems. They extend simple types with a new constructor $\cap$ in such a way that a program $t$ is typable with $\sigma \cap \tau$ if $t$ is typable with both types $\sigma$ and $\tau$ independently. Intersection types were first introduced as \emph{models} capturing computational properties of functional programming in a broader sense~\cite{CDC78}. For example, termination of different evaluation strategies can be characterized by typability in some appropriate intersection type system: a program $t$ is terminating if and only if $t$ is typable. Originally, intersection enjoys associativity, commutativity, and in particular idempotency (\ie\ $\sigma \cap \sigma = \sigma$). By switching to a \emph{non-idempotent} intersection constructor, one naturally comes to represent types by multisets, which is why they are called multi-types. Just like their idempotent precursors, multi-types still allow for a characterization of several operational properties of programs, but they also grant a substantial improvement: they provide  quantitative measures about these properties. For example, it is still possible to prove that a program is terminating if and only if it is typable, but now an {\it upper bound} or {\it exact measure} for the time needed for its evaluation length can be derived from the typing derivation of the program. This shift of perspective, from idempotent to non-idempotent types, goes beyond lowering the logical complexity of the proof: the quantitative information provided by typing derivations in the non-idempotent setting unveils crucial quantitative relations between typing (static) and reduction (dynamic) of programs.

{\bf Upper Bounds and Exact Split Measures.} Multi-types are extensively used to reason about programming languages from a quantitative point of view, as pioneered by de Carvalho \cite{deCarvalho2007,deCarvalho2018}. For example, they are able to provide \emph{upper bounds}, in the sense that the evaluation length of a program $t$ {\it plus} the size of its result (called {\it normal form}) can be bounded by the size of the type derivation of $t$. A major drawback of this approach, however, is that the size of normal forms can be exponentially bigger than the length of the evaluation reaching those normal forms. This means that bounding the sum of these two natural numbers at the same time is too rough, and not very relevant from a quantitative point of view. Fortunately, it is possible to extract better measures from a multi-type system. A crucial point to obtain \emph{exact measures}, instead of upper bounds, is to consider minimal type derivations, called \emph{tight derivations}. Moreover, using appropriate refined tight systems it is also possible to obtain \emph{independent} measures (called exact \emph{split} measures) for {\it length} and for {\it size}. More precisely, the quantitative typing systems\footnote{In this paper, by quantitative types we mean non-idempotent intersection types. Another meaning can be found in~\cite{Atkey18}.} are now equipped with constants and counters, together with an appropriate notion of tightness, which encodes minimality of type derivations. For any tight type derivation $\Phi$ of a program $t$ with counters $b$ and $d$, it is now possible to show that $t$ evaluates to a normal form of size $d$ in exactly $b$ steps. Therefore, the type system is not only \emph{sound}, \ie\ it is able to {\it guess} the number of steps to normal form as well as the size of this normal form, but the opposite direction providing \emph{completeness} of the approach also holds.

{\bf Contribution.} The focus of this paper is on effectful computations, such as reading and writing on a global memory able to hold values in cells. Taking inspiration from the monadic approach adopted in~\cite{deLiguoroT21}, we design a tight quantitative type system that provides exact split measures. More precisely, our system is not only capable of discriminating between length of evaluation to normal form and size of the normal form, but the measure corresponding to the length of the evaluation is split into two different natural numbers: the first one corresponds to the length of standard computation ($\beta$-reduction) and the second one to the number of memory accesses. We show that the system is sound \ie\ for any tight type derivation $\Phi$ of $t$ ending with counters $(b,m,d)$, the term $t$ is normalisable by performing $b$ evaluation steps and $m$ memory accesses, yielding a normal form having size $d$. The opposite direction, giving completeness of the model, is also proved.

In order to  gradually present the material, we first develop the technique for a weak (open) call-by-value (CBV) calculus, which can be seen as a contribution per se, and then we encapsulate these preliminary ideas in the  general framework of the language with global state.

{\bf Summary.} \cref{s:wocbv} illustrates the technique on a weak (open) CBV calculus. We then lift the technique to the $\lam$-calculus with global state in \cref{s:lambda-calculus-with-state} by following the same methodology. More precisely, \cref{s:syntax-gs} introduces the $\lamcc$-calculus, \cref{s:typing-system-gs}  defines a  quantitative type system $\sysgs$. Soundness and completeness of $\sysgs$ w.r.t. $\lamcc$ are proved in \cref{s:sound-complete-lambda-gs}. We conclude and discuss related work in \cref{s:conclusion}. In~\cref{sec:proofs}, we include some additional propositions and lemmas, as well as all the proofs in full detail.

{\bf Preliminary General Notations.} We start with some general notations. Given a (one-step) reduction relation $\red[\rel]$, $\redn[\rel]$ denotes the reflexive-transitive closure of $\red[\rel]$. We write $t \rred^b u$ for a reduction sequence from $t$ to $u$ of length $b$. A term $t$ is said to be (1) in \defn{$\rel$-normal form} (written $t \not \red[\rel]$) iff there is no $u$ such that $t \red[\rel] u$, and (2) \defn{$\rel$-normalizing} iff there is some $\rel$-normal form $u$ such that $t \redn[\rel] u$. The reduction relation $\rel$ is normalizing iff every term is $\rel$-normalizing.

\section{Weak Open CBV}
\label{s:wocbv}

In this section we first introduce the technique of tight typing on a simple language without effects, the weak open CBV. \cref{s:syntax-wocbv} defines the syntax and operational semantics of the language, \cref{s:types-for-wocbv} presents the tight typing system $\syscbv$ and discusses soundness and completeness of $\syscbv$ w.r.t. the CBV language.

\subsection{Syntax and Operational Semantics}
\label{s:syntax-wocbv}

Weak open CBV is based on two principles: reduction is \emph{weak} (not performed inside abstractions), and terms are \emph{open} (may contain free variables). \defn{Value}, \defn{terms} and \defn{weak contexts} are given by the following grammars, respectively:
\[ v, w  ::=  x \mid \lam x.t \qquad t, u, p ::= v \mid t u \qquad \W ::= \square \mid \W t \mid t \W \]

We write $\val$ for the set of all values. Symbol $\id$ is used to denote the identity function $\lam z.z$.

The sets of \defn{free} and \defn{bound} variables of terms and the notion of $\alpha$-conversion are defined as usual. A term $t$ is said to be \defn{closed} if $t$ does not contain any free variable, and \defn{open} otherwise. The \defn{size of a term $t$}, denoted $\size{t}$, is given by: $\size{x} = \size{\lam x.t} = 0$; and $\size{t u} = 1 + \size{t} + \size{u}$. Since our reduction relation is weak, \ie, reduction does not occur in the body of abstractions, we assign size zero to abstractions.

We now introduce the operational semantics of our language, which models the core behavior of  programming languages such as OCaml, where CBV evaluation is \emph{weak}. The \defn{deterministic reduction relation} (written $\dred$), is given by the following rules:
\[ \begin{array}{l@{\sep}l@{\sep}l}
        \begin{prooftree}
        \hypo{\phantom{DDDD}}
        \infer1[(\ruleBeta)]{(\lam x.t) v \dred t \subs{x}{v}}
    \end{prooftree} & 
    \begin{prooftree}
        \hypo{t \dred t'}
        \infer1[(\ruleAppL)]{t u \dred t' u}
    \end{prooftree} & 
    \begin{prooftree}
        \hypo{t \not\dred}
        \hypo{u \dred u'}
        \infer2[(\ruleAppR)]{t u \dred t u'}
    \end{prooftree}
\end{array} \]

\defn{Terms in $\dred$-normal form} can be characterized by the following grammars: $\normal  ::=  \val  \mid \neutral$ and $\neutral ::=  x \ \normal \mid \normal \ \neutral \mid \neutral \ \normal$.

\begin{restatable}[]{proposition}{propcharnfs}
    \label{prop:char-nfs}
    Let $t$ be a term. Then  $t \in \normal$  iff $t \not\dred$. 
\end{restatable}

In closed CBV~\cite{Plotkin1975} (only reducing closed terms), abstractions are the only normal forms, but in open CBV, the following terms turn out to be also acceptable normal forms: $x y$, $x (\lam y.y (\lam z.z))$ and $(\lam x.x) (y (\lam z.z))$.

\subsection{A Quantitative Type System for the Weak Open CBV}
\label{s:types-for-wocbv}

The \emph{untyped} $\lam$-calculus can be interpreted as a \emph{typed} calculus with a single type $D$, where $D = D \ta D$~\cite{Treglia22}. Applying Girard's~\cite{Girard87} \textit{``boring''} CBV translation of intuitionistic logic into linear logic, we get $D =\ !D \multimap\ !D$~\cite{Accattoli15}. Type system $\syscbv$ is built having this equation in mind.

The \defn{set of types} is given by the following grammar:
\[ \begin{array}{lrcl}
    \textbf{(Tight Constants)} & \tightt & ::= & \tvar \mid \tabs \mid \tneutral \\
    \textbf{(Value Types)} & \sig & ::= & \tvar \mid \tabs \mid \M \mid \M \ta \tau \\
    \textbf{(Multi-Types)} & \M & ::= & \mul{\sig_i}_{\iI} \ \text{where $I$ is a finite set} \\
    \textbf{(Types)} & \tau & ::= & \tneutral \mid \sig
\end{array} \]

Tight constants are minimal types assigned to terms reducing to normal forms ($\tvar$ for persistent variables, $\tabs$ for abstractions or variables that are going to be replaced by abstractions, and $\tneutral$ for neutral terms). Given an arbitrary tight constant $\tightt_0$, we write $\nott{\tightt_0}$ to denote all the other tight constants in $\tightt$ different from $\tightt_0$. Multi-types are multisets of value types. A \defn{(typing) environment}, written $\Gam, \Del$, is a function from variables to multi-types, assigning the empty multi-type $\emul$  to all but a finite set of variables. The domain of $\Gam$ is $\dom{\Gam} \defeq \{x \mid \Gam(x) \not= \emul\}$. The \defn{union} of environments, written $\Gam + \Del$, is defined by $(\Gam + \Del)(x) = \Gam(x) \sqcup \Del(x)$, where $\sqcup$ denotes \defn{multiset union}. An example is $(x : \mul{\sig_1}, y : \mul{\sig_2}) + (x : \mul{\sig_1}, z : \mul{\sig_2}) = (x : \mul{\sig_1, \sig_1}, y : \mul{\sig_2}, z : \mul{\sig_2})$. This notion is extended to a finite union of environments, written $+_{\iI} \Gam_i$ (the empty environment is obtained when $I = \eset$). We write $\Gam \sm x$ for the  environment $(\Gam \sm x)(x) = \emul$ and $(\Gam \sm x)(y) = \Gam(y)$ if $y \not= x$ and we  write $\Gam; x : \M$ for $\Gam + (x : \M)$, when $x \not\in \dom{\Gam}$. Notice that $\Gam$ and $\Gam; x:\emul$ are the same environment. 

A \defn{judgement} has the form $\seqi{\Gam}{t}{\tau}{(b,s)}$, where $b$ and $s$ are two natural numbers, representing, respectively,  the number of $\beta$-steps needed to normalize $t$, and the size of the normal form of $t$. The \defn{typing system $\syscbv$} is defined by the rules in~\cref{fig:typingruleslamop}. We write $\tr \seqi{\Gam}{t}{\tau}{(b,s)}$ if there is a (tree) \defn{type derivation} of the judgement $\seqi{\Gam}{t}{\tau}{(b,s)}$ using the rules of system $\syscbv$. The term $t$ is \defn{$\syscbv$-typable} (we may omit the name $\syscbv$) iff there is an environment  $\Gam$, a type $\tau$ and counters $(b,s)$ such that $\tr \seqi{\Gam}{t}{\tau}{(b,s)}$. We use letters $\Phi, \Psi, \dots$ to name type derivations, by writing for example $\Phi \tr \seqi{\Gam}{t}{\tau}{(b,s)}$.

\begin{figure}[h!]
    \[ \begin{array}{c}
        \begin{prooftree}
            \hypo{\phantom{DDDDD}}
            \infer1[(\ruleAx)]{\seqi{x : \mul{\sig}}{x}{\sig}{(0,0)}}
        \end{prooftree}
        \sep 
        \begin{prooftree}
            \hypo{\seqi{\Gam}{t}{\tau}{(b,s)}}
            \infer1[(\ruleLam)]{\seqi{\Gam \sm x}{\lam x.t}{\Gam(x) \ta \tau}{(b, s)}}
        \end{prooftree}
        \\[.5cm]
        \begin{prooftree}
            \hypo{\seqi{\Gam}{t}{\M \ta \tau}{(b,s)}}
            \hypo{\seqi{\Del}{u}{\M}{(b',s')}}
            \infer2[(\ruleApp)]{\seqi{\Gam + \Del}{t u}{\tau}{(1+b+b',s+s')}}
        \end{prooftree} 
        \sep
        \begin{prooftree}
            \hypo{(\seqi{\Gam_i}{v}{\sig_i}{(b_i,s_i)})_{\iI}}
            \infer1[(\ruleMany)]{\seqi{+_{\iI} \Gam_i}{v}{\mul{\sig_i}_{\iI}}{(+_{\iI} b_i, +_{\iI} s_i)}}
        \end{prooftree}
        \\[.5cm]
        \begin{prooftree}
            \hypo{\phantom{UUUUU}}
            \infer1[(\ruleLamP)]{\seqi{}{\lam x.t}{\tabs}{(0,0)}}
        \end{prooftree}
        \\[.5cm]
        \begin{prooftree}
            \hypo{\seqi{\Gam}{t}{\nott{\tabs}}{(b,s)}}
            \hypo{\seqi{\Del}{u}{\tightt}{(b',s')}}
            \infer2[(\ruleAppPOne)]{\seqi{\Gam + \Del}{t u}{\tneutral}{(b+b',1+s+s')}}
        \end{prooftree}
        \sep
        \begin{prooftree}
            \hypo{\seqi{\Gam}{t}{\tightt}{(b,s)}}
            \hypo{\seqi{\Del}{u}{\tneutral}{(b',s')}}
            \infer2[(\ruleAppPTwo)]{\seqi{\Gam + \Del}{t u}{\tneutral}{(b+b',1+s+s')}}
        \end{prooftree}
    \end{array} \]
    \caption{Typing Rules of System $\syscbv$}
    \label{fig:typingruleslamop}
  \end{figure}
  
Notice that in rule (\ruleAx) of \cref{fig:typingruleslamop} variables can only be assigned value types $\sig$ (in particular no type $\tneutral$): this is because they can only be substituted by values. Due to this fact, multi-types only contain value types. Regarding typing rules (\ruleAx), (\ruleLam), (\ruleApp), and (\ruleMany), they are the usual rules for non-idempotent intersection types~\cite{Bucciarelli2017}. Rules (\ruleLamP), (\ruleAppPOne), and (\ruleAppPTwo) are used to type \emph{persistent} symbols, \ie\ symbols that are not going to be \emph{consumed} during evaluation. More specifically, rule (\ruleLamP) types abstractions (with type $\tabs$) that are normal regardless of the typability of its body. Rule (\ruleAppPOne) types applications that will never reduce to an abstraction on the left (thus of any tight constant that is not $\tabs$, \ie\ $\nott{\tabs}$), while any term reducing to a normal form is allowed on the right (of tight constant $\tightt$). Rule (\ruleAppPTwo) also types applications, but when values will never be obtained on the right (only neutral terms of type $\tneutral$). Rule (\ruleAx) is also used to type persistent variables, in  particular when $\sig \in \{ \tvar, \tabs\}$.

A \defn{type} $\tau$ is \defn{tight} if $\tau \in \tightt$. We write $\tightp{\M}$, if every $\sig \in \M$ is tight. A \defn{type environment} $\Gam$ is \defn{tight} if it assigns tight multi-types to all variables. A \defn{type derivation} $\Phi \tr \seqi{\Gam}{t}{\tau}{(b,s)}$ is \defn{tight} if $\Gam$ and $\tau$ are both tight.

\begin{example}
    \label{ex:plotkintypederiv}
    Let $t = (\lam x.(xx) (y y)) (\lam z.z)$. Let $\Phi$ be the following typing derivation:
    {\small \[ \begin{prooftree}
        \infer0[(\ruleAx)]{\seqi{x : \mul{\mul{\tabs} \ta \tabs}}{x}{\mul{\tabs} \ta \tabs}{(0,0)}}
        \infer0[(\ruleAx)]{\seqi{x : \mul{\tabs}}{x}{\tabs}{(0,0)}}
        \infer1[(\ruleMany)]{\seqi{x : \mul{\tabs}}{x}{\mul{\tabs}}{(0,0)}}
        \infer2[(\ruleApp)]{\seqi{x : \mul{\mul{\tabs} \ta \tabs, \tabs}}{xx}{\tabs}{(1,0)}}
    \end{prooftree} \]}
    And $\Psi$ be the following typing derivation:
    {\small \[ \begin{prooftree}
        \hypo{\Phi}
        \infer0[(\ruleAx)]{\seqi{y : \mul{\tvar}}{y}{\tvar}{(0,0)}}
        \infer0[(\ruleAx)]{\seqi{y : \mul{\tvar}}{y}{\tvar}{(0,0)}}
        \infer2[(\ruleAppPOne)]{\seqi{y : \mul{\tvar, \tvar}}{yy}{\tneutral}{(0,1)}}
        \infer2[(\ruleAppPTwo)]{\seqi{x : \mul{\mul{\tabs} \ta \tabs, \tabs}, y : \mul{\tvar, \tvar}}{(xx) (yy)}{\tneutral}{(1,1)}}
        \infer1[(\ruleLam)]{\seqi{y : \mul{\tvar, \tvar}}{\lam x.(xx) (yy)}{\mul{\mul{\tabs} \ta \tabs, \tabs} \ta \tneutral}{(1,2)}}
    \end{prooftree} \]}
    Then, we can build the following tight typing derivation $\Phi_t$ for $t$:
    {\small \[ \begin{prooftree}
        \hypo{\Psi}
        \infer0[(\ruleLamP)]{\seqi{z : \mul{\tabs}}{z}{\tabs}{(0,0)}}
        \infer1[(\ruleLam)]{\seqi{}{\lam z.z}{\mul{\tabs} \ta \tabs}{(0,0)}}
        \infer0[(\ruleLamP)]{\seqi{}{\lam z.z}{\tabs}{(0,0)}}
       \infer2[(\ruleMany)]{\seqi{}{\lam z.z}{\mul{\mul{\tabs} \ta \tabs, \tabs}}{(0,0)}}
        \infer2[(\ruleApp)]{\seqi{y : \mul{\tvar, \tvar}}{(\lam x.(xx)(yy)) (\lam z.z)}{\tneutral}{(2,2)}}
    \end{prooftree} \]}
\end{example}

The type system $\syscbv$ can be shown to be \emph{sound} and \emph{complete} w.r.t. the operational semantics $\dred$ introduced in \cref{s:syntax-wocbv}. Soundness means that not only  a \emph{tightly} typable term $t$ is terminating, but also that the \emph{tight}   type derivation of $t$ gives exact and split measures concerning the reduction sequence from $t$ to normal form. More precisely, if $\Phi \tr \seqi{\Gam}{t}{\tau}{(b,s)}$ is tight, then there exists $u \in \normal$ such that  $t \drred^b u$ with $\size{u} = s$. Dually for \emph{completeness}. Because we are going to show this kind of properties for the more sophisticated language with global state (\cref{s:sound-complete-lambda-gs}), we do not give here technical details of them. However, we highlight these properties on our previous example.   Consider again term $t$ in \cref{ex:plotkintypederiv} and its tight derivation $\Phi_t$ with counters $(b,s)=(2,2)$. Counter $b$ is different from $0$, so $t \notin \normal$, but $t$ normalizes in two $\betav$-steps ($b = 2$) to a normal form having size $s = 2$. Indeed, $(\lam x.(xx)(yy))(\lam z.z) \red[\betav] ((\lam z.z)(\lam z.z))(yy) \red[\betav] (\lam z.z)(yy)$ and $\size{(\lam z.z)(yy)}=2$.

\section{A \texorpdfstring{$\lambda$}{Lambda}-Calculus with Global State}
\label{s:lambda-calculus-with-state}

Based on the preliminary presentation of \cref{s:wocbv}, we now introduce a $\lam$-calculus with global state following a CBV strategy. \cref{s:syntax-gs} defines the syntax and operational semantics of the $\lam$-calculus with global state. \cref{s:typing-system-gs} presents the tight typing system $\sysgs$, and \cref{s:sound-complete-lambda-gs} shows soundness and completeness.

\subsection{Syntax and Operational Semantics}
\label{s:syntax-gs}

Let $l$ be a location drawn from some set of location names. \defn{Values}, \defn{terms}, \defn{states} and \defn{configurations} of $\lamcc$ are defined respectively as follows:
\[ \begin{array}{rl@{\hspace{0.6cm}}rl}
    v, w & ::= x \mid \lam x.t & t, u, p & ::= v \mid v t \mid \get{l}{x}{t} \mid \set{l}{v}{t} \\
    s, q & ::= \estate \mid \upd{l}{v}{s} & c & ::= (t, s)
\end{array} \]

Notice that applications are restricted to the form $vt$. This, combined with the use of a deterministic reduction strategy based on weak contexts, ensures that composition of effects is well behaved. Indeed, this kind of restriction is usual in computational calculi~\cite{Moggi1989,Sabry1997,deLiguoroT21,Faggian2021}.

Intuitively, operation $\get{l}{x}{t}$ interacts with the global state by retrieving the value stored in location $l$ and binding it to variable $x$ of the continuation $t$. And operation $\set{l}{v}{t}$ interacts with the state by storing value $v$ in location $l$ and (possibly) overwriting whatever was previously stored there, and then returns $t$.

The  size function is extended to states and configurations: $\size{s}:=0$, and $\size{(t,s)}: = \size{t}$. The update constructor is commutative in the following sense:
\[ \begin{array}{rclr}
        \upd{l}{v}{\upd{l'}{w}{s}} & \eqstate & \upd{l'}{w}{\upd{l}{v}{s}} \text{ if $l \not= l'$}
\end{array} \]
We denote by $\equivstate$ the equivalence relation generated by the axiom $\eqstate$. \label{def:domainS} We write \defn{$l \in \dom{s}$}, if $s \equivstate \upd{l}{v}{q}$, for some value $v$ and state $q$.
Moreover, these $v$ and $q$ are \emph{unique}. For example, if $l_1 \neq l_2$, then $s_1 = \upd{l_1}{v_1}{\upd{l_2}{v_2}{q}} \equivstate \upd{l_2}{v_2}{\upd{l_1}{v_1}{q}} = s_2 $, but $\upd{l_1}{v_1}{\upd{l_1}{v_2}{s}} \not\equivstate \upd{l_1}{v_2}{\upd{l_1}{v_1}{s}}$.
As a consequence, whenever we want to access the content of a particular location in a state, we can simply assume that the location is at the top of the state.

The operational semantics of the $\lamcc$-calculus is given on configurations. The \defn{deterministic reduction relation} $\gsred$ is defined to be the union of the rules $\red[\gname]\ (\gname \in \{\beta_v, \getname, \setname\})$ below. We write $(t,s) \gsrred^{(b,m)} (u,q)$ if $(t,s)$ reduces to $(u,q)$ in $b$ $\beta_v$-steps and $m$ $\getname/\setname$-steps.
\[ \begin{array}{c}
  \begin{prooftree}
    \hypo{\phantom{BUUUUUUUUU}}
    \infer1[(\ruleBeta)]{((\lam x.t) v,s) \redbetav (t \subs{x}{v},s)}
  \end{prooftree} 
  \sep
  \begin{prooftree}
    \hypo{s \equivstate \upd{l}{v}{q}}
    \infer1[(\ruleGet)]{(\get{l}{x}{t}, s) \redget (t \subs{x}{v}, s)}
  \end{prooftree}
  \\[.5cm]
  \begin{prooftree}
    \hypo{(t, s) \red[\gname] (u, q) \sep \gname \in \{\beta_v, \getname, \setname\}}
    \infer1[(\ruleAppR)]{(v t, s) \red[\gname] (v u, q)}
  \end{prooftree}
  \sep
  \begin{prooftree}
    \hypo{\phantom{BUUUUUUUU}}
    \infer1[(\ruleSet)]{(\set{l}{v}{t}, s) \redset (t, \upd{l}{v}{s})}
  \end{prooftree}
\end{array} \]

Note that in reduction rule (\ruleAppR), the $\gname$ appearing as the name of the reduction rule in the premise is the same as the one appearing in the reduction rule in the conclusion.

\newpage

\begin{example}
  \label{ex:globalstate}
  Consider the configuration $c_0 = ((\lam x.\get{l}{y}{yx}) (\set{l}{\id}{z}), \estate)$. Then we can reach an irreducible configuration as follows:
  \[ \begin{array}{c}
    ((\lam x.\get{l}{y}{yx}) (\set{l}{\id}{z}), \estate) \redget ((\lam x.\get{l}{y}{yx}) z, \upd{l}{\id}{\estate})
    \\
    \redbetav (\get{l}{y}{yz}, \upd{l}{\id}{\estate}) \redget (\id z, \upd{l}{\id}{\estate}) \redbetav (z, \upd{l}{\id}{\estate})
  \end{array} \]
\end{example}

A configuration $(t,s)$ is said to be \defn{blocked} if either $t = \get{l}{x}{u}$ and $l \not\in \dom{s}$; or $t = v u$ and $(u,s)$ is blocked. A configuration is \defn{unblocked} if it is not blocked.
As an example, $(\get{l}{x}{x}, \estate)$ is obviously blocked. As a consequence, the following configuration reduces to a blocked one: $((\lam y.y\ \get{l}{x}{x}) z, \estate) \red (z\ \get{l}{x}{x}, \estate)$.
This suggests a notion of \defn{final configuration}: $(t,s)$ is \defn{\final}\ if either $(t, s)$ is blocked; or $t \in \normal$, where \defn{neutral} and \defn{normal} terms are given respectively by the grammars $\neutral ::= x \ \normal \mid (\lam x.t) \ \neutral$ and $\normal ::= \val  \mid \neutral$.

\begin{restatable}[]{proposition}{propnormalifffinal}
    \label{prop:normal-iff-final}
    Let $(t,s)$ be a configuration. Then $(t,s)$ is \final\ iff $(t,s) \not\ra$.
\end{restatable}

Notice that when $(t,s)$ is  an unblocked final configuration, then $t \in \normal$. These are the configurations  captured by the typing system $\sysgs$ in \cref{s:typing-system-gs}. 
Consider the final configurations $c_0=(\get{l}{x}{x}, \estate)$, $c_1=(z\ \get{l}{x}{x}, \estate)$, $c_2= (y, s)$ and $c_3=((\lam x.x) (yz), s)$. Then $c_0$ and $c_1$ are blocked, while $c_2$ and $c_3$ are unblocked.

\subsection{A Quantitative Type System for the \texorpdfstring{$\lamcc$}{LambdaCC}-calculus}
\label{s:typing-system-gs}

We now introduce the quantitative type system $\sysgs$ for $\lamcc$. To deal with global states, we extend the language of types with the notions of state, configuration and monadic types. To do this, we translate linear arrow types according to Moggi's~\cite{Moggi1989} CBV interpretation of reflexive objects in the category of $\lam_c$-models: $D =\ !D \multimap\ !D$ becomes $D =\ !D \multimap\ T(!D)$, where $T$ is a monad. Type system $\sysgs$ was built having this equation in mind, similarly to what was done in~\cite{GTV23}.

The \defn{set of types} is given by the following grammar:
\[ \begin{array}{lrcl}
    \textbf{(Tight Constants)} & \tightt & ::= & \tvar \mid \vl \mid \tneutral \\
    \textbf{(Value Types)} & \sig & ::= & \tvar \mid \vl \mid \M \mid \M \ta \del \\
    \textbf{(Multi-types)} & \M & ::= & \mul{\sig_i}_{\iI} \ \text{where $I$ is a finite set} \\
    \textbf{(Liftable Types)} & \mu & ::= & \tvar \mid \vl \mid \M \\
    \textbf{(Types)} & \tau & ::= & \tneutral \mid \sig \\
    \textbf{(State Types)} & \stype & ::= & \conj{(l_i : \M_i)}_{\iI} \mbox{ where all } l_i \mbox{ are distinct} \\
    \textbf{(Configuration Types)} & \ctype & ::= & \conftype{\tau}{\stype} \\
    \textbf{(Monadic Types)} & \del & ::= & \comptype{\stype}{\ctype} \\
\end{array} \]

In system $\sysgs$, the minimal types to be assigned to normal forms distinguish between variables  ($\tvar$), abstractions ($\vl$), and neutral terms ($\tneutral$). A \defn{multi-type} is a multi-set of value types. A \defn{state type} is a partial function mapping labels to (possibly empty) multi-types. A \defn{configuration type} is a product type, where the first component is a type and the second is a state type. A \defn{monadic type} associates a state type to a configuration type.  We use symbol $\gtype$ to denote a value type or a monadic type. \defn{Typing environments} and operations over types are defined in the same way as in system $\syscbv$.

The \defn{domain} of a state type $\stype$ is the set of all its labels, \ie\ $\dom{\stype} := \{ l \mid (l:\M) \in \stype \}$. Also, when $l \in \dom{\stype}$, \ie\ $(l : \M) \in \stype$, we write $\stype(l)$ to denote $\M$. The \defn{union of state types} is defined as follows: 
\begin{align*}
    (\stype \splus \stype')(l) = & \text{ if $(l : \M) \in \stype$ then (if $(l : \M') \in \stype'$ then $\M \sqcup \M'$ else $\M$)} \\ 
                            & \text{ else (if $(l : \M') \in \stype'$ then $\M'$ else \undefined)}
\end{align*}

\begin{example}
    Let $\stype = \{(l_1 : \mul{\sig_1, \sig_2}), (l_2 : \mul{\sig_1})\} \splus \{(l_2 : \mul{\sig_1, \sig_2}), (l_3 : \mul{\sig_3})\}$. Then,
    $\stype(l_1) = \mul{\sig_1,\sig_2}$,
    $\stype(l_2) = \mul{\sig_1, \sig_1, \sig_2}$,
    $\stype(l_3) = \mul{\sig_3}$,
    and $\stype(l) = \undefined$, 
    assuming $l \not= l_i$, for $i \in \{1,2,3\}$.
\end{example}
  
Notice that $\dom{\stype\splus \stype'} = \dom{\stype} \cup \dom{\stype'} $. Also  $\conj{(l : \emul)} \splus \stype \not= \stype$, if $l \not\in \dom{\stype}$, while $x:\emul; \Gam  = \Gam$. Indeed, typing environments are total functions, where variables mapped to $\emul$ do not occur in  typed programs. In contrast, states are partial functions, where labels  mapped to $\emul$ correspond to positions in memory that are accessed (by get or set), but ignored/discarded by the typed program. We use $\conj{(l : \M)}; \stype$ for $\conj{(l : \M)} \splus \stype$ if $l \not\in \dom{\stype}$.

A \defn{term type judgement} {(resp. \defn{state type judgment} and \defn{configuration type judgment})} has the form $\seqi{\Gam}{t}{\gtype}{(b,m,d)}$ (resp. $\seqi{\Gam}{s}{\stype}{(b,m,d)}$ and $\seqi{\Gam}{(t,s)}{\ctype}{(b,m,d)}$) where $b,m,d$ are three natural numbers, the first and second representing, respectively, the number of $\beta$-steps and $\getname/\setname$-steps needed to normalize $t$, and the third representing the size of the normal form of $t$. The \defn{typing system $\sysgs$} is defined by the rules in~\cref{fig:typingruleslamcc}. We write $\tr \mathcal{J}$ if there is a type derivation of the judgement $\mathcal{J}$ using the rules of system $\sysgs$. The term $t$  (resp. state $s$, configuration $(t,s)$) is \defn{$\sysgs$-typable} iff there is an environment  $\Gam$, a type $\gtype$ (resp. $\stype$, $\ctype$) and counters $(b,m,d)$ such that $\tr \seqi{\Gam}{t}{\gtype}{(b,m,d)}$ (resp. $\tr \seqi{\Gam}{s}{\stype}{(b,m,d)}$, $\tr \seqi{\Gam}{(t,s)}{\ctype}{(b,m,d)}$). As before, we use letters $\Phi, \Psi, \dots$ to name type derivations.

Rules (\ruleAx), (\ruleLam), (\ruleMany), and (\ruleApp) are essentially the same as in~\cref{fig:typingruleslamop}, but with types lifted to monadic types (\ie\ \emph{decorated} with state types). 
Rule (\ruleApp) assumes a value type associated to a value $v$ on the left premise and a monadic type associated to a term $t$  on the right premise. To type the  application $vt$, it is necessary to match both the value type $\M$ inside the type of $t$ with the input value type of $v$, and the output state type $\stype'$ of $t$ with the input state type of $v$. Rule (\ruleLift) is used to lift multi-types or tight constants $\tvar$ and $\tabs$ (the type of values) to monadic types. Rules (\ruleGet) and (\ruleSet) are used to type operations over the state. Rule (\ruleEmp) types empty states, rule (\ruleUpd) types states, and (\ruleConf) types configurations.

A \defn{type} $\tau$ is \defn{tight}, if $\tau \in \tightt$. We write \defn{$\tightp{\M}$} if every $\rdel \in \M$ is tight. A \defn{state type $\stype$} is \defn{tight} if $\tightp{\stype(l)}$ holds for all $l \in \dom{\stype}$. A \defn{configuration type $\conftype{\tau}{\stype}$} is \defn{tight}, if $\tau$ and $\stype$ are tight. A monadic type $\comptype{\stype}{\ctype}$ is \defn{tight}, if $\ctype$ is tight. The notion of tightness of type derivations is defined in the same way as in system $\syscbv$, \ie\ a \defn{type derivation} $\Phi$ is \defn{tight} if the type environment and the type of the conclusion of $\Phi$ are tight.

\begin{figure}[h]
    \[ \begin{array}{c}
        \mbox{\bf Rules for Terms}
        \\[.2cm]
        \begin{prooftree}
            \hypo{\phantom{AAAAA}}
            \infer1[(\ruleAx)]{\seqi{x : \mul{\rdel}}{x}{\rdel}{(0,0,0)}}
        \end{prooftree}
        \sep
        \begin{prooftree}
            \hypo{\seqi{\Gam}{v}{\mu}{(b,m,d)}}
            \infer1[(\ruleLift)]{\seqi{\Gam}{v}{\tcomptype{\stype}{\mu}{\stype}}{(b,m,d)}}
        \end{prooftree}
        \\[.5cm]
        \hspace{-.4cm}
        \begin{prooftree}
            \hypo{\seqi{\Gam}{t}{\comptype{\stype}{\ctype}}{(b,m,d)}}
            \infer1[(\ruleLam)]{\seqi{\Gam \sm x}{\lam x.t}{\Gam(x) \ta (\comptype{\stype}{\ctype})}{(b,m,d)}}
        \end{prooftree}
        \sep
        \begin{prooftree}
            \hypo{(\seqi{\Gam_i}{v}{\rdel_i}{(b_i,m_i,d_i)})_{\iI}}
            \infer1[(\ruleMany)]{\seqi{+_{\iI} \Gam_i}{v}{\mul{\rdel_i}_{\iI}}{(+_{\iI} b_i, +_{\iI} m_i, +_{\iI} d_i)}}
        \end{prooftree}
        \\[.5cm]
        \begin{prooftree}
            \hypo{\seqi{\Gam}{v}{\M \ta (\comptype{\stype'}{\ctype})}{(b,m,d)}}
            \hypo{\seqi{\Del}{t}{\tcomptype{\stype}{\M}{\stype'}}{(b',m',d')}}
            \infer2[(\ruleApp)]{\seqi{\Gam + \Del}{v t}{\comptype{\stype}{\ctype}}{(1+b+b',m+m',d+d')}}
        \end{prooftree}
        \\[.5cm]
        \begin{prooftree}
            \hypo{\seqi{\Gam}{t}{\comptype{\stype}{\ctype}}{(b,m,d)}}
            \infer1[(\ruleGet)]{\seqi{\Gam \sm x}{\get{l}{x}{t}}{\comptype{\conj{(l : \Gam(x))} \splus \stype}{\ctype}}{(b,1+m,d)}}
        \end{prooftree}
        \\[.5cm]
        \begin{prooftree}
            \hypo{\seqi{\Gam}{v}{\M}{(b,m,d)}}
            \hypo{\seqi{\Del}{t}{\comptype{\conj{(l : \M)}; \stype}{\kap}}{(b',m',d')}}
            \infer2[(\ruleSet)]{\seqi{\Gam + \Del}{\set{l}{v}{t}}{\comptype{\stype}{\kap}}{(b+b',1+m+m',d+d')}}
        \end{prooftree}
        \\[.5cm]
        \begin{prooftree}
            \hypo{\phantom{text}}
            \infer1[(\ruleLamP)]{\seqi{}{\lam x.t}{\vl}{(0,0,0)}}
        \end{prooftree}
        \\[.5cm]
        \hspace{-.4cm}
        \begin{prooftree}
            \hypo{\seqi{\Gam}{t}{\tcomptype{\stype}{\tightt}{\stype'}}{(b,m,d)}}
            \infer1[(\ruleAppPOne)]{\seqi{(x : \mul{\tvar}) + \Gam}{x t}{\tcomptype{\stype}{\tneutral}{\stype'}}{(b,m,1+d)}}
        \end{prooftree}
        \sep
        \begin{prooftree}
            \hypo{\seqi{\Gam}{u}{\tcomptype{\stype}{\tneutral}{\stype'}}{(b,m,d)}}
            \infer1[(\ruleAppPTwo)]{\seqi{\Gam}{(\lam x.t) u}{\tcomptype{\stype}{\tneutral}{\stype'}}{(b,m,1+d)}}
        \end{prooftree}
        \\[.5cm]     
        \mbox{\bf Rules for States}  
        \\[.2cm]
        \begin{prooftree}
            \hypo{\phantom{text}}
            \infer1[(\ruleEmp)]{\seqi{}{\estate}{\eset}{(0,0,0)}}
        \end{prooftree}
        \sep
        \begin{prooftree}
            \hypo{\seqi{\Gam}{v}{\M}{(b,m,d)}}
            \hypo{\seqi{\Del}{s}{\stype}{(b',m',d')}}
            \infer2[(\ruleUpd)]{\seqi{\Gam + \Del}{\upd{l}{v}{s}}{\conj{(l : \M)}; \stype}{(b+b',m+m',d+d')}}
        \end{prooftree}    
        \\[.5cm]
        \mbox{\bf Rule for Configurations}
        \\[.2cm]
        \begin{prooftree}
            \hypo{\seqi{\Gam}{t}{\comptype{\stype}{\ctype}}{(b,m,d)}}
            \hypo{\seqi{\Del}{s}{\stype}{(b',m',d')}}
            \infer2[(\ruleConf)]{\seqi{\Gam + \Del}{(t, s)}{\ctype}{(b+b',m+m',d+d')}}
        \end{prooftree}
    \end{array} \]
    \caption{Typing Rules for $\lamcc$.}
    \label{fig:typingruleslamcc}
\end{figure}

\begin{example}
    Consider configuration $c_0$ from~\cref{ex:globalstate}. Let $\M = \mul{\mul{\tvar} \ta \tcomptype{\eset}{\tvar}{\eset}}$, and $\Phi$ be the following typing derivation:
    {\small \[ \begin{prooftree}
        \infer0[(\ruleAx)]{\seqi{y : \M}{y}{\mul{\tvar} \ta \tcomptype{\eset}{\tvar}{\eset}}{(0,0,0)}}
        \infer0[(\ruleAx)]{\seqi{x : \mul{\tvar}}{x}{\tvar}{(0,0,0)}}
        \infer1[(\ruleMany)]{\seqi{x : \mul{\tvar}}{x}{\mul{\tvar}}{(0,0,0)}}
        \infer1[(\ruleLift)]{\seqi{x : \mul{\tvar}}{x}{\tcomptype{\eset}{\mul{\tvar}}{\eset}}{(0,0,0)}}
        \infer2[(\ruleApp)]{\seqi{y : \M, x : \mul{\tvar}}{yx}{\tcomptype{\eset}{\tvar}{\eset}}{(1,0,0)}}
        \infer1[(\ruleGet)]{\seqi{x : \mul{\tvar}}{\get{l}{y}{yx}}{\tcomptype{\conj{(l : \M)}}{\tvar}{\eset}}{(1,1,0)}}
        \infer1[(\ruleLam)]{\seqi{}{\lam x.\get{l}{y}{yx}}{\mul{\tvar} \ta (\tcomptype{\conj{(l : \M)}}{\tvar}{\eset})}{(1,1,0)}}
    \end{prooftree} \]}
    And $\Phi'$ be the following typing derivation:
    {\small \[ \hspace*{-7pt} \begin{prooftree}
            \infer0[(\ruleAx)]{\seqi{x : \mul{\tvar}}{x}{\tvar}{(0,0,0)}}
            \infer1[(\ruleLift)]{\seqi{x : \mul{\tvar}}{x}{\tcomptype{\eset}{\tvar}{\eset}}{(0,0,0)}}
            \infer1[(\ruleLam)]{\seqi{}{\id}{\mul{\tvar} \ta \tcomptype{\eset}{\tvar}{\eset}}{(0,0,0)}}
            \infer1[(\ruleMany)]{\seqi{}{\id}{\M}{(0,0,0)}}
            \infer0[(\ruleAx)]{\seqi{z : \mul{\tvar}}{z}{\tvar}{(0,0,0)}}
            \infer1[(\ruleMany)]{\seqi{z : \mul{\tvar}}{z}{\mul{\tvar}}{(0,0,0)}}
            \infer1[(\ruleLift)]{\seqi{z : \mul{\tvar}}{z}{\tcomptype{\conj{(l : \M)}}{\mul{\tvar}}{\conj{(l : \M)}}}{(0,0,0)}}
            \infer2[(\ruleSet)]{\seqi{z : \mul{\tvar}}{\set{l}{\id}{z}}{\tcomptype{\eset}{\mul{\tvar}}{\conj{(l : \M)}}}{(0,1,0)}}
    \end{prooftree} \]}
    Then we can build the following tight typing derivation $\Phi_c$ for $c$:
    {\small \[ \begin{prooftree}
        \hypo{\Phi}
        \hypo{\Phi'}
        \infer2[(\ruleApp)]{\seqi{z : \mul{\tvar}}{(\lam x.\get{l}{y}{yx})(\set{l}{\id}{z})}{\tcomptype{\eset}{\tvar}{\eset}}{(2,2,0)}}
        \infer0[(\ruleEmp)]{\seqi{}{\estate}{\eset}{(0,0,0)}}
        \infer2[(\ruleConf)]{\seqi{z : \mul{\tvar}}{((\lam x.\get{l}{y}{yx})(\set{l}{\id}{z}), \estate)}{\conftype{\tvar}{\eset}}{(2,2,0)}}
    \end{prooftree} \]}
    We will come back to this example at the end of~\cref{s:sound-complete-lambda-gs}.
\end{example}

\subsection{Soundness and Completeness}
\label{s:sound-complete-lambda-gs}

In this section, we show the main properties of the type system $\sysgs$ with respect to the operational semantics of the $\lam$-calculus with global state introduced in \cref{s:syntax-gs}. The properties of type system $\sysgs$ are similar to the ones for $\syscbv$, but now with respect to configurations instead of terms. \emph{Soundness} does not only state that a (tightly) typable configuration $(t,s)$ is terminating, but also gives exact (and split) measures concerning the reduction sequence from $(t,s)$ to a final form. \emph{Completeness} guarantees that a terminating configuration $(t,s)$ is tightly typable, where the measures of the associated reduction sequence of $(t,s)$ to final form are reflected in the counters of the resulting type derivation of $(t,s)$. This is the first work providing a model for a language with global memory being able to count the number of memory accesses.

We start by noting that type system $\sysgs$ does not type blocked configurations, which is exactly the notion that we want to capture.

\begin{restatable}[]{proposition}{proptypedunblock}
    \label{prop:typed-unblock}
    If $\Phi \tr \seqi{\Gam}{(t,s)}{\ctype}{(b,m,d)}$, then $(t,s)$ is unblocked.
\end{restatable}

We also show that counters capture the notion of normal form correctly, both for terms and states.

\begin{restatable}[]{lemma}{lemzerocounters} \mbox{}
  \label{lem:zero-counters-size-store}
  \begin{enumerate}
    \item \label{lem:zero-counters} Let $\Phi \tr \seqi{\Gam}{t}{\del}{(0,0,d)}$ be tight. Then, (1) $t \in \normal$ and (2) $d = \size{t}$.
    \item \label{lem:zero-size-store} Let $\Phi \tr \seqi{\Del}{s}{\stype}{(0,0,d)}$ be tight. Then $d = 0$.
  \end{enumerate}
\end{restatable}

In fact, we can show the following stronger property with respect to the counters for the number of $\betav$- and $\getname/\setname$-steps.

\begin{restatable}[]{lemma}{lemzeronfs}
    \label{lem:zero-nfs}
    Let $\Phi \tr \seqi{\Gam}{t}{\del}{(b,m,d)}$ be tight. Then, $b = m = 0$ iff $t \in \normal$.
\end{restatable}

The following property is essential for tight type systems~\cite{Accattoli2020}, and it shows that tightness of types spreads throughout type derivations of neutral terms, just as long as the environments are tight. 

\begin{restatable}[{\bf Tight Spreading}]{lemma}{lemcomtightspreading}
    \label{lem:comp-tight-spreading}
    Let $\Phi \tr \seqi{\Gam}{t}{\tcomptype{\stype}{\tau}{\stype'}}{(b,m,d)}$, such that $\Gam$ is tight. If  $t \in \neutral$, then $\tau \in \tightt$.
\end{restatable}

The two following properties ensure tight typability of final configurations. For that we need to be able to \emph{tightly} type any state, as well as any normal form. In fact, normal forms do not depend on a particular state since they are irreducible, so we can type them with any state type.

\begin{restatable}[{\bf Typability of States and Normal Forms}]{lemma}{typstates} \mbox{}
  \label{lem:typestatesnfs}
  \begin{enumerate}
      \item \label{lem:typ-states} Let $s$ be a state. Then, there exists $\Phi\ \tr \seqi{}{s}{\stype}{(0,0,0)}$ tight.
      \item \label{lem:comp-typ-nfs} Let  $t \in \normal$. Then for any tight $\stype$ there exists $\Phi \tr \seqi{\Gam}{t}{\tcomptype{\stype}{\tightt}{\stype}}{(0,0,d)}$ tight s.t. $d = \size{t}$.
  \end{enumerate}
\end{restatable}

Finally, we state the usual basic properties.

\begin{restatable}[{\bf Substitution} and {\bf Anti-Substitution}]{lemma}{lemcompsubsantisubs} \mbox{}
    \label{lem:comp-subs-antisubs}    
    \begin{enumerate}
        \item {\bf (Substitution)} \label{lem:comp-subs} If $\Phi_t \tr \seqi{\Gam_t; x : \M}{t}{\del}{(b_t,m_t,d_t)}$ and $\Phi_v \tr \seqi{\Gam_v}{v}{\M}{(b_v,m_v,d_v)}$, then $\Phi_{t \subs{x}{v}} \tr \seqi{\Gam_t + \Gam_v}{t \subs{x}{v}}{\del}{(b_t+b_v,m_t+m_v,d_t+d_v)}$.
        \item {\bf (Anti-Substitution)} \label{lem:comp-antisubs} If $\Phi_{t \subs{x}{v}} \tr \seqi{\Gam_{t \subs{x}{v}}}{t \subs{x}{v}}{\del}{(b,m,d)}$, then $\Phi_t \tr \seqi{\Gam_t; x : \M}{t}{\del}{(b_t,m_t,d_t)}$ and $\Phi_v \tr \seqi{\Gam_v}{v}{\M}{(b_v,m_v,d_v)}$, such that $\Gam_{t \subs{x}{v}} = \Gam_t + \Gam_v$, $b = b_t+b_v$, $m = m_t+m_v$, and $d = d_t + d_v$.
    \end{enumerate}
\end{restatable}

\begin{restatable}[{\bf Split Exact Subject Reduction} and {\bf Expansion}]{lemma}{lemexactredexp}
  \label{lem-exact-red-exp} \mbox{}
  \begin{enumerate}
    \item {\bf (Subject Reduction)} \label{lem:subj-comp-red} Let $(t,s) \red[\gname] (u,q)$. If $\Phi \tr \seqi{\Gam}{(t,s)}{\ctype}{(b,m,d)}$ is tight, then $\Phi' \tr \seqi{\Gam}{(u,q)}{\ctype}{(b',m',d)}$, where $\gname =\beta$ implies $b' = b - 1$ and $m' = m$, while $\gname \in \{\getname, \setname\}$ implies $b'=b$ and  $m' = m - 1$.
    \item {\bf (Subject Expansion)} \label{lem:comp-subj-exp} Let $(t,s) \red[\gname] (u,q)$. If $\Phi' \tr \seqi{\Gam}{(u,q)}{\ctype}{(b',m',d)}$ is tight, then $\Phi \tr \seqi{\Gam}{(t,s)}{\ctype}{(b,m,d)}$, where $\gname =\beta$ implies $b' = b - 1$ and $m' = m$, while $\gname \in \{\getname, \setname\}$ implies $b'=b$ and  $m' = m - 1$.
  \end{enumerate}
\end{restatable}

Soundness (resp. completeness) is based on exact subject reduction (resp. expansion), in turn based on the previous substitution (resp. anti-substitution) lemma.

\begin{restatable}[{\bf Quantitative Soundness} and {\bf Completeness}]{theorem}{compsoundness} \mbox{}
    \begin{enumerate}
        \item {\bf (Soundness)} If $\Phi \tr \seqi{\Gam}{(t,s)}{\kap}{(b,m,d)}$ tight, then there exists $(u,q)$ such that $u \in \normal$ and $(t,s) \gsrred^{(b,m)} (u,q)$ with $b$ $\beta$-steps, $m$ $\getname/\setname$-steps, and $\size{(u,q)} = d$.
        \item {\bf (Completeness)} If $(t,s) \rra^{(b,m,d)} (u,q)$ and $u \in \normal$, then there exists $\Phi \tr \seqi{\Gam}{(t,s)}{\ctype}{(b,m,\size{(u,q)})}$ tight.
    \end{enumerate}
\end{restatable}

\begin{example}
  Consider again configuration $c_0$ from~\cref{ex:globalstate} and its associated tight derivation $\Phi_{c_0}$. The first two counters of $\Phi_c$ are different from $0$: this means that $c$ is not a final configuration, but normalizes in two $\betav$-step ($b = 2$) and two $\getname/\setname$-steps ($m = 2$), to a final configuration having size $d = 0 = \size{z} = \size{(z, \upd{l}{I}{\estate})}$.
\end{example}

\section{Conclusion and Related Work}
\label{s:conclusion}

This paper provides a foundational step into the development of quantitative models for programming languages with effects. We focus on a simple language with global memory access capabilities. Due to the inherent lack of confluence in such framework we fix a particular evaluation strategy following a (weak) CBV approach. We provide a type system for our language that is able to (both) extract and discriminate between (exact) measures for the length of evaluation, number of memory accesses and size of normal forms. This study provides a valuable insight into time and space analysis of languages with global memory, with respect to length of evaluation and the size of normal forms, respectively.

In future work we would like to explore effectful computations involving global memory in a more general framework being able to capture different models of computation, such as the CBPV~\cite{Levy99} or the bang calculus~\cite{BucciarelliKRV20}. Furthermore, we would like to apply our quantitative techniques to other  effects that can be found in programming languages, such as non-termination, exceptions, non-determinism, and I/O.

{\bf Related Work.}  Several papers proposed quantitative approaches for different notions of CBV (without effects). But none of them exploits the idea of exact \emph{and} split tight typing. Indeed, the first non-idempotent intersection type system for Plotkin's CBV is~\cite{Ehrhard12}, where reduction is allowed under abstractions, and terms are considered to be closed.  This work was further extended to~\cite{CarraroG14}, where commutation rules are added to the calculus. None of these contributions extracts quantitative bounds from the type derivations. A calculus for open CBV is proposed in~\cite{AccattoliG18}, where \emph{fireball} --normal forms-- can be either erased or duplicated. Quantitative results are obtained, but no split measures. Other similar approaches appear in~\cite{Guerrieri19}. A logical characterization of CBV solvability is given in~\cite{AccattoliG22}, the resulting non-idempotent system gives quantitative information of the \emph{solvable} associated reduction relation. A similar notion of solvability for CBV for generalized applications was studied in~\cite{KesnerP22}, together with a logical characterization provided by a quantitative system. Other non-idempotent systems for CBV were proposed~\cite{Manzonetto2019,Kerinecetal21}, but they are defective in the sense that they do not enjoy subject reduction and expansion. Split measures for (strong) open CBV are developed in~\cite{KesnerV22}.

In~\cite{Dezani-CiancagliniGR09}, a system with universally quantified intersection and reference types is introduced for a language belonging to the ML-family. However, intersections are idempotent and only (qualitative) soundness is proved.

More recently, there has been a lot of work involving probabilistic versions of the lambda calculus. In~\cite{FaggianR19}, extensions of the lambda calculus with
a probabilistic choice operator are introduced. However, no quantitative results are provided. In~\cite{BreuvartL18}, monadic intersection types are used to obtain a (non-exact) quantitative model for a probabilistic calculus identical to the one in~\cite{FaggianR19}.

Concerning (exact) quantitative models for programming languages with global state, the state of the art is still underexplored. Some sound but not complete approaches are given in~\cite{Benton2009,Davies2000}, and quantitative results are not provided. Our work is inspired by a recent idempotent (thus only qualitative and not quantitative) model for CBV with global memory proposed by~\cite{deLiguoroT21}. This work was further extended in~\cite{GTV23} to a more generic framework of algebraic effectful computation, still the model does not provide any quantitative information about the evaluation of programs and the size of their results.


\renewcommand{\em}[1]{{\it #1}}

\bibliographystyle{splncs04}

\bibliography{refs}

\appendix
\newcommand{\maybehide}[1]{#1}

\section{Proofs}
\label{sec:proofs}

\subsection{Weak Open CBV}

\subsubsection{General Lemmas}

\begin{proposition}[{\bf Diamond}]
    \label{prop:diamond}
The relation $\redcbv$ enjoys the diamond property: if $t \redcbv t_i\ (i=1,2)$ and $t_1 \neq t_2$, then there exists $t_3$ such that $t_i \redcbv t_3\ i=1,2$.
\end{proposition}

\propcharnfs*

\maybehide{\begin{proof}
    We are going to show this proposition by splitting the original statement into the two following ones:
    \begin{enumerate}
        \item \label{prop:char-nfs:1} $t \not\dred$ and $\neg\isvalue{t}$ iff $t \in \neutral$.
        \item \label{prop:char-nfs:2} $t \not\dred$ iff $t \in \normal$.
    \end{enumerate}
    The proof now follows by simultaneous induction over both these statements:
    \begin{itemize}
        \item[$\Ra$)] By induction over $t$: 
        \begin{enumerate}
            \item Let $t \not\dred$ and $\neg\isvalue{t}$. We want to show that $t \in \neutral$:
            \begin{itemize}
                \item Case $t = x$ or $t = \lam x.u$. Then $\neg\isvalue{t}$ does not hold. Therefore, the statement holds vacuously.
                \item Case $t = u p$. Since $u p \not\dred$, then, in particular, it must be the case that either $\neg\isabs{u}$ or $\neg\isvalue{p}$ must hold, according to rule (\ruleBeta):
                \begin{itemize}
                    \item Assume $\neg\isabs{u}$ holds. It must be the case that $u \not\dred$, according to rule (\ruleAppL). And it also must be the case that $p \not\dred$, according to rule (\ruleAppR). Therefore, $p \in \normal$, by the \ih (\cref{prop:char-nfs}.\ref{prop:char-nfs:2}). Now, we have to consider $u$, which can be a variable, or not:
                    \begin{itemize}
                        \item Case $u = x$. Then $u p \in x \ \normal \in \neutral$.
                        \item Case $u$ is not a variable. Then $\neg\isvalue{u}$ holds. Therefore, we have $u \in \neutral$, by the \ih (\cref{prop:char-nfs}.\ref{prop:char-nfs:1}). Thus, $u p \in \neutral \ \normal \in \neutral$.
                    \end{itemize}
                    \item Assume $\neg\isvalue{p}$ holds. Then it must be the case that $u \not\dred$, according to rule (\ruleAppL). And that $p \not\dred$, according to rule (\ruleAppR). Therefore, $u \in \normal$, by the \ih (\ref{prop:char-nfs}.\ref{prop:char-nfs:2}), and $p \in \neutral$, by the \ih (\cref{prop:char-nfs}.\ref{prop:char-nfs:1}). Thus, $u p \in \normal \ \neutral \in \neutral$.
                \end{itemize}
            \end{itemize}
            \item Let $t \not\dred$. We want to show that $t \in \normal$:
            \begin{itemize}
                \item Case $t \in \val$. Then, clearly $t \in \normal$.
                \item Case $t \not\in \val$. Then, $\neg\isvalue{t}$ holds. Therefore, $t \in \neutral$, by \cref{prop:char-nfs}.\ref{prop:char-nfs:1}. Thus, in particular, $t \in \normal$.
            \end{itemize}
        \end{enumerate}
        \item[$\La$)] By induction over $t \in \normal$:
        \begin{enumerate}
            \item Let $t \in \neutral$. We want to show that $t \not\dred$ and $\neg\isvalue{t}$:
            \begin{itemize}
                \item Case $t = u p \in x \ \normal$. Then $u = x$ and $p \in \normal$. Since $u = x$, then both rules (\ruleBeta) and (\ruleAppL) cannot be applied. Since $p \in \normal$, then $p \not\dred$, by the \ih (\cref{prop:char-nfs}.\ref{prop:char-nfs:2}). Therefore, rule (\ruleAppR) also cannot be applied. Thus, $u p \not\dred$. And we can conclude, since $\neg\isvalue{u p}$ clearly holds.
                \item Case $t = u p \in \normal \ \neutral$. Then $u \in \normal$ and $p \in \neutral$. Since $u \in \normal$, then $u \not\dred$, by the \ih (\cref{prop:char-nfs}.\ref{prop:char-nfs:2}). Since $p \in \neutral$, then $p \not\dred$ and $\neg\isvalue{p}$ holds, by the \ih (\cref{prop:char-nfs}.\ref{prop:char-nfs:1}). Since $\neg\isvalue{p}$, then rule (\ruleBeta) cannot be applied. Since $u \not\dred$ and $p \not\dred$, then rules (\ruleAppL) and (\ruleAppR) cannot be applied. Therefore, $u p \not\dred$. And we can conclude since $\neg\isvalue{u p}$ clearly holds.
                \item Case $t = u p \in \neutral \ \normal$. Then $u \in \neutral$ and $p \in \neutral$. Since $u \in \neutral$, then $u \not\dred$ and $\neg\isvalue{u}$ holds, by the \ih (\cref{prop:char-nfs}.\ref{prop:char-nfs:1}). Since $p \in \normal$, then $p \not\dred$, by the \ih (\cref{prop:char-nfs}.\ref{prop:char-nfs:2}). Since $\neg\isvalue{u}$, then rule (\ruleBeta) cannot be applied. Since $u \not\dred$ and $p \not\dred$, then rules (\ruleAppL) and (\ruleAppR) cannot be applied. Therefore $u p \not\dred$. And we can conclude since $\neg\isvalue{u p}$ clearly holds.
            \end{itemize}
            \item Let $t \in \normal$. We want to show that $t \not\dred$:
            \begin{itemize}
                \item Case $t \in \val$. Then, clearly $t \not\dred$.
                \item Case $t \not\in \val$. Then, $t \in \neutral$, by definition. Thus, $t \not\dred$ holds, by~\cref{prop:char-nfs}.\ref{prop:char-nfs:1}.
            \end{itemize}
        \end{enumerate}
    \end{itemize}
\end{proof}
}
  
\begin{lemma}[Relevance]
    Let $\Phi \tr \seqi{\Gam}{t}{\tau}{(b,s)}$. Then $\dom{\Gam} \subseteq \fv{t}$.
\end{lemma}

\maybehide{\begin{proof}
    The proof following by induction over $\Phi$. Case $\Phi$ ends with rule (\ruleAx) or (\ruleLamP), then $\Phi$ is clearly relevant. The other cases following easily from the \ih.
\end{proof}}

\subsubsection{Soundness (Auxiliary Lemmas)}

\begin{lemma}
    \label{lem:values-not-neutral}
    Let $\Phi \tr \seqi{\Gam}{t}{\tau}{(b,s)}$. If $t \in \val$, then $\tau \not= \tneutral$.
\end{lemma}

\maybehide{\begin{proof}
    By case analysis on the form of $t \in \val$:
    \begin{itemize}
        \item Case $t = x$. Then we have to consider two additional cases according to the last rule used in $\Phi$:
        \begin{itemize}
            \item Case $\Phi$ ends with rule (\ruleAx), then $\tau$ is of the form $\sig \not= \tneutral$.
            \item Case $\Phi$ ends with rule (\ruleMany), then $\tau$ is of the form $\M \not= \tneutral$.
        \end{itemize}
        \item Case $t = \lam x.t$. Then we have to consider three additional cases according to the last rule used in $\Phi$:
        \begin{itemize}
            \item Case $\Phi$ ends with rule (\ruleLam), then $\tau$ is of the form $\M \ta \del \not= \tneutral$.
            \item Case $\Phi$ ends with rule (\ruleMany), then $\tau$ is of the form $\M \not= \tneutral$.
            \item Case $\Phi$ ends with rule (\ruleLamP), then $\tau = \tabs \not= \tneutral$.
        \end{itemize}
    \end{itemize}
\end{proof}}

\begin{lemma}
    \label{lem:notabs-implies-negabs}
    If $\Phi \tr \seqi{\Gam}{t}{\tau}{(b,s)}$, such that $\Gam$ is tight. If $\tau \in \nott{\tabs}$, then $\neg\isabs{t}$.
\end{lemma}

\maybehide{\begin{proof}
    By induction over $\Phi$:
    \begin{itemize}
        \item Case $\Phi$ ends with rule (\ruleAx), (\ruleApp), (\ruleAppPOne), or (\ruleAppPTwo), then $\neg\isabs{t}$ holds by definition.
        \item Case $\Phi$ ends with rule (\ruleLam), (\ruleMany), or (\ruleLamP),  then $\tau \not\in \nott{\tabs}$. Therefore, these cases do not apply.
    \end{itemize}
\end{proof}}

\begin{lemma}[{\bf Zero Steps and Normal Forms}]
    \label{lem:zero-steps-nfs}
    Let $\Phi \tr \seqi{\Gam}{t}{\tau}{(b,s)}$ be tight. $b = 0$ iff $t \in \normal$.
\end{lemma}

\maybehide{\begin{proof} \mbox{}
    \begin{itemize}
        \item[$\Ra$)] We want to show that, if $b = 0$, then $t \in \normal$. For this, we are going to split the original statement into the two following ones:
        \begin{enumerate}
            \item \label{lem:zero-steps-nfs:1} Let $\Phi \tr \seqi{\Gam}{t}{\tau}{(0,s)}$ be tight and $\neg\isvalue{t}$, then $t \in \neutral$.
            \item \label{lem:zero-steps-nfs:2} Let $\Phi \tr \seqi{\Gam}{t}{\tau}{(0,s)}$ be tight, then $t \in \normal$.
        \end{enumerate}
        The proof now follows by simultaneous induction over both these statements:
        \begin{enumerate}
            \item Let $\Phi \tr \seqi{\Gam}{t}{\tau}{(0,s)}$ be tight and $\neg\isvalue{t}$:
            \begin{itemize}
                \item Case $\Phi$ ends with rule (\ruleAx), (\ruleLam), (\ruleMany), or (\ruleLamP), then $\isvalue{t}$ holds. Therefore, these cases do not apply.
                \item Case $\Phi$ ends with rule (\ruleApp), then $b > 0$. Therefore, this case does not apply.
                \item Case $\Phi$ ends with rule (\ruleAppPOne), then $t$ is of the form $up$ and $\Phi$ is of the following form:
                \[ \begin{prooftree}
                    \hypo{\Phi_u \tr \seqi{\Gam_u}{u}{\nott{\tabs}}{(0,s_u)}}
                    \hypo{\Phi_p \tr \seqi{\Gam_p}{p}{\tightt}{(0,s_p)}}
                    \infer2[(\ruleAppPOne)]{\seqi{\Gam_u + \Gam_p}{up}{\tneutral}{(0,1+s_u+s_p)}}
                \end{prooftree} \]
                where $\tau = \tneutral$, $\Gam = \Gam_u + \Gam_p$ is tight, and $s = 1 + s_u + s_p$. Moreover, $\Gam_u$ and $\Gam_p$ are tight. By the \ih (\cref{lem:zero-steps-nfs}.\ref{lem:zero-steps-nfs:2}) over $\Phi_u$ and $\Phi_p$, we have that $u, p \in \normal$. By~\cref{lem:notabs-implies-negabs}, we have that $\neg\isabs{u}$. Therefore, either $u$ is a variable or $u \in \neutral$ by definition. So, in both cases, we can conclude that $u p \in \neutral$.
                \item Case $\Phi$ ends with rule (\ruleAppPTwo), then $t$ is of the form $up$ and $\Phi$ is of the following form:
                \[ \begin{prooftree}
                    \hypo{\Phi_u \tr \seqi{\Gam_u}{u}{\tightt}{(0,s_u)}}
                    \hypo{\Phi_p \tr \seqi{\Gam_p}{p}{\tneutral}{(0,s_p)}}
                    \infer2[(\ruleAppPTwo)]{\seqi{\Gam_u + \Gam_p}{up}{\tneutral}{(0,1+s_u+s_p)}}
                \end{prooftree} \]
                where $\tau = \tneutral$, $\Gam = \Gam_u + \Gam_p$, and $s = 1 + s_u + s_p$. Moreover, $\Gam_u$ and $\Gam_p$ are tight. By the \ih (\cref{lem:zero-steps-nfs}.\ref{lem:zero-steps-nfs:2}) over $\Phi_u$, we have that $u \in \normal$. By applying~\cref{lem:values-not-neutral} to $\Phi_p$, we have that $\neg\isvalue{p}$. By the \ih (\cref{lem:zero-steps-nfs}.\ref{lem:zero-steps-nfs:1}) over $\Phi_p$, we have that $p \in \neutral$. So, in both cases, we can conclude that $up \in \neutral$.
            \end{itemize}
            \item Let $\Phi \tr \seqi{\Gam}{t}{\tau}{(0,s)}$ be tight:
            \begin{itemize}
                \item Case $\Phi$ ends with rule (\ruleAx), (\ruleLam), or (\ruleLamP). Then, clearly $t \in \val$, so we can conclude immediately.
                \item Case $\Phi$ ends with rule (\ruleMany), then $\tau$ is of the form $\M \not\in \tightt$. Therefore, this case does not apply.
                \item In all the remaining cases $\neg\isvalue{t}$ holds. Then $t \in \neutral$, by \cref{lem:zero-steps-nfs}.\ref{lem:zero-steps-nfs:1}, so $t \in \normal$.
            \end{itemize}
        \end{enumerate}
        \item[$\La)$] We want to show that, if $t \in \normal$, then $b = 0$. The proof follows by induction over $t \in \normal$:
        \begin{enumerate}
            \item Case $t \in \neutral$. Then we have to consider the following additional cases:
            \begin{itemize}
                \item Case $t = xp$, such that $p \in \normal$. Then there are three additional cases to consider:
                \begin{itemize}
                    \item Case $\Phi$ ends with (\ruleApp), then it must be of the following form:
                    \[ \begin{prooftree}
                        \hypo{\seqi{x : \mul{\M \ta \tau}}{x}{\M \ta \tau}{(0,0)}}
                        \hypo{\Phi_p \tr \seqi{\Gam_p}{p}{\M}{(b_p,s_p)}}
                        \infer2[(\ruleApp)]{\seqi{(x : \mul{\M \ta \tau}) + \Gam_p}{xp}{\tau}{(1+b_p,s_p)}}
                    \end{prooftree} \]
                    where $\Gam = (x : \mul{\M \ta \tau}) + \Gam_p$ is tight, $b = 1+b_p$, and $s = s_p$. But, $\mul{\M \ta \tau}$ is not tight, since $\M \ta \tau \not\in \tightt$. Therefore, this case does apply.
                    \item Case $\Phi$ ends with (\ruleAppPOne), then $\Phi$ must be of the following form:
                    \[ \begin{prooftree}
                        \hypo{\seqi{(x : \mul{\tvar})}{x}{\tvar}{(0,0)}}
                        \hypo{\Phi_p \tr \seqi{\Gam_p}{p}{\tightt}{(b_p,s_p)}}
                        \infer2[(\ruleAppPOne)]{\seqi{(x : \mul{\tvar}) + \Gam_p}{up}{\tneutral}{(b_p,1+s_p)}}
                    \end{prooftree} \]
                    where $\tau = \tneutral$, $\Gam = (x : \mul{\tvar}) + \Gam_p$ is tight, $b = b_p$, and $s = 1 + s_p$. Moreover, $\Gam_p$ is tight. By the \ih over $\Phi_p$, we have that $b_p = 0$. So we can conclude with $b = b_p = 0$.
                    \item Case $\Phi$ ends with (\ruleAppPTwo). This case is very similar to the case where $\Phi$ ends with rule (\ruleAppPOne).
                \end{itemize}
                \item Case $t = up$, such that $u \in \normal$ and $p \in \neutral$. Then there are three additional cases to consider:
                \begin{itemize}
                    \item Case $\Phi$ ends with (\ruleApp), then it must be of the following form:
                    \[ \begin{prooftree}
                        \hypo{\seqi{\Gam_u}{u}{\M \ta \tau}{(b_u,s_u)}}
                        \hypo{\Phi_p \tr \seqi{\Gam_p}{p}{\M}{(b_p,s_p)}}
                        \infer2[(\ruleApp)]{\seqi{\Gam_u + \Gam_p}{up}{\tau}{(1+b_u+b_p,s_u+s_p)}}
                    \end{prooftree} \]
                    where $\tau = \tau$, $\Gam = \Gam_u + \Gam_p$ is tight, $b = 1 + b_u + b_p$, and $s = s_u + s_p$. By~\cref{lem:tight-spreading}.\ref{lem:tight-spreading:2}, we have that $\M \in \tightt$, which is a contradiction. Therefore, this case does not apply.
                    \item Case $\Phi$ ends with (\ruleAppPOne) or (\ruleAppPTwo). These cases are very similar to the corresponding cases when $t = x p$, such that $p \in \normal$.
                \end{itemize}
                \item Case $t = up$, such that $u \in \neutral$ and $p \in \normal$. Then there are three cases to consider:
                \begin{itemize}
                    \item Case $\Phi$ ends with (\ruleApp), then it must be of the following form:
                    \[ \begin{prooftree}
                        \hypo{\seqi{\Gam_u}{u}{\M \ta \tau}{(b_u,s_u)}}
                        \hypo{\Phi_p \tr \seqi{\Gam_p}{p}{\M}{(b_p,s_p)}}
                        \infer2[(\ruleApp)]{\seqi{\Gam_u + \Gam_p}{up}{\tau}{(1+b_u+b_p,s_u+s_p)}}
                    \end{prooftree} \]
                    where $\tau = \tau$, $\Gam = \Gam_u + \Gam_p$ is tight, $b = 1 + b_u + b_p$, and $s = s_u + s_p$. By~\cref{lem:tight-spreading}.\ref{lem:tight-spreading:2} over $u \in \neutral$, we have that $\M \ta \tau \in \tightt$, which is a contradiction. Therefore, this case does not apply.
                    \item Case $\Phi$ ends with (\ruleAppPOne) or (\ruleAppPTwo). These cases are very similar to corresponding cases when $t = x p$, such that $p \in \normal$, or $t = up$, such that $u \in \normal$ and $p \in \neutral$.
                \end{itemize}
            \end{itemize}
            \item Case $t \in \normal$. Then we can consider the two following additional cases:
            \begin{itemize}
                \item Case $t \in \val$. Then $\Phi$ must end with (\ruleAx), (\ruleLam), (\ruleMany), or (\ruleLamP). With the exception of the case where $\Phi$ ends with rule (\ruleMany), we can conclude $b = 0$ immediately for every other case, by definition. Case $\Phi$ ends with rule (\ruleMany), then $\tau$ is of the form $\M \not\in \tightt$. Therefore, this case does not apply.
                \item Case $t \not\in \val$. Then, $t \in \neutral$, by definition. Therefore, $b = 0$, by \cref{lem:zero-steps-nfs}.\ref{lem:zero-steps-nfs:1}.
            \end{itemize}
        \end{enumerate}
    \end{itemize}
\end{proof}}

\begin{lemma}
    \label{lem:corr-size-counter}
    Let $\Phi \tr \seqi{\Gam}{t}{\tau}{(b,s)}$ be tight. If $b = 0$ then $s = \size{t}$.
\end{lemma}

\maybehide{\begin{proof}
    The proof follows by induction over $\Phi$:
    \begin{itemize}
        \item Case $\Phi$ ends with rule (\ruleAx) or (\ruleLamP). Then $t \in \val$ and $s = 0$. So we can conclude with $\size{t} = 0 = s$.
        \item Case $\Phi$ ends with rule (\ruleLam). Then $\tau$ is of the form $\Gam_u(x) \ta \del \not\in \tightt$, so this case does not apply.
        \item Case $\Phi$ ends with rule (\ruleApp). Then $b > 0$, so this case does not apply.
        \item Case $\Phi$ ends with rule (\ruleMany). Then $\tau$ is of the form $\M \not\in \tightt$, so this case does not apply.
        \item Case $\Phi$ ends with rule (\ruleAppPOne). Then $t = up$ and $\Phi$ must be of the following form:
        \[ \begin{prooftree}
            \hypo{\Phi_u \tr \seqi{\Gam_u}{u}{\nott {\tabs}}{(0,s_u)}}
            \hypo{\Phi_p \tr \seqi{\Gam_p}{p}{\tightt}{(0,s_p)}}
            \infer2[(\ruleAppPOne)]{\seqi{\Gam_u + \Gam_p}{up}{\tneutral}{(0,1+s_u+s_p)}}
        \end{prooftree} \]
        where $\tau = \tneutral$, $\Gam = \Gam_u + \Gam_p$, and $s = 1 + s_u + s_p$. Moreover, $\Gam_u$ and $\Gam_p$ are tight. By the \ih over $\Phi_u$ and $\Phi_p$, we have $s_u = \size{u}$ and $s_p = \size{p}$. So we can conclude with $s = 1 + \size{u} + \size{p} = \size{up}$.
        \item Case $\Phi$ ends with rule (\ruleAppPTwo). This case is very similar to the case where $\Phi$ ends with rule (\ruleAppPOne).
    \end{itemize}
\end{proof}}

\begin{lemma}[{\bf Split for Values}]
    \label{lem:split-values}
    Let $\Phi_v \tr \seqi{\Gam}{v}{\M}{(b,s)}$, such that $\M = \sqcup_{\iI} \M_i$. Then, there exist ($\Phi^i_v \tr \seqi{\Gam_i}{v}{\M_i}{(b_i,s_i)})_{\iI}$, such that $\Gam = +_{\iI} \Gam_i$, $b = +_{\iI} b_i$, and $s = +_{\iI} s_i$.
\end{lemma}

\maybehide{\begin{proof}
    We start by noting that $\Phi_v$ must end with the rule ($\ruleMany$). Therefore, we have $\Gam = +_{\jJ} \Gam_j$, $\M = \mul{\sig_j}_{\jJ}$, $b = +_{\jJ} b_j$, $s = +_{\jJ} s_j$, and $(\Phi^j_v \tr \seqi{\Gam_j}{v}{\sig_j}{(b_j,s_j)})_{\jJ}$, for some $J$. Let $\M_i = \mul{\sig_k}_{\kK_i}$, for each $\iI$, such that $J = +_{\iI} K_i$. Then, by using rule ($\ruleMany$), we can build $\Phi^i_v \tr \seqi{\Gam_i}{v}{\M_i}{(b_i, s_i)}$, for each $\iI$, such that $\Gam_i = +_{\kK_i} \Gam_k$, $b_i = +_{\kK_i} b_k$, and $s_i = +_{\kK_i} s_k$. So we can conclude with $\Gam = +_{\jJ} \Gam_j = +_{\iI} (+_{\kK_i} \Gam_k) = +_{\iI} \Gam_i$, $b = +_{\jJ} b_j = +_{\iI} (+_{\kK_i} b_k) = +_{\iI} b_i$, and $s = +_{\jJ} s_j = +_{\iI} (+_{\kK_i} s_k) = +_{\iI} s_i$.
\end{proof}}

\subsubsection{Completeness (Auxiliary Lemmas)}

\begin{lemma}[{\bf Tight Spreading}]
    \label{lem:tight-spreading}
    Let $\Phi \tr \seqi{\Gam}{t}{\tau}{(b,s)}$, such that $\Gam$ is tight:
    \begin{enumerate}
        \item \label{lem:tight-spreading:1} If $b = 0$ and $\tau$ is not an arrow type or a multi-type, then $\tau \in \tightt$.
        \item \label{lem:tight-spreading:2} If $t \in \neutral$, then $\tau \in \tightt$.
    \end{enumerate}
\end{lemma}

\maybehide{\begin{proof} \mbox{}
    \begin{enumerate}
        \item We want to show that, if $b = 0$ and $\tau$ is not an arrow type or a multiset type, then $\tau \in \tightt$. The proof follows by induction over $\Phi$:
        \begin{itemize}
            \item Case $\Phi$ ends with rule ($\ruleAx$), then it is of the following form:
            \[ \begin{prooftree}
                \infer0[(\ruleAx)]{\seqi{x : \mul{\sig}}{x}{\sig}{(0,0)}}
            \end{prooftree} \]
            such that $\tau = \sig$, $\Gam = x : \mul{\sig}$, and $s = 0$. If $x : \mul{\sig}$ is tight, then $\sig \in \{\tabs, \tvar\}$. Therefore, we can conclude with $\sig \in \{\tabs, \tvar\} \subset \tightt$.
            \item Case $\Phi$ ends with rule (\ruleLam), then $\tau$ is an arrow type. Therefore, this case does not apply.
            \item Case $\Phi$ ends with rule (\ruleApp), then $b > 0$. Therefore, this case does not apply.
            \item Case $\Phi$ ends with rule (\ruleMany), then $\tau$ is a multiset type. Therefore, this case does not apply.
            \item Case $\Phi$ ends with rule (\ruleLamP), then $\tau = \tabs \in \tightt$. 
            \item Case $\Phi$ ends with rules (\ruleAppPOne) or (\ruleAppPTwo), then $\tau = \tneutral \in \tightt$.
        \end{itemize}
        \item We want to show that, if $t \in \neutral$, then $\tau \in \tightt$. By induction over $t \in \neutral$:
        \begin{itemize}
            \item Case $t = xp$, such that $p \in \normal$. Then we have to consider the following three cases depending on the last rule in $\Phi$:
            \begin{itemize}
                \item Case $\Phi$ ends with rule (\ruleApp), then it must be of the following form:
                \[ \begin{prooftree}
                    \hypo{\seqi{x : \mul{\M \ta \tau}}{x}{\M \ta \del}{(0,0)}}
                    \hypo{\Phi_p \tr \seqi{\Gam_p}{p}{\M}{(b_p,s_p)}}
                    \infer2[(\ruleApp)]{\seqi{(x : \mul{\M \ta \tau}) + \Gam_p}{xp}{\del}{(1+b_p,s_p)}}
                \end{prooftree} \]
                where $\Gam = (x : \mul{\M \ta \del}) + \Gam_p$ is tight, $b = 1+b_p$, and $s = s_p$. But, $\mul{\M \ta \del}$ is not tight, since $\M \ta \del \not\in \tightt$. Therefore, this case does apply.
                \item Case $\Phi$ ends with rule (\ruleAppPOne) or (\ruleAppPTwo). Then $\tau = \tneutral \in \tightt$, so we can conclude immediately.
            \end{itemize}
            \item Case $t = up$, such that $u \in \normal$ and $p \in \neutral$. Then we have to consider the following three cases depending on the last rule in $\Phi$:
            \begin{itemize}
                \item Case $\Phi$ ends with rule (\ruleApp), then it must be of the following form:
                \[ \begin{prooftree}
                    \hypo{\Phi_u \tr \seqi{\Gam_u}{u}{\M \ta \tau}{(b_u, s_u)}}
                    \hypo{\Phi_p \tr \seqi{\Gam_p}{p}{\M}{(b_p, s_p)}}
                    \infer2[(\ruleApp)]{\seqi{\Gam_u + \Gam_p}{up}{\tau}{(1+b_u+b_p, s_u+s_p)}}
                \end{prooftree} \]
                where $\Gam = \Gam_u + \Gam_p$ is tight, $b = 1 + b_u + b_p$, and $s = s_u + s_p$. Moreover, $\Gam_p$ is tight. By the \ih over $\Phi_p$, we have that $\M \in \tightt$, which is a contradiction. Therefore, this case does not apply.
                \item Case $\Phi$ ends with rule (\ruleAppPOne) or (\ruleAppPTwo). Then $\tau = \tneutral \in \tightt$, so we can conclude immediately.
            \end{itemize}
            \item Case $t = up$, such that $u \in \neutral$ and $p \in \normal$. Then we have to consider the following three cases depending on the last rule in $\Phi$:
            \begin{itemize}
                \item Case $\Phi$ ends with rule (\ruleApp), then it must be of the following form:
                \[ \begin{prooftree}
                    \hypo{\Phi_u \tr \seqi{\Gam_u}{u}{\M \ta \tau}{(b_u, s_u)}}n
                    \hypo{\Phi_p \tr \seqi{\Gam_p}{p}{\M}{(b_p, s_p)}}
                    \infer2[(\ruleApp)]{\seqi{\Gam_u + \Gam_p}{up}{\tau}{(1+b_u+b_p, s_u+s_p)}}
                \end{prooftree} \]
                where $\Gam = \Gam_u + \Gam_p$ is tight, $b = 1 + b_u + b_p$, and $s = s_u + s_p$. Moreover, $\Gam_p$ is tight. By the \ih over $\Phi_p$, we have that $\M \in \tightt$, which is a contradiction. Therefore, this case does not apply.
                \item Case $\Phi$ ends with rule (\ruleAppPOne) or (\ruleAppPTwo). Then $\tau = \tneutral \in \tightt$, so we can conclude immediately.
            \end{itemize}
        \end{itemize}
    \end{enumerate}
\end{proof}}

\begin{lemma}[{\bf Typability of Normal Forms}]
    \label{lem:typ-nfs}
    If $t \in \normal$, then there exists a tight derivation $\Phi \tr \seqi{\Gam}{t}{\tau}{(b,s)}$, such that $s = \size{t}$.
\end{lemma}

\maybehide{To show this proposition we are going to need to split the original statement into the two following ones:
\begin{enumerate}
    \item \label{prop:typ-nfs:1} If $t \in \neutral$, then there exists a tight derivation $\Phi \tr \seqi{\Gam}{t}{\tneutral}{(b,s)}$, such that $s = \size{t}$.
    \item \label{prop:typ-nfs:2} If $t \in \normal$, then there exists a tight derivation $\Phi \tr \seqi{\Gam}{t}{\tightt}{(b,s)}$, such that $s = \size{t}$.
\end{enumerate}
The proof follows by simultaneous induction over both these statements:
\begin{enumerate}
    \item Let $t \in \neutral$. We want to show that there exists a tight derivation $\Phi \tr \seqi{\Gam}{t}{\tneutral}{(b,s)}$:
    \begin{itemize}
        \item Case $t = up \in x \ \normal$. Then $u = x$ and $p \in \normal$. Therefore, there exists a tight derivation $\Phi_p \tr \seqi{\Gam_p}{p}{\tightt}{(b_p,s_p)}$, by the \ih (\cref{lem:typ-nfs}.\ref{prop:typ-nfs:2}), such that $\size{p} = s_p$. Thus, we can build $\Phi$ as follows:
        \[ \begin{prooftree}
            \infer0[(\ruleAx)]{\seqi{x : \mul{\tvar}}{x}{\tvar}{(0,0)}}
            \hypo{\Phi_p \tr \seqi{\Gam_p}{p}{\tightt}{(b_p,s_p)}}
            \infer2[(\ruleAppPOne)]{\seqi{ x : \mul{\tvar} + \Gam_p}{x p}{\tneutral}{(b_p,1+s_p)}}
        \end{prooftree} \]
        And we can conclude with $\Gam = x : \mul{\tvar} + \Gam_p$, $b = b_p$, and $s = 1+s_p = 1 + \size{x} + \size{p} = \size{xp}$.
        \item Case $t = up \in \normal \ \neutral$. Then $u \in \normal$ and $p \in \neutral$. Therefore, there exists a tight derivation $\Phi_u \tr \seqi{\Gam_u}{u}{\tightt}{(b_u,s_u)}$, such that $\size{u} = s_u$, by the \ih (\cref{lem:typ-nfs}.\ref{prop:typ-nfs:2}), and there exists a tight derivation $\Phi_p \tr \seqi{\Gam_p}{p}{\tneutral}{(b_p,s_p)}$, such that $\size{p} = s_p$ by the \ih (\cref{lem:typ-nfs}.\ref{prop:typ-nfs:1}). Thus, we can build $\Phi$ as follows:
        \[ \begin{prooftree}
            \hypo{\Phi_u \tr \seqi{\Gam_u}{u}{\tightt}{(b_u,s_u)}}
            \hypo{\Phi_p \tr \seqi{\Gam_p}{p}{\tneutral}{(b_p,s_p)}}
            \infer2[(\ruleAppPTwo)]{\seqi{\Gam_u + \Gam_p}{up}{\tneutral}{(b_u+b_p,1+s_u+s_p)}}
        \end{prooftree} \]
        And we can conclude with $\Gam = \Gam_u + \Gam_p$, $b = b_u+b_p$, and $s = 1+s_u+s_p = 1 + \size{u} + \size{p} = \size{up}$.
        \item Case $t = up \in \neutral \ \normal$. Then $u \in \neutral$ and $p \in \normal$. Therefore, there exists a tight derivation $\Phi_u \tr \seqi{\Gam_u}{u}{\tneutral}{(b_u,s_u)}$, such that $\size{u} = s_u$, by the \ih (\cref{lem:typ-nfs}.\ref{prop:typ-nfs:1}), and there exists a tight derivation $\Phi_p \tr \seqi{\Gam_p}{p}{\tightt}{(b_p,s_p)}$, such that $\size{p} = s_p$, by the \ih (\cref{lem:typ-nfs}.\ref{prop:typ-nfs:2}). Thus, we can build $\Phi$ as follows:
        \[ \begin{prooftree}
            \hypo{\Phi_u \tr \seqi{\Gam_u}{u}{\tneutral}{(b_u,s_u)}}
            \hypo{\Phi_p \tr \seqi{\Gam_p}{p}{\tightt}{(b_p,s_p)}}
            \infer2[(\ruleAppPOne)]{\seqi{\Gam_u + \Gam_p}{up}{\tneutral}{(b_u+b_p,1+s_u+s_p)}}
        \end{prooftree} \]
        And we can conclude with $\Gam = \Gam_u + \Gam_p$, $b = b_u+b_p$, and $s = 1 + s_u + s_p = 1 + \size{u} + \size{p} = \size{up}$.
    \end{itemize}
    \item Case $t \in \normal$. We want to show that there exists a tight derivation $\Phi \tr \seqi{\Gam}{t}{\tightt}{(b,s)}$:
    \begin{itemize}
        \item Case $t = x$. Then we can build $\Phi$ as follows:
        \[ \begin{prooftree}
            \infer0[(\ruleAx)]{\seqi{x : \mul{\sig}}{x}{\sig}{(0,0)}}
        \end{prooftree} \]
        by picking $\sig \in \{\tabs, \tvar\}$. And we can conclude with $\Gam = \eset$, $b = 0$, and $s = 0 = \size{x}$.
        \item Case $t = \lam x.u$. Then we can build $\Phi$ as follows:
        \[ \begin{prooftree}
            \infer0[(\ruleLamP)]{\seqi{}{\lam x.u}{\tabs}{(0,0)}}
        \end{prooftree} \]
        And we can conclude with $\Gam = \eset$, $b = 0$, and $s = 0 = \size{\lam x.u}$.
        \item The remaining cases are for when $t \in \neutral$, so they are subsumed by previous cases.
    \end{itemize}
\end{enumerate} 
}

\begin{lemma}[{\bf Merge for Values}]
    \label{lem:merge-values}
    Let $(\Phi^i_v \tr \seqi{\Gam_i}{v}{\M_i}{(b_i,s_i)})_{\iI}$. Then, there exists $\Phi_v \tr \seqi{\Gam}{v}{\M}{(b,s)}$, such that $\Gam = +_{\iI} \Gam_i$, $\M = +_{\iI} \M_i$, $b = +_{\iI} b_i$, and $s = +_{\iI}$.
\end{lemma}

\maybehide{\begin{proof}
    We start by noting that each $\Phi^i_v$ must end with the rule ($\ruleMany$). Therefore, for each $\iI$, we have $\Gam_i = +_{\kK_i} \Gam_k$, $\M_i = \mul{\sig_k}_{\kK_i}$, such that $b_i = +_{\kK_i} b_k$ and $s_i = +_{\kK_i} s_k$, and the following derivations $(\Phi^k_v \tr \seqi{\Gam_k}{v}{\sig_k}{(b_k,s_k)})_{\kK_i}$. Let $J = +_{\iI} K_i$ and $\M = \mul{\sig_j}_{\jJ} = \mul{\sig_k}_{\kK_i, \iI}$. We can use rule ($\ruleMany$) to build $\Phi_v \tr \seqi{\Gam}{v}{\M}{(+_{\jJ} b_j, +_{\jJ} s_j)}$. So we can conclude with $\Gam = +_{\jJ} \Gam_j = +_{\iI} (+_{\kK_i} \Gam_k) = +_{\iI} \Gam_i$, $b = +_{\jJ} b_j = +_{\iI} (+_{\kK_i} b_k) = +_{\iI} b_i$, and $s = +_{\jJ} s_j = +_{\iI} (+_{\kK_i} s_k) = +_{\iI} s_i$.
\end{proof}}

\subsubsection{Soundness and Completeness (Main Results)}

\begin{lemma}[{\bf Substitution and Anti-Substitution}]
    \label{lem:subsantisubs}
    \begin{enumerate} \mbox{}
        \item \label{lem:subs} Let $\Phi_t \tr \seqi{\Gam_t; x : \M}{t}{\tau}{(b_t,s_t)}$ and $\Phi_v \tr \seqi{\Gam_v}{v}{\M}{(b_v,s_v)}$, then there exists $\Phi_{t \subs{x}{v}} \tr \seqi{\Gam_t + \Gam_v}{t \subs{x}{v}}{\tau}{(b_t+b_v,s_t+s_v)}$.
        \item \label{lem:antisubs} Let $\Phi_{t \subs{x}{v}} \tr \seqi{\Gam_{t \subs{x}{v}}}{t \subs{x}{v}}{\tau}{(b,s)}$. Then, there exist $\Phi_t \tr \seqi{\Gam_t; x : \M}{t}{\tau}{(b_t,s_t)}$ and $\Phi_v \tr \seqi{\Gam_v}{v}{\M}{(b_v,s_v)}$, such that $\Gam_{t \subs{x}{v}} = \Gam_t + \Gam_v$, $b = b_t + b_v$, and $s = s_t + s_v$.
    \end{enumerate}
\end{lemma}

\maybehide{\begin{proof} \mbox{}
    \begin{enumerate}
        \item 
    The proof follows by induction over $\Phi_t$:
    \begin{itemize}
        \item Case $\Phi_t$ ends with rule (\ruleAx). Then $t$ must be a variable and we need to consider two cases:
        \begin{itemize}
            \item Assume $t = y = x$. Then $\Gam_t = \eset$, $\tau = \M$, $t \subs{x}{v} = v$, $b_t = 0$, and $s_t = 0$. So we can take $\Phi_{t \subs{x}{v}} = \Phi_v$ and conclude with $\Gam_t + \Gam_v = \Gam_v$, $b_t + b_v = b_v$, and $s_t + s_v = s_v$.
            \item Assume $t = y \not= x$. Then $\M = \emul$, $\Gam_v = \eset$, $t \subs{x}{v} = t$, $b_v = 0$, and $s_v = 0$. So we can take $\Phi_{t \subs{x}{v}} = \Phi_t$ and conclude with $\Gam_t + \Gam_v = \Gam_t$, $b_t + b_v = b_t$, and $s_t + s_v = s_t$.
        \end{itemize}
        \item Case $\Phi_t$ ends with rule (\ruleLam). Then $t$ must be of the form $\lam y.u$ and $\Phi_t$ must be of the following form (by $\alpha$-conversion):
        \[ \begin{prooftree}
            \hypo{\Phi_u \tr \seqi{\Gam; x : \M}{u}{\tau'}{(b_t,s_t)}}
            \infer1[(\ruleLam)]{\seqi{(\Gam \sm y); x : \M}{\lam y.u}{\Gam(y) \ta \tau'}{(b_t, s_t)}}
        \end{prooftree} \]
        where $\tau = \Gam(y) \ta \tau'$ and $\Gam_t = (\Gam \sm y)$. By the \ih, we have the following derivation $\Phi_{u \subs{x}{v}} \tr \seqi{\Gam + \Gam_v}{u \subs{x}{v}}{\tau}{(b_t + b_v, s_t + s_v)}$. Therefore, we can construct $\Phi_{t \subs{x}{v}}$ as follows:
        \[ \begin{prooftree}
            \hypo{\Phi_{u \subs{x}{v}} \tr \seqi{\Gam + \Gam_v}{u \subs{x}{v}}{\tau'}{(b_t + b_v, s_t + s_v)}}
            \infer1[(\ruleLam)]{\seqi{(\Gam + \Gam_v) \sm y}{(\lam y.u) \subs{x}{v}}{\Gam(y) \ta \tau'}{(b_t + b_v, s_t + s_v)}}
        \end{prooftree} \]
        And we can conclude with $(\Gam + \Gam_v) \sm y = (\Gam \sm y) + \Gam_v = \Gam_t + \Gam_v$, by $\alpha$-conversion.
        \item Case $\Phi_t$ ends with rule ($\ruleApp$). Then $t$ must be of the form $up$ and $\Phi_t$ must be of the following form:
        \[ \begin{prooftree}
            \hypo{\Phi_u \tr \seqi{\Gam; x : \M_1}{u}{\M' \ta \tau}{(b_u, s_u)}}
            \hypo{\Phi_p \tr \seqi{\Del; x : \M_2}{p}{\M'}{(b_p,s_p)}}
            \infer2[(\ruleApp)]{\seqi{(\Gam + \Del); x : \M_1 \sqcup \M_2}{up}{\tau}{(1+b_u+b_p, s_u+s_p)}}
        \end{prooftree} \]
        where $\Gam_t = (\Gam + \Del)$, $\M = \M_1 \sqcup \M_2$, $b_t = 1 + b_u + b_p$, and $s_t = s_u + s_p$. By~\cref{lem:split-values}, we know there exist the following derivations $(\Phi^i_v \tr \seqi{\Gam^i_v}{v}{\M_i}{(b_i,s_i)})_{i \in \{1,2\}}$, such that $\Gam_v = \Gam^1_v + \Gam^2_v$, $b_v = b_1 + b_2$, and $s_v = s_1 + s_2$. By the \ih, we know there exist $\Phi_{u \subs{x}{v}} \tr \seqi{\Gam + \Gam^1_v}{u \subs{x}{v}}{\M' \ta \tau}{(b_u+b_1, s_u+s_1)}$ and $\Phi_{p \subs{x}{v}} \tr \seqi{\Del + \Gam^2_v}{p \subs{x}{v}}{\M'}{(b_p + b_2, s_p + s_2)}$. So we can construct $\Phi_{t \subs{x}{v}}$ as follows:
        \[ \begin{prooftree}
            \hypo{\Phi_{u \subs{x}{v}} \tr \seqi{\Gam + \Gam^1_v}{u \subs{x}{v}}{\M' \ta \tau}{(b_u+b_1, s_u+s_1)}}
            \hypo{\Phi_{p \subs{x}{v}} \tr \seqi{\Del + \Gam^2_v}{p \subs{x}{v}}{\M'}{(b_p+b_2,s_p+s_2)}}
            \infer2[(\ruleApp)]{\seqi{(\Gam + \Del) + (\Gam^1_v + \Gam^2_v)}{(u p) \subs{x}{v}}{\tau}{(1+b_u+b_p+b_1+b_2, s_u + s_p + s_1 + s_2)}}
        \end{prooftree} \]
        And we can conclude with $\Gam_t + \Gam_v = (\Gam + \Del) + (\Gam^1_v + \Gam^2_v)$, $b_t + b_v = 1 + b_u + b_p + b_1 + b_2$, and $s_t + s_v = s_u + s_p + s_1 + s_2$.
        \item Case $\Phi_t$ ends with rule ($\ruleMany$). Then $t$ must be of the form $w$ and $\Phi$ must be of the following form:
        \[ \begin{prooftree}
            \hypo{(\Phi^i_w \tr \seqi{\Gam_i; x : \M_i}{w}{\sig_i}{(b_i,s_i)})_{\iI}}
            \infer1[(\ruleMany)]{\seqi{+_{\iI} \Gam_i; x : \sqcup_{\iI} \M_i}{w}{\mul{\sig_i}_{\iI}}{(+_{\iI} b_i, +_{\iI} s_i)}}
        \end{prooftree} \]
        where $\tau = \mul{\sig_i}_{\iI}$, $\Gam_t = +_{\iI} \Gam_i$, $b_t = +_{\iI} b_i$, and $s_t = +_{\iI} s_i$. By~\cref{lem:split-values}, we have the following derivations $(\Phi^i_v \tr \seqi{\Gam^i_v}{v}{\M_i}{(b^i_v, s^i_v)})_{\iI}$, such that $\Gam_v = +_{\iI} \Gam^i_v$, $b_v = +_{\iI} b^i_v$, and $s_v = +_{\iI} s^i_v$. By the \ih over each $\Phi^i_w$, we have $(\Phi^i_{w \subs{x}{v}} \tr \seqi{\Gam_i + \Gam^i_v}{w \subs{x}{v}}{\sig_i}{(b_i + b^i_v, s_i + s^i_v)})_{\iI}$. Therefore, we can construct $\Phi_{t \subs{x}{v}}$ as follows:
        \[ \begin{prooftree}
            \hypo{(\Phi^i_{w \subs{x}{v}} \tr \seqi{\Gam_i + \Gam^i_v}{w \subs{x}{v}}{\sig_i}{(b_i + b^i_v, s_i + s^i_v)})_{\iI}}
            \infer1[(\ruleMany)]{\seqi{+_{\iI} (\Gam_i + \Gam^i_v)}{w \subs{x}{v}}{\mul{\sig_i}_{\iI}}{(+_{\iI} (b_i + b^i_v), +_{\iI} (s_i + s^i_v))}}
        \end{prooftree} \]
        And we can conclude with $\Gam_t + \Gam_v = +_{\iI} \Gam_i +_{\iI} \Gam^i_v = +_{\iI} (\Gam_i + \Gam^i_v)$, $b_t + b_v = +_{\iI} b_i +_{\iI} b^i_v = +_{\iI} (b_i + b^i_v)$, and $s_t + s_v = +_{\iI} s_i +_{\iI} s^i_v = +_{\iI} (s_i + s^i_v)$.
        \item Case $\Phi_t$ ends with rule (\ruleLamP). Then $t$ must be of the form $\lam y.u$, $\Gam_t = \eset$, $\tau = \tabs$, $\M = \emul$, $\Gam_v = \eset$, $t \subs{x}{v} = \lam y.(u \subs{x}{v}) = (\lam y.u) \subs{x}{v}$, $b_t = b_v = 0$, and $s_t = s_v = 0$. So we can construct $\Phi_{t \subs{x}{v}}$ as follows:
        \[ \begin{prooftree}
            \infer0[(\ruleLamP)]{\seqi{}{(\lam y.u) \subs{x}{v}}{\tabs}{(0,0)}}
        \end{prooftree} \]
        And conclude with $\Gam_t + \Gam_v = \eset$, $b_t + b_v = 0$, and $s_t + s_v = 0$.
        \item Case $\Phi_t$ ends with rule (\ruleAppPOne) or (\ruleAppPTwo), the proof is very similar to when $\Phi_t$ ends with rule (\ruleApp).
    \end{itemize}

        \item 
    The proof follows by induction over $t$:
    \begin{itemize}
        \item Case $t = y$. Then we have to consider two cases:
        \begin{itemize}
            \item Case $t = y \not= x$. Then, $t \subs{x}{v} = y$. Let $\Gam_v = \eset$, $\M = \emul$, $b_v = 0$, and $s_v = 0$. Then, $\Phi_v$ is derivable using rule ($\ruleMany$). We also take $\Phi_t = \Phi_{t \subs{x}{v}}$, so that, in particular $\Gam_t = \Gam_{t \subs{x}{v}}$. Then, we conclude with $\Gam_{t \subs{x}{v}} = \Gam_t + \Gam_v = \Gam_t$, $b = b_t + b_v = b_t$, and $s = s_t + s_v = s_t$.
            \item Case $t = y = x$. Then, $t \subs{x}{v} = v$. Let $\Gam_t = \eset$, $b_t = 0$, and $s_t = 0$. Now, we have to consider two cases depending on the last rule used in $\Phi_{t \subs{x}{v}}$: 
            \begin{itemize}
                \item Case $\Phi_{t \subs{x}{v}}$ ends with rule ($\ruleAx$), then $\tau = \sig$. Let $\Gam_v = \Gam_{t \subs{x}{v}}$, $\M = \mul{\sig}$, $b_v = b$, and $s_v = s$. Then, we can build derivation $\Phi_v$ as follows:
                \[ \begin{prooftree}
                    \hypo{\Phi_{t \subs{x}{v}} \tr \seqi{\Gam_{t \subs{x}{v}}}{v}{\sig}{(b,s)}}
                    \infer1[(\ruleMany)]{\seqi{\Gam_{t \subs{x}{v}}}{v}{\mul{\sig}}{(b,s)}}
                \end{prooftree} \]
                Let $\Gam_t = \eset$, $b_t = 0$, and $s_t = 0$. Then, $\Phi_t \tr \seqi{x : \mul{\sig}}{x}{\sig}{(0,0)}$ is given by rule ($\ruleAx$). So we can conclude with $\Gam_{t \subs{x}{v}} = \Gam_v = \Gam_t + \Gam_v$, $b = b_v = b_t + b_v$, and $s = s_v = s_t + s_v$.
                \item Case $\Phi_{t \subs{x}{v}}$ ends with rule ($\ruleMany$), then $\tau = \mul{\sig_i}_{\iI}$, for some $I$. Let $\Gam_t = \eset$, and $\M = \mul{\sig_i}_{\iI}$. Then, we can build $\Phi_t$ as follows:
                \[ \begin{prooftree}
                    \infer0[(\ruleAx)]{(\seqi{x : \mul{\sig_i}}{x}{\sig_i}{(0,0)})_{\iI}}
                    \infer1[(\ruleMany)]{\seqi{x : \mul{\sig_i}_{\iI}}{x}{\mul{\sig_i}_{\iI}}{(0,0)}}
                \end{prooftree} \] 
                Then, we can take $\Phi_v = \Phi_{t \subs{x}{v}}$, so that $\Gam_v = \Gam_{t \subs{x}{v}}$, $b_v = b$, and $s_v = s$. And we can conclude $\Gam_{t \subs{x}{v}} = \Gam_v = \Gam_t + \Gam_v$, $b = b_v = b_t + b_v$, and $s = s_v = s_t + s_v$.
            \end{itemize}
        \end{itemize}
        \item Case $t = \lam y.u$. Then $t \subs{x}{v} = (\lam y.u) \subs{x}{v} = \lam y.(u \subs{x}{v})$ and we have to consider three cases:
        \begin{itemize}
            \item Case $\Phi_{t \subs{x}{v}}$ ends with rule (\ruleLam), then it must be of the following form:
            \[ \begin{prooftree}
                \hypo{\Phi_{u \subs{x}{v}} \tr \seqi{\Gam_{u \subs{x}{v}}; y : \M'}{u \subs{x}{v}}{\tau'}{(b, s)}}
                \infer1[(\ruleLam)]{\seqi{\Gam_{u \subs{x}{v}}}{\lam y.(u \subs{x}{v})}{\M' \ta \tau'}{(b, s)}}
            \end{prooftree} \]
            where $\tau = \M' \ta \tau'$, and $\Gam_{t \subs{x}{v}} = \Gam_{u \subs{x}{v}}$. By the \ih, we have the following derivations $\Phi_u \tr \seqi{\Gam_u; y: \M'; x : \M}{u}{\del}{(b_u, s_u)}$ and $\Phi_v \tr \seqi{\Gam_v}{v}{\M}{(b_v, s_v)}$, such that $\Gam_{u \subs{x}{v}} = \Gam_u + \Gam_v$, $b = b_u + b_v$, and $s = s_u + s_v$. And we can build $\Phi_{\lam y.u}$ as follows:
            \[ \begin{prooftree}
                \hypo{\Phi_u \tr \seqi{\Gam_u; y : \M'; x : \M}{u}{\tau'}{(b_u, s_u)}}
                \infer1[(\ruleLam)]{\seqi{\Gam_u; x : \M}{\lam y.u}{\M' \ta \tau'}{(b_u, s_u)}}
            \end{prooftree} \]
            So we can pick $\Phi_t = \Phi_{\lam y.u}$, and conclude with $\Gam_{t \subs{x}{v}} = \Gam_{u \subs{x}{v}} = \Gam_u + \Gam_v$, $b = b_u + b_v$, and $s = s_u + s_v$.
            \item Case $\Phi_{t \subs{x}{v}}$ ends with rule (\ruleLamP), then is must be of the following form:
            \[ \begin{prooftree}
                \infer0[(\ruleLamP)]{\seqi{}{\lam y.(u \subs{x}{v})}{\tabs}{(0,0)}}
            \end{prooftree} \]
            where $\tau = \tabs$, $\Gam_{t \subs{x}{v}} = \eset$, $b = 0$, and $s = 0$. Let $\Gam_t = \eset$, $\M = \emul$, $b_t = 0$, and $s_t = 0$. Then, we can build $\Phi_t$ as follows:
            \[ \begin{prooftree}
                \infer0[(\ruleLamP)]{\seqi{}{\lam y.u}{\tabs}{(0,0)}}
            \end{prooftree} \]
            Let $\Gam_v = \eset$, $b_v = 0$, and $s_v = 0$. Then $\Phi_v$ can be constructed by using rule (\ruleMany) with no premises. So we can conclude with $\Gam_{t \subs{x}{v}} = \eset = \Gam_t + \Gam_v$, and $b = 0 = b_t + b_v$, and $s = 0 = s_t + s_v$.
            \item Case $\Phi_{t \subs{x}{v}}$ ends with rule ($\ruleMany$). Then $t \subs{x}{v}$ and $t$ are values, and $\Phi_{t \subs{x}{v}}$ must be of the following form:
            \[ \begin{prooftree}
                \hypo{(\Phi_i \tr \seqi{\Gam_i}{t \subs{x}{v}}{\sig_i}{(b_i,s_i)})_{\iI}}
                \infer1[(\ruleMany)]{\seqi{+_{\iI} \Gam_i}{t \subs{x}{v}}{\mul{\sig_i}_{\iI}}{(+_{\iI} b_i, +_{\iI} s_i)}}
            \end{prooftree} \]
            where $\tau = \mul{\sig_i}_{\iI}$, $\Gam_{t \subs{x}{v}} = +_{\iI} \Gam_i$, $b = +_{\iI} b_i$, and $s = +_{\iI} s_i$. By the \ih over each $\Phi_i$, we have the following derivations $\Phi^i_t \tr \seqi{\Gam^i_t; x : \M_i}{t}{\sig_i}{(b^i_t, s^i_t)}$ and $\Phi^i_v \tr \seqi{\Gam^i_v}{v}{\M_i}{(b^i_v, s^i_v)}$, such that $\Gam_i = \Gam^i_t + \Gam^i_v$, $b_i = b^i_t + b^i_v$, and $s_i = s^i_t + s^i_v$,for each $\iI$. So we can build $\Phi_t$ as follows:
            \[ \begin{prooftree}
                \hypo{(\Phi^i_t \tr \seqi{\Gam^i_t; x : \M_i}{t}{\sig_i}{(b^i_t, s^i_t)})_{\iI}}
                \infer1[(\ruleMany)]{\seqi{+_{\iI} \Gam^i_t; x : \sqcup_{\iI} \M_i}{t}{\mul{\sig_i}_{\iI}}{(+_{\iI} b^i_t, +_{\iI} s^i_t)}}
            \end{prooftree} \]
            such that $\Gam_t = +_{\iI} \Gam^i_t$, $\M = \sqcup_{\iI} \M_i$, $b_t = +_{\iI} b^i_t$, and $s_t = +_{\iI} s^i_t$. By~\cref{lem:merge-values}, we can take the following derivation $\Phi_v \tr \seqi{+_{\iI} \Gam^i_v}{v}{\M}{(+_{\iI} b^i_v, +_{\iI} s^i_v)}$. And we can conclude with $\Gam_{t \subs{x}{v}} = +_{\iI} \Gam_i = +_{\iI} (\Gam^i_t + \Gam^i_v) = +_{\iI} \Gam^i_t +_{\iI} \Gam^i_v = \Gam_t + \Gam_v$, $b = +_{\iI} b_i = +_{\iI} (b^i_t + b^i_v) = +_{\iI} b^i_t +_{\iI} b^i_v = b_t + b_v$, and $s = +_{\iI} s_i = +_{\iI} (s^i_t + s^i_v) = +_{\iI} s^i_t +_{\iI} s^i_v = s_t + s_v$.
        \end{itemize}
        \item Case $t = up$. Then $t \subs{x}{v} = (u \subs{x}{v}) (p \subs{x}{v})$ and we have to consider three cases:
        \begin{itemize}
            \item Case $\Phi_{t \subs{x}{v}}$ ends with ($\ruleApp$), then it must be of the following form:
            \[ \begin{prooftree}
                \hypo{\Phi_{u \subs{x}{v}} \tr \seqi{\Gam_{u \subs{x}{v}}}{u \subs{x}{v}}{\M' \ta \tau}{(b', s')}}
                \hypo{\Phi_{p \subs{x}{v}} \tr \seqi{\Gam_{p \subs{x}{v}}}{p \subs{x}{v}}{\M'}{(b'', s'')}}
                \infer2[(\ruleApp)]{\seqi{\Gam_{u \subs{x}{v}} + \Gam_{p \subs{x}{v}}}{(u \subs{x}{v})(p \subs{x}{v})}{\tau}{(1+b'+b'', s'+s'')}}
            \end{prooftree} \]
            where $\Gam_{t \subs{x}{v}} = \Gam_{u \subs{x}{v}} + \Gam_{p \subs{x}{v}}$, $b = 1+b'+b''$, and $s = s' + s''$. By the \ih over $\Phi_{u \subs{x}{v}}$, we have the following derivations $\Phi_u \tr \seqi{\Gam_u; x : \M_1}{u}{\M' \ta \tau}{(b_u,s_u)}$ and $\Phi^1_v \tr \seqi{\Gam^1_v}{v}{\M_1}{(b^1_v,s^1_v)}$, such that $\Gam_{u \subs{x}{v}} = \Gam_u + \Gam^1_v$, $b' = b_u + b^1_v$, and $s' = s_u + s^1_v$. And by the \ih over $\Phi_{p \subs{x}{v}}$, we have the following derivation $\Phi_{p} \tr \seqi{\Gam_p; x : \M_2}{p}{\M'}{(b_p,s_p)}$ and $\Phi^2_v \tr \seqi{\Gam^2_v}{v}{\M_2}{(b^2_v,s^2_v)}$, such that $\Gam_{p \subs{x}{v}} = \Gam_p + \Gam^2_v$, $b'' = b_p + b^2_v$, and $s'' = s_p + s^2_v$. By~\cref{lem:merge-values}, we can take the following derivation $\Phi_v \tr \seqi{\Gam^1_v + \Gam^2_v}{v}{\M_1 \sqcup \M_2}{(b^1_v+b^2_v, s^1_v + s^2_v)}$, such that $\Gam_v = \Gam^1_v + \Gam^2_v$, $b_v = b^1_v + b^2_v$, and $s_v = s^1_v + s^2_v$. And we can build $\Phi_{up}$ as follows:
            \[ \begin{prooftree}
                \hypo{\Phi_u \tr \seqi{\Gam_u; x : \mul{\sig_i}_{\iI_1}}{u}{\M' \ta \tau}{(b_u,s_u)}}
                \hypo{\Phi_{p} \tr \seqi{\Gam_p; x : \mul{\sig_i}_{\iI_2}}{p}{\M'}{(b_p,s_p)}}
                \infer2[(\ruleApp)]{\seqi{(\Gam_u + \Gam_p); x : \mul{\sig_i}_{\iI}}{up}{\tau}{(1+b_u+b_p, s_u+s_p)}}
            \end{prooftree} \]
            such that $\Gam_t = \Gam_u + \Gam_p$, $b_t = 1+b_u + b_p$, and $s_t = s_u + s_p$. So we can pick $\Phi_t = \Phi_{up}$, and conclude with $\Gam_{t \subs{x}{v}} = \Gam_{u \subs{x}{v}} + \Gam_{p \subs{x}{v}} = \Gam_u + \Gam^1_v + \Gam_p + \Gam^2_v = (\Gam_u + \Gam_p) + (\Gam^1_v + \Gam^2_v) = \Gam_t + \Gam_v$, $b = 1+b'+b'' = 1+b_u + b^1_v + b_p + b^2_v = 1 + (b_u + b_p) + (b^1_v + b^2_v) = b_t + b_v$, and $s = s_u + s^1_v + s_p + s^2_v = (s_u + s_p) + (s^1_v + s^2_v) = s_t + s_v$.
            \item Case $\Phi_{t \subs{x}{v}}$ ends with (\ruleAppPOne) and (\ruleAppPTwo). These cases are very similar to the case where $\Phi_{t \subs{x}{v}}$ ends with rule (\ruleApp).
        \end{itemize}
    \end{itemize}
    \end{enumerate}
\end{proof}}

\begin{lemma}[{\bf Split Exact Subject Reduction and Expansion}]
    \label{lem:subjred-subjexp} \mbox{}
    \begin{enumerate} 
        \item \label{lem:subj-red} Let $\Phi_t \tr \seqi{\Gam}{t}{\tau}{(b,s)}$ be tight. If $t \dred t'$, then there exists $\Phi_{t'} \tr \seqi{\Gam}{t'}{\tau}{(b-1,s)}$.
        \item \label{lem:subj-exp} Let $\Phi_{t'} \tr \seqi{\Gam}{t'}{\tau}{(b,s)}$ be tight. If $t \dred t'$, then there exists $\Phi_t \tr \seqi{\Gam}{t}{\tau}{(b+1, s)}$.
    \end{enumerate}
\end{lemma}

\maybehide{\begin{proof} \mbox{}
    \begin{enumerate}
        \item 
    We will actually prove the following stronger version of the statement, which allows us to reason inductively:

    Let $\Phi_t \tr \seqi{\Gam}{t}{\tau}{(b,s)}$, such that $\Gam$ is tight, and either $\tau$ is tight or $\neg\isvalue{t}$. If $t \dred t'$, then there exists $\Phi_{t'} \tr \seqi{\Gam}{t'}{\tau}{(b-1,s)}$.

    The proof now follows by induction over $\dred$:
    \begin{itemize}
        \item Case $t = (\lam x.u) v \dred u \subs{x}{v} = t'$.Assume that $\Phi_t$ ends with rule (\ruleAppPOne). Then $\lam x.u$ must be assigned type $\nott{\tabs}$, which is not possible by~\cref{lem:notabs-implies-negabs}. Now, assume that $\Phi_t$ ends with rule (\ruleAppPTwo). Then $v$ must be assigned typed $\tneutral$, which is not possible by~\cref{lem:values-not-neutral}. Therefore, $\Phi_t$ must be of the following form:
        \[ \begin{prooftree}
            \hypo{\Phi_u \tr \seqi{\Gam_u; x : \M}{u}{\tau}{(b_u,s_u)}}
            \infer1[(\ruleLam)]{\seqi{\Gam_u}{(\lam x.u)}{\M \ta \tau}{(b_u, s_u)}}
            \hypo{\Phi_{v} \tr \seqi{\Gam_v}{v}{\M}{(b_v,s_v)}}
            \infer2[(\ruleApp)]{\seqi{\Gam_u + \Gam_{v}}{(\lam x.u) v}{\tau}{(1+b_u+b_v, s_u+s_v)}}
        \end{prooftree} \]
        where $\tau \in \tightt$, $\Gam = \Gam_u + \Gam_v$ is tight, $b = 1 + b_u + b_v$, and $s = s_u + s_v$. By~\cref{lem:subsantisubs}.\ref{lem:subs}, we know there exists the following derivation $\Phi_{u \subs{x}{v}} \tr \seqi{\Gam_u + \Gam_v}{u \subs{x}{v}}{\tau}{(b_u+b_v,s_u+s_v)}$. So we can take $\Phi_{t'} = \Phi_{u \subs{x}{v}}$ and conclude with $b - 1 = b_u + b_v$.
        \item Case $t = up \dred u'p = t'$, such that $u \dred u'$. Then $\Phi_t$ must either end with (\ruleApp), (\ruleAppPOne), or (\ruleAppPTwo):
        \begin{itemize}
            \item Case $\Phi_t$ ends with rule (\ruleApp), then it must be of the following form:
            \[ \begin{prooftree}
                \hypo{\Phi_u \tr \seqi{\Gam_u}{u}{\M \ta \tau}{(b_u,s_u)}}
                \hypo{\Phi_p \tr \seqi{\Gam_p}{p}{\M}{(b_p,s_p)}}
                \infer2[(\ruleApp)]{\seqi{\Gam_u + \Gam_p}{up}{\tau}{(1 +b_u+b_p,s_u+s_p)}}
            \end{prooftree} \]
            where $\tau = \tau \in \tightt$, $\Gam = \Gam_u + \Gam_p$ is tight, $b = 1+b_u + b_p$, and $s = s_u + s_p$. Since $u \dred u'$, it is clear that $\neg\isvalue{u}$ holds. Moreover, $\Gam_u$ is necessarily tight. Therefore, by the \ih, there exists $\Phi_{u'} \tr \seqi{\Gam_u}{u'}{\M \ta \tau}{(b_u-1, s_u)}$. Thus, we can build $\Phi_{t'}$ as follows:
            \[ \begin{prooftree}
                \hypo{\Phi_{u'} \tr \seqi{\Gam_u}{u'}{\M \ta \tau}{(b_u-1, s_u)}}
                \hypo{\Phi_p \tr \seqi{\Gam_p}{p}{\M}{(b_p,s_p)}}
                \infer2[(\ruleApp)]{\seqi{\Gam_u + \Gam_p}{u'p}{\tau}{(b_u+b_p,s_u+s_p)}}
            \end{prooftree} \]
            And we can conclude with $b - 1= b_u + b_p$.
            \item Case $\Phi_t$ ends with rule (\ruleAppPOne) or (\ruleAppPTwo), the proof are similar to the one where $\Phi_t$ ends with rule (\ruleApp).
        \end{itemize}
        \item Case $t = up \dred up' = t'$, such that $u \not\dred$ and $p \dred p'$. Then $\Phi_t$ must either end with (\ruleApp), (\ruleAppPOne), or (\ruleAppPTwo):
        \begin{itemize}
            \item Case $\Phi_t$ ends with rule (\ruleApp), then it must be of the following form:
            \[ \begin{prooftree}
                \hypo{\Phi_u \tr \seqi{\Gam_u}{u}{\M \ta \tau}{(b_u,s_u)}}
                \hypo{\Phi_p \tr \seqi{\Gam_p}{p}{\M}{(b_p,s_p)}}
                \infer2[(\ruleApp)]{\seqi{\Gam_u + \Gam_p}{up}{\tau}{(1+b_u+b_p,s_u+s_p)}}
            \end{prooftree} \]
            where $\tau \in \tightt$, $\Gam = \Gam_u + \Gam_p$ is tight, $b = 1 + b_u + b_p$, and $s = s_u + s_p$. Since $p \dred p'$, it is clear that $\neg\isvalue{p}$. Moreover, $\Gam_p$ is necessarily tight. Therefore, by the \ih, we know there exists the following derivation $\Phi_{p'} \tr \seqi{\Gam_p}{p'}{\M}{(b_p-1, s_p)}$. Thus, we can build $\Phi_{t'}$ as follows:
            \[ \begin{prooftree}
                \hypo{\Phi_u \tr \seqi{\Gam_u}{u}{\M \ta \tau}{(b_u, s_u)}}
                \hypo{\Phi_{p'} \tr \seqi{\Gam_p}{p'}{\M}{(b_p-1,s_p)}}
                \infer2[(\ruleApp)]{\seqi{\Gam_u + \Gam_p}{up'}{\tau}{(b_u+b_p,s_u+s_p)}}
            \end{prooftree} \]
            And we can conclude with $b - 1 = b_u + b_p$.
            \item Case $\Phi_t$ ends with rule (\ruleAppPOne) or (\ruleAppPTwo), the proofs are similar to the ones where $\Phi_t$ ends with rule (\ruleApp).
        \end{itemize}
    \end{itemize}

        \item 
    Just like for~\cref{lem:subjred-subjexp}.\ref{lem:subj-red}, we will actually prove the following stronger version of the statement, which allows us to reason inductively:

    Let $\Phi_{t'} \tr \seqi{\Gam}{t'}{\tau}{(b,s)}$, such that $\Gam$ is tight, and either ($\tau \in \tightt$ or $\neg\isvalue{t}$). If $t \dred t'$, then there exists $\Phi_t \tr \seqi{\Gam}{t}{\tau}{(b+1,s)}$.
    
    The proof now follows by induction over $\dred$:
    \begin{itemize}
        \item Case $t = (\lam x.u) v \dred u \subs{x}{v} = t'$. Then $\Phi_{t'} \tr \seqi{\Gam}{u \subs{x}{v}}{\tau}{(b,s)}$ and, by~\cref{lem:subsantisubs}.\ref{lem:antisubs}, there exist the following derivations $\Phi_u \tr \seqi{\Gam_u; x : \M}{u}{\tau}{(b_u, s_u)}$ and $\Phi_v \tr \seqi{\Gam_v}{v}{\M}{(b_v,s_v)}$, such that $\tau \in \tightt$, $\Gam = \Gam_u + \Gam_v$ is tight, $b = b_u + b_v$, and $s = s_u + s_v$. So we can build $\Phi_t$ as follows:
        \[ \begin{prooftree}
            \hypo{\Phi_u \tr \seqi{\Gam_u; x : \M}{u}{\tau}{(b_u, s_u)}}
            \infer1[(\ruleLam)]{\seqi{\Gam_u}{\lam x.u}{\M \ta \tau}{(b_u,s_u)}}
            \hypo{\Phi_v \tr \seqi{\Gam_v}{v}{\M}{(b_v,s_v)}}
            \infer2[(\ruleApp)]{\seqi{\Gam_u + \Gam_v}{(\lam x.u)v}{\tau}{(1+b_u+b_v, s_u+s_v)}}
        \end{prooftree} \]
        And we can conclude with $b + 1 = 1 + b_u + b_v$.
        \item Case $t = up \dred u'p = t'$, such that $u \dred u'$. Then $\Phi_{t'}$ must either end with (\ruleApp), (\ruleAppPOne), or (\ruleAppPTwo):
        \begin{itemize}
            \item Case $\Phi_{t'}$ ends with rule (\ruleApp), then it must be of the following form:
            \[ \begin{prooftree}
                \hypo{\Phi_{u'} \tr \seqi{\Gam_u}{u'}{\M' \ta \tau}{(b_u, s_u)}}
                \hypo{\Phi_p \tr \seqi{\Gam_p}{p}{\M'}{(b_p, s_p)}}
                \infer2[(\ruleApp)]{\seqi{\Gam_u + \Gam_p}{u'p}{\tau}{(1 + b_u + b_p, s_u + s_p)}}
            \end{prooftree} \]
            where $\tau \in \tightt$, $\Gam = \Gam_u + \Gam_p$ it tight, $b = 1 + b_u + b_p$, and $s = s_u + s_p$. Since $u \dred u'$, it is clear that $\neg\isvalue{u}$. Moreover, $\Gam_p$ is tight. Therefore, by the \ih, there exists the following derivation $\Phi_u \tr \seqi{\Gam_u}{u}{\M' \ta \tau}{(b_u + 1, s_u)}$. Thus, we can build $\Phi_{t'}$ as follows:
            \[ \begin{prooftree}
                \hypo{\Phi_u \tr \seqi{\Gam_u}{u}{\M' \ta \tau}{(b_u + 1, s_u)}}
                \hypo{\Phi_p \tr \seqi{\Gam_p}{p}{\M'}{(b_p, s_p)}}
                \infer2[(\ruleApp)]{\seqi{\Gam_u + \Gam_p}{up}{\tau}{(1 + b_u + 1 + b_p, s_u + s_p)}}
            \end{prooftree} \]
            And we can conclude with $b + 1 = (1 + b_u + b_p) + 1 = 1 + b_u + 1 + b_p$.
            \item Case $\Phi_{t'}$ ends with rule (\ruleAppPOne) or (\ruleAppPTwo), the proofs are similar to the one where $\Phi_{t'}$ ends with rule (\ruleApp).
        \end{itemize}
        \item Case $t = up \dred up' = t'$, such that $p \dred p'$. Then $\Phi_{t'}$ must either ends with (\ruleApp), (\ruleAppPOne), or (\ruleAppPTwo):
        \begin{itemize}
            \item Case $\Phi_{t'}$ ends with rule ($\ruleApp$), then it must be of the following form:
            \[ \begin{prooftree}
                \hypo{\Phi_u \tr \seqi{\Gam_u}{u}{\M' \ta \tau}{(b_u, s_u)}}
                \hypo{\Phi_{p'} \tr \seqi{\Gam_p}{p'}{\M'}{(b_p, s_p)}}
                \infer2[(\ruleApp)]{\seqi{\Gam_u + \Gam_p}{u p'}{\tau}{(1 + b_u + b_p, s_u + s_p)}}
            \end{prooftree} \]
            where $\tau \in \tightt$, $\Gam = \Gam_u + \Gam_{p'}$ is tight, $b = 1 + b_u + b_p$, $s_t = s_u + s_p$. Since $p \dred p'$, it is clear that $\neg\isvalue{p}$ holds. Moreover, $\Gam_p$ is tight. Therefore, by the \ih, we have the following derivation $\Phi_p \tr \seqi{\Gam_p}{p}{\M' \ta \tau}{(b_p + 1, s_p)}$. Thus, we can build $\Phi_{t'}$ as follows:
            \[ \begin{prooftree}
                \hypo{\Phi_u \tr \seqi{\Gam}{u}{\M' \ta \tau}{(b_u, s_u)}}
                \hypo{\Phi_p \tr \seqi{\Gam_p}{p}{\M'}{(b_p + 1, s_p)}}
                \infer2[(\ruleApp)]{\seqi{\Gam_u + \Gam_p}{up}{\tau}{(1 + b_u + b_p + 1, s_u + s_p)}}
            \end{prooftree} \]
            And we can conclude with $b + 1 = (1 + b_u + b_p) + 1 = 1 + b_u + b_p + 1$.
            \item Case $\Phi_{t'}$ ends with rule (\ruleAppPOne) or (\ruleAppPTwo), the proofs are similar to the one where $\Phi_{t'}$ ends with rule (\ruleApp).
        \end{itemize}   
    \end{itemize}
    \end{enumerate}
\end{proof}}

\begin{theorem}[{\bf Quantitative Soundness and Completeness}]
    \label{thm:soundnesscompleteness}
  \item \label{thm:soundness} If $\Phi \tr \seqi{\Gam}{t}{\tau}{(b,s)}$ is tight, then there exists $u \in \normal$ such that  $t \drred^b u$ with $\size{u} = s$.
  \item \label{thm:completeness} If $t \drred^b u$ with  $u \in \normal$, then there exists a tight type derivation $\Phi_t \tr \seqi{\Gam}{t}{\tau}{(b, \size{u})}$.
\end{theorem}

\maybehide{\begin{proof} \mbox{}
    \begin{enumerate} 
        \item 
    The proof follows by induction over $b$:
    \begin{itemize}
        \item Case $b = 0$. Then $t \in \normal$, by~\cref{lem:zero-steps-nfs}. And $d = \size{t}$, by~\cref{lem:corr-size-counter}. So we can conclude with $u = t$.
        \item Case $b > 0$. Then $t \not\in \normal$, by~\cref{lem:zero-steps-nfs}. Therefore, there exists $t'$ such that $t \dred t'$, by~\cref{prop:char-nfs}. By\cref{lem:subjred-subjexp}.\ref{lem:subj-red}, there exists $\Phi_{t'} \tr \seqi{\Gam}{t'}{\tau}{(b-1, s)}$. By the \ih, there exists $u \in \normal$, such that $t' \drred^{b-1} u$, such that $d = \size{u}$. So we can conclude with $t \dred t' \drred^{b-1} u$, which means that $t \drred^b u$, as expected.
    \end{itemize}
        \item 
    The proof follows by induction over $b$:
    \begin{itemize}
        \item Case $b = 0$. Then $t = u$, which means that $t \in \normal$. Therefore, we can conclude by~\cref{lem:typ-nfs}.
        \item Case $b > 0$. Then there exists $t'$, such that $t \dred t' \drred^{b-1} u$. By the \ih, there exists a tight derivation $\Phi_{t'} \tr \seqi{\Gam}{t'}{\tau}{(b-1, \size{u})}$. By\cref{lem:subjred-subjexp}.\ref{lem:subj-exp}, there exists a tight derivation $\Phi \tr \seqi{\Gam}{t}{\tau}{(b, \size{u})}$. So, we can conclude.
    \end{itemize}
    \end{enumerate}
\end{proof}}

\subsection{A \texorpdfstring{$\lambda$}{Lambda}-Calculus with Global State}

\subsubsection{General Lemmas}

\propnormalifffinal*

\maybehide{\begin{proof}
    \begin{itemize}
    \item[$\Ra$)] Let $(t, s)$ be \final. We consider two cases:
      \begin{itemize}
            \item Case $(t,s)$ is blocked. We reason by induction on blocked configurations. \begin{itemize}
                \item Case $(t,s) = (\get{l}{x}{u}, s)$, such that $l \not\in \dom{s}$. Then $(t, s) \not\ra$ is straightforward.
                \item Case $(t,s) = (v u, s)$ and $(u,s)$ is blocked.
                  Then by the \ih, we have that $(u,s) \not\ra$. Therefore, $(v u, s) \not\ra$ holds.
            \end{itemize}
          \item Case $t \in \normal$. We reason by induction on
            $\normal$. \begin{itemize}
                \item Case $t=v \in \val$. Then $(v,s) \not\ra$  is straightforward.
                \item Case $t \in \neutral$. Then $t = v u$ and we have to consider two different  cases: \begin{itemize}
                  \item Case  $v= x$ and $u \in \normal$. Then by the \ih, we have $(u,s) \not\ra$. Therefore, $(v u, s) \not\ra$ holds.
                    \item Case $v = (\lam x.p)$ and $u \in \neutral$. Then $u \in \normal$, and by the \ih, we have that $(u,s) \not\ra$. Therefore $(v u,s) \not\ra$ holds.
                \end{itemize}
            \end{itemize}
        \end{itemize}
      \item[$\La$)] Let $t \not \ra$. We reason by
        induction on $t$: \begin{itemize}
            \item Case $t = v$. Then $t \in \normal$. Therefore $(t,s)$ is \final.
            \item Case $t = v u$. Since $(v u, s) \not\ra$, then $(u,s) \not\ra$. By the \ih, we have $(u,s)$ \final. Now, we reason
              by cases: \begin{itemize}
                \item Case $(u, s)$ is blocked. Then, $(v u, s)$ is blocked by definition. 
                \item Case $u \in \normal$. Then we have two cases: \begin{itemize}
                    \item Case $u \in \neutral$. Then $vu \in \normal$. Therefore,  $(t,s)$ is \final.
                    \item Case $u \in \val$ and $v = \lam x.p$. Then $((\lam x.p) u, s) \ra (p \subs{x}{u}, s)$, which yields a contradiction with the hypothesis $t=vu\not\ra$. Thus, this case does not apply.
                \end{itemize}
            \end{itemize}
          \item Case $t = \get{l}{x}{u}$. Since $(\get{l}{x}{u},s) \not\ra$, then $l \not\in \dom{s}$. Therefore, $(\get{l}{x}{u},s)$ is blocked, which implies
$(t,s)$ is \final. 
            \item Case $t = \set{l}{v}{u}$. Then $(\set{l}{v}{u}, s) \ra (u, \upd{l}{v}{s})$, which yields to a contraction with the hypothesis  $t\not\ra$. 
              Therefore, this case does not apply.
        \end{itemize}
    \end{itemize}
\end{proof}
} 

\proptypedunblock*

\maybehide{\begin{proof}
    By induction on $t$: \begin{itemize}
        \item Case $t \in \val $ or $t = \set{l}{v}{t}$. Then the conclusion trivially holds, since clearly $(t,s)$ is not a blocked configuration.
        \item Case $t = \get{l}{x}{t}$. We have two  cases: \begin{itemize}
            \item Case $l \in \dom{s}$. Then $(t,s)$ is clearly unblocked.
            \item Case $l \not\in \dom{s}$. Let $\stype_0 = \conj{(l : \Gam(x))} \splus \stype$. Since $t = \get{l}{x}{u}$, then $\Phi$ must be of the following form:
            \[ \begin{prooftree}
                \hypo{\Phi \tr \seqi{\Gam_u \sm x}{\get{l}{x}{u}}{\comptype{
                    \stype_0}{\ctype}}{(b_u,m_u,d_u)}}
                \hypo{\Phi_s \tr \seqi{\Del}{s}{\stype_0}{(b_s,m_s,d_s)}}
                \infer2[(\ruleConf)]{\seqi{(\Gam_u \sm x) + \Del}{(\get{l}{x}{t}, s)}{\ctype}{(b_u+b_s,1+m_u+m_s,d_u+d_s)}}
              \end{prooftree} \] 
              where $\Gam = \Gam_u \sm x$, $b = b_u+b_s$, $m = 1+m_u+m_s$, and $d = d_u + d_s$. Thus, $l \in \dom{\conj{(l : \Gam_u(x))} \splus \stype}$, and so by
            \cref{lem:states-and-state-types} we have 
          $l \in \dom{s}$, which gives a contradiction with the hypothesis $l \not\in \dom{s}$. Therefore, this case does not apply,
        \end{itemize}
        \item Case $t = v u$. Assume $\Phi_v \tr \seqi{\Gam_v}{v}{\M \ta (\comptype{\stype'}{\ctype})}{(b_v,m_v,d_v)}$ and $\Phi_u \tr \seqi{\Gam_u}{u}{\tcomptype{\stype}{\M}{\stype'}}{(b_u,m_u,d_u)}$. Then $\Phi$ must be of the following form:
        \[ \begin{prooftree}
            \hypo{\Phi_v}
            \hypo{\Phi_u}
            \infer2[(\ruleApp)]{\seqi{\Gam_v + \Gam_u}{v u}{\comptype{\stype}{\ctype}}{(1+b_v+b_u,m_v+m_u,d_v+d_u)}}
            \hypo{\Phi_s \tr \seqi{\Del}{s}{\stype}{(b_s,m_s,d_s)}}
            \infer2[(\ruleConf)]{\seqi{(\Gam_v + \Gam_u) + \Del}{(v u, s)}{\ctype}{(1+b_v + b_u + b_s, m_v + m_u + m_s, d_v + d_u + d_s)}}
        \end{prooftree} \]
        where $\Gam = (\Gam_v + \Gam_u) + \Del$, $b = 1+b_v + b_u + b_s$, $m = m_v + m_u + m_s$, and $d = d_v + d_u + d_s$. Thus, we can build the following derivation for $(u,s)$:
        \[ \begin{prooftree}
            \hypo{\Phi_u \tr \seqi{\Gam_u}{u}{\tcomptype{\stype}{\M}{\stype'}}{(b_u,m_u,d_u)}}
            \hypo{\Phi_s \tr \seqi{\Del}{s}{\stype}{(b_s,m_s,d_s)}}
            \infer2[(\ruleConf)]{\seqi{\Gam + \Del}{(u,s)}{\conftype{\M}{\stype'}}{(b_u+b_s,m_u+m_s,d_u+d_s)}}
        \end{prooftree} \]
        By the \ih, we have that $(u,s)$ is unblocked. Therefore, $(v u, s)$ also unblocked.
    \end{itemize}
\end{proof}} 

\begin{lemma}[Relevance]
    Let $\Phi \tr \seqi{\Gam}{t}{\gtype}{(b,m,d)}$ (resp. $\Phi' \tr \seqi{\Gam}{s}{\stype}{(b',m',d')}$). Then $\dom{\Gam} \subseteq \fv{t}$ (resp. $\dom{\Gam} \subseteq \fv{s}$).
\end{lemma}

\maybehide{\begin{proof}
    The proof following by induction over $\Phi$ (resp. $\Phi'$). Case $\Phi$ (resp. $\Phi'$) ends with rule (\ruleAx) or (\ruleLamP) (resp. rule (\ruleEmp)), then $\Phi$ (resp. $\Phi'$) is clearly relevant. The other cases follow easily from the \ih.
\end{proof}}

\subsubsection{Soundness Lemmas (Auxiliary Lemmas)}

\lemzerocounters*

\maybehide{\begin{proof} \mbox{}
    \begin{enumerate}
        \item 
  We replace the statement by the following three ones.
  \begin{enumerate}
    \item[(1.1)] If $\del = \tcomptype{\stype}{\tneutral}{\stype'}$, then $t \in \neutral$.
    \item[(1.2)] If $\del = \tcomptype{\stype}{\tightt}{\stype'}$, then $t \in \normal$.
    \item[(2)] $d = \size{t}$.
  \end{enumerate}
  We reason by simultaneous induction on tight derivations.
  \begin{itemize}
    \item Case $\Phi$ ends with (\ruleAx). This case does not apply since the resulting type is not a monadic type.
    \item Case $\Phi$ ends with (\ruleLam). This case does not apply since the resulting type is not a monadic type.
    \item Case $\Phi$ ends with (\ruleApp). Then the first counter in the conclusion of the derivation is necessarily greater  than $0$ and thus this case does not apply. 
    \item Case $\Phi$ ends with (\ruleMany). This case does not apply since the resulting type is not a monadic type.
    \item Case $\Phi$ ends with (\ruleLift). Then we have to consider three cases, depending on the form of $\mu$: 
      \begin{itemize} 
        \item If $\mu = \M$, then $\Phi$ cannot be tight. 
        \item If $\mu = \tvar$, then $t = x$, and $s = 0$. The condition of case (1.1) is not possible by construction. In case (1.2) we can conclude $x \in \val \subseteq \normal$. The statement (2) $d=0 = \size{x}$ is straightforward. 
        \item If $\mu = \tabs$, then $t = \lam x.u$ and $d = 0$. so that $t=\lam x.u$ and $d=0$. The condition of case (1.1) is not possible by construction. In case (1.2) we can conclude $\lam x.u \in \val \subseteq \normal$. The statement (2) $d=0 = \size{\lam x.u}$ is straightforward.
      \end{itemize}
    \item Case $\Phi$ ends with (\ruleGet). Then the second counter in the conclusion of the derivation is necessarily greater than $0$ and thus this case does not apply. 
    \item Case $\Phi$ ends with (\ruleSet). Then the second counter in the conclusion of the derivation is necessarily greater than $0$ and thus this case does not apply.
  
    \item Case $\Phi$ ends with (\ruleLamP). This case does not apply since the resulting type is not a monadic type.
    \item Case $\Phi$ ends with (\ruleAppPOne), so that $t = xu$ and $d = 1+d'$. If the condition of case (1.1) holds for $t$, that means that the condition of case (1.2) holds for $u$. By the \ih (1.2) $u \in \normal$ so that $xu \in \neutral$. The condition of case (1.2) holds for $t$ only for  $\tightt=\tneutral$, and then $xu \in \neutral$ holds by case (1.1), which implies $xu \in \normal$ by definition. To show statement (2), we apply the \ih (2) to $u$ and obtain $d'=\size{u}$, then $d=1+d' = 1 + \size{u} = \size{t}$.
    \item Case $\Phi$ ends with (\ruleAppPTwo), so that $t = (\lam x. p)u$ and $d = 1+d'$. If the condition of case (1.1) holds for $t$, that means that
    the condition of case (1.1) holds for $u$. By the \ih (1.1) $u \in \neutral$ so that $t \in \neutral$. The condition of case (1.2) holds for $t$ only for  $\tightt=\tneutral$, and then $t \in \neutral$ holds by case (1.1), which implies $t  \in \normal$ by definition. To show statement (3), we apply the \ih (3) to $u$ and obtain $d'=\size{u}$, then $d=1+d' = 1 + \size{u} = \size{t}$.
  \end{itemize}
        \item     By induction over $\Phi$: \begin{itemize}
        \item Case $\Phi$ ends with ($\ruleEmp$). Then it must be of the following form:
        \[ \begin{prooftree}
            \infer0[(\ruleEmp)]{\seqi{}{\estate}{\eset}{(0,0,0)}}
        \end{prooftree} \]
        where $d = 0$. So we can conclude.
        \item Case $\Phi$ ends with ($\ruleUpd$). Then it must be of the following form:
        \[ \begin{prooftree}
            \hypo{\Phi_v \tr \seqi{\Gam_v}{v}{\M}{(0,0,d_v)}}
            \hypo{\Phi_q \tr \seqi{\Del_q}{q}{\stype_q}{(0,0,d_q)}}
            \infer2[(\ruleUpd)]{\seqi{\Gam_v + \Del_q}{\upd{l}{v}{q}}{\conj{l : \M}; \stype_q}{(0,0,d_v+d_q)}}
        \end{prooftree} \]
        where $d = d_v + d_q$. Then $\Phi_v$ must be of the form:  
        \[ \begin{prooftree}
            \hypo{(\seqi{\Gam_v^i}{v}{\rdel_i}{(0,0,d_v^i)})_{\iI}}
            \infer1[(\ruleMany)]{\seqi{+_{\iI} \Gam_v^i}{v}{\mul{\rdel_i}_{\iI}}{(0, 0, +_{\iI} d_v^i)}}
        \end{prooftree} \]
        where $\Gam_v =+_{\iI} \Gam_v^i$, $d_v = +_{\iI} d_v^i$, and $\M = \mul{\rdel_i}_{\iI}$. Given that $\conj{l : \M}; \stype_q$ is tight, then $\tightp{\M}$, and so  $\rdel_i$ is tight, for each $\iI$. Then by point (1) of Lemma~\ref{lem:zero-counters}, $\size {v}=d_v^i$ for each $\iI$. But since $v\in \val$ then its size is $0$, which means $d_v^i=0$ for each $\iI$, therefore $d_v = +_{\iI} d_v^i = 0$.  Furthermore, $d_q = 0$, by the \ih\  Therefore we can conclude $d = d_v + d_q = +_{\iI} d^i_v + 0 = 0 + 0 = 0$.
    \end{itemize}

    \end{enumerate}
\end{proof}} 

\begin{lemma}
    \label{lem:zero-counters-normal}
    Let $\Phi \tr \seqi{\Gam}{t}{\del}{(0,0,d)}$ be tight. If $t \in \normal$, then $\del = \stype \ra \tightt \tim \stype'$ and $\stype =\stype'$.
\end{lemma}

\maybehide{\begin{proof} 
  By induction on $t \in \normal$. We consider two cases:
  \begin{itemize}
    \item Case $t \in \val$. Then such a typing derivation can only end with rule (\ruleAx) followed by rule (\ruleLift) or (\ruleLamP)followed by rule (\ruleLift), in which cases the statement is obvious.
    \item Case $t = vu \in \neutral$. Since the first counter of the derivation is $0$, $\Phi$ can only end with a persistent rule (\ruleAppPOne) or (\ruleAppPTwo). In both cases, we can conclude by applying the \ih to $u \in \normal$ or $u \in \neutral$ and their type derivations, which gives  $\stype = \stype'$.
  \end{itemize}
\end{proof}} 

\lemzeronfs*

\maybehide{\begin{proof}\
  \begin{itemize}
    \item[$\Ra$)] By point (1) of~\cref{lem:zero-counters}.
    \item[$\La$)] By induction on $t$: \begin{itemize}
    \item Case $t \in \val$. There are six cases to consider for $\Phi$:
    \begin{itemize}    
      \item $\Phi$ ends with (\ruleAx). This case does not apply since the resulting type is not a monadic type. 
      \item $\Phi$ ends with (\ruleLam). This case does not apply since the resulting type is not a monadic type.
      \item $\Phi$ ends with (\ruleMany). This case does not apply since the resulting type is not a monadic type.
      \item $\Phi$ ends with (\ruleLift). Then we have to consider three cases, depending on the form of $\mu$: 
      \begin{itemize} 
        \item If $\mu = \M$, this case does not apply, since $\del = \tcomptype{\stype}{\M}{\stype'}$, but $\M \not\in \tightt$. 
        \item If $\mu = \tvar$, then the premise of $\Phi$ must end with rule (\ruleAx). Therefore, $\Phi \tr \seqi{x:\mul{\tvar}}{x}{\tcomptype{\stype}{\tvar}{\stype}}{(0,0,0)}$, with $\stype$ tight, and the conclusion holds trivially.
        \item If $\mu = \tabs$, then the premise of $\Phi$ must end with rule (\ruleLamP). Therefore, $\Phi \tr \seqi{x:\mul{\nott{\tneutral}}}{x}{\tcomptype{\stype}{\nott{\tneutral}}{\stype}}{(0,0,0)}$, with $\stype$ tight, and the conclusion holds trivially.
      \end{itemize}

      \item $\Phi$ ends with (\ruleLamP). This case does not apply since the resulting type is not a monadic type.
    \end{itemize}
    \item Case $t = xu$. Then $u \in \normal$, by definition and there are two cases to consider for $\Phi$:
    \begin{itemize}
      \item If $\Phi$ ends with (\ruleApp). Then $\Phi_u \tr \seqi{\Gam_u}{u}{\tcomptype{\stype}{\M}{\stype'}}{(b_u,m_u,d_u)}$, $\Phi_x \tr \seqi{x : \M \ta (\comptype{\stype'}{\ctype})}{x}{\M \ta (\comptype{\stype'}{\ctype})}{(b_x,m_x,d_x)}$, such that $\Gam =  (x:\mul{\M \ta (\comptype{\stype'}{\ctype})}) + \Gam_u$ is tight. Absurd, since $\M \ta (\comptype{\stype'}{\ctype})$ is not tight, therefore this case does not apply.
      \item If $\Phi$ ends with (\ruleAppPOne). Then $\Phi_u \tr \seqi{\Gam_u}{u}{\tcomptype{\stype}{\tightt}{\stype'}}{(b_u,m_u,d_u)}$, such that $\Gam = (x: \mul{\tvar})+\Gam_u$ is tight, $b = b_u$, $m =m_u$, $d = d_u+ 1$, and $\stype'$ is tight. By the \ih\ on $u$, we have $b_u=m_u=0$, therefore $b = m = 0$.
      \end{itemize}
      \item Case $t = (\lam x.p) u$. Then $u \in \neutral$, by definition and there are two cases to consider for $\Phi$:
      \begin{itemize}
        \item If $\Phi$ ends with (\ruleApp). Then $\Phi_u \tr \seqi{\Gam_u}{u}{\tcomptype{\stype}{\M}{\stype'}}{(b_u,m_u,d_u)}$, $\Phi_{\lam x.p} \tr \seqi{\Gam_{\lam x.p}}{\lam x.p}{\M \ta (\comptype{\stype'}{\ctype})}{(b_p,m_p,d_p)}$, such that $\Gam = \Gam_u + \Gam_{\lambda x.p}$ is tight, $b = 1+b_l+b_u$, $m = m_l+m_u$, $d = d_l+ d_m$. Since $\Gam_u$ is tight and $u\in\neutral$, by~\cref{lem:comp-tight-spreading}, $\M \in \tightt$, which is absurd. Therefore, this case does not apply.
        \item If $\Phi$ ends with (\ruleAppPTwo). Then $\Phi_u \tr \seqi{\Gam_u}{u}{\tcomptype{\stype}{\tneutral}{\stype'}}{(b_u,m_u,d_u)}$, such that $\Gam = \Gam_u$ is tight, $b = b_u$, $m=m_u$, $d = d_u+ 1$ and $\stype'$ is tight. By the \ih\ on $u$, we have $b_u=m_u=0$. Therefore $b = m = 0$.
      \end{itemize}
    \end{itemize}
  \end{itemize}
\end{proof}
}

\begin{lemma}
    \label{lem:states-and-state-types}
    Let $\Phi \tr \seqi{\Del}{s}{\stype}{(b,m,d)}$. If $l \in \dom{\stype}$, then $l \in \dom{s}$.
\end{lemma}
  
\maybehide{\begin{proof}
    We proceed by proving the following stronger version of the statement: 
    
    Let $\Phi_s \tr \seqi{\Del_s}{s}{\stype_s}{(b_s,m_s,d_s)}$. If $l \in \dom{\stype_s}$, then $s \equivstate \upd{l}{v}{q}$, for some value $v$ and store $q$.
    
    The proof follows by induction on $\Phi_s$: 
    \begin{itemize}
        \item Case $\Phi_s$ ends with ($\ruleEmp$). Then the conclusion is vacuously true.
        \item Case $\Phi_s$ ends with ($\ruleUpd$). Then $\Phi_s$ is of the following form: 
        \[ \begin{prooftree}
            \hypo{\Phi_v \tr \seqi{\Gam_v}{v}{\M}{(b_v,m_v,d_v)}}
            \hypo{\Phi_q \tr \seqi{\Del_q}{q}{\stype_q}{(b_q,m_q,d_q)}}
            \infer2[(\ruleUpd)]{\seqi{\Gam_v + \Del_q}{\upd{l'}{v}{q}}{\conj{l' : \M}; \stype_q}{(b_v+b_q,m_v+m_q,d_v+d_q)}}
        \end{prooftree} \]
        where $\Del_s = \Gam_v + \Del_q$, $s = \upd{l'}{v}{q}$, $\stype_s = \conj{l' : \M}; \stype_q$, $b_s = b_v + b_q$, $m_s = m_v + m_q$, and $d_s = d_v + d_q$. Now we consider two  cases: 
        \begin{itemize}
            \item Case $l = l'$. Then we are done.
            \item Case $l \not= l'$. Since we are assuming that $l \in \dom{\stype_s}$, then it must be case that $l \in \dom{\stype_q}$. But, then by the \ih, we have $q \equivstate \upd{l}{w}{q'}$, for some value $w$ and store $q'$. Therefore, $s \equivstate \upd{l'}{v}{\upd{l}{w}{q'}} \equivstate \upd{l}{w}{\upd{l'}{v}{q'}}$.
        \end{itemize}
    \end{itemize}
    The correctness of the original statement now follows easily from the fact that, clearly, if $s \equivstate \upd{l}{v}{q}$, then $l \in \dom{s}$, by Definition~\ref{def:domainS}.
\end{proof}
} 

\begin{lemma}[{\bf Split Lemma}] \mbox{} 
    \label{lem:split-values-stores}
    \begin{enumerate}
        \item {(\bf Values)} \label{lem:com-split-values}  Let $\Phi_v \tr \seqi{\Gam}{v}{\M}{(b,m,d)}$, such that $\M = \sqcup_{\iI} \M_i$. Then, there exist ($\Phi^i_v \tr \seqi{\Gam_i}{v}{\M_i}{(b_i,m_i,d_i)})_{\iI}$, such that $\Gam = +_{\iI} \Gam_i$, $b = +_{\iI} b_i$, $m = +_{\iI} m_i$, and $d = +_{\iI} d_i$.
        \item {\bf (States)} \label{lem:split-state} Let $\Phi_s \tr \seqi{\Gam}{s}{\stype}{(b,m,d)}$, such that $l \in \dom{\stype}$. Then, $s \equivstate \upd{l}{v}{q}$, $\Phi_v \tr \seqi{\Gam_v}{v}{\stype(l)}{(b_v,m_v,d_v)}$ and $\Phi_q \tr \seqi{\Gam_q}{q}{\stype'}{(b_q,m_q,d_q)}$, such that $\Gam = \Gam_v + \Gam_q$, $\stype = \conj{(l : \stype(l))}; \stype'$, $b = b_v+b_q$, $m = m_v+m_q$, and $d = d_v + d_q$.
    \end{enumerate}
\end{lemma}

\maybehide{\begin{proof}
    The proof for values is very similar to the corresponding proof for $\lam_s$, so we are only going to show the split lemma for states.
    The proof follows by induction on the structure of $s$: \begin{itemize}
        \item Case $s = \estate$. Then the statement is vacuously true.
        \item Case $s = \upd{l'}{w}{q'}$. Then $\Phi_s$ is of the form: 
        \[ \begin{prooftree}
            \hypo{\Phi_{w} \tr \seqi{\Gam_{w}}{w}{\M}{(b_w,m_w,d_w)}}
            \hypo{\Phi_{q'} \tr \seqi{\Gam_{q'}}{q'}{\stype_{q'}}{(b_{q'}, m_{q'},d_{q'})}}
            \infer2[(\ruleUpd)]{\seqi{\Gam_{w} + \Gam_{q'}}{\upd{l'}{w}{q'}}{\conj{(l' : \M)}; \stype_{q'}}{(b_w+b_{q'},m_w+m_{q'},d_w+d_{q'})}}
        \end{prooftree} \] where $\Gam = \Gam_{w} + \Gam_{q'}$, $\stype = \conj{(l' : \M)}; \stype_{q'}$, $b = b_w + b_{q'}$, $m = m_w + m_{q'}$, and $d = d_w + d_{q'}$. 
        We  consider two cases: \begin{itemize}
            \item Case $l' = l$. Then we simply take $v = w$ and $q = q'$ and we are done.
            \item Case $l' \not= l$.  Since $l \in \dom{\conj{(l' : \M)}; \stype_{q'}}$ and $l' \not= l$, then $l \in \dom{\stype_{q'}}$. By applying the \ih to $q'$, we have that  $q' \equivstate \upd{l}{w'}{q''}$, $\Phi_{w'} \tr \seqi{\Gam_{w'}}{w'}{\stype_{q'}(l)}{(b_{w'},m_{w'},d_{w'})}$ and $\Phi_{q''} \tr \seqi{\Gam_{q''}}{q''}{\stype_{q''}}{(b_{q''},m_{q''},d_{q''})}$, such that $\Gam_{q'} = \Gam_{w'} + \Gam_{q''}$, $\stype_{q'} = \conj{(l : \stype_{q'}(l))}; \stype_{q''}$, $b_{q'} = b_{w'} + b_{q''}$, $m_{q'} = m_{w'} + m_{q''}$, and $d_{q'} = d_{w'} + d_{q''}$. But $s = \upd{l'}{w}{\upd{l}{w'}{q''}} \equivstate \upd{l}{w'}{\upd{l'}{w}{q''}}$, so we can take $v = w'$, $q = \upd{l'}{w}{q''}$, and consider $\Phi_q$ to be the following derivation:
            \[ \begin{prooftree}
                \hypo{\Phi_{w} \tr \seqi{\Gam_{w}}{w}{\M}{(b_w,m_w,d_w)}}
                \hypo{\Phi_{q''} \tr \seqi{\Gam_{q''}}{q''}{\stype_{q''}}{(b_{q''}, m_{q''}, d_{q''})}}
                \infer2[(\ruleUpd)]{\seqi{\Gam_{w} + \Gam_{q''}}{\upd{l'}{w}{q''}}{\conj{(l' : \M)}; \stype_{q''}}{(b_w+b_{q''}, m_w + m_{q''}, d_w+d_{q''})}}
            \end{prooftree} \] where  $\Gam_q = \Gam_{w} + \Gam_{q''}$ and $\stype_q=\conj{(l' : \M)}; \stype_{q''}$. We can then conclude with the following observations:
            \begin{itemize}
            \item $\Gam_v + \Gam_q = \Gam_{w'} +\Gam_{w} + \Gam_{q''} =
              \Gam_{w} + \Gam_{q'} = \Gam$,
                \item Since $\stype = \conj{(l' : \M)}; \stype_{q'}$ and $l' \not= l$, then $\stype(l) = \stype_{q'}(l)$ and
                \begin{align*}
                    \stype = \conj{(l' : \M)}; \stype_{q'} & = \conj{(l': \M)}; \conj{(l : \stype_{q'}(l))}; \stype_{q''} \\
                    & = \conj{(l : \stype_{q'}(l))}; \stype_{q} \\
                    & = \conj{(l : \stype(l))}; \stype_q
                \end{align*}
              \item $b_v + b_q= b_{w'} + b_{w} + b_{q''}=
                 b_w + b_{q'} = b$, $m_v + m_q= m_{w'} + m_{w} + m_{q''}=
                 m_w + m_{q'} = b$ and
                 $d_v + d_q= d_{w'} + d_{w} + d_{q''}=
                 d_w + d_{q'} = d$.
            \end{itemize} 
        \end{itemize}
    \end{itemize}
\end{proof}
} 

\begin{lemma}
    \label{lem:comp-values-not-neutral}
    Let $\Phi \tr \seqi{\Gam}{t}{\tcomptype{\stype}{\tau}{\stype'}}{(b,m,d)}$. If $t \in \val$, then $\tau \neq \tneutral$.
\end{lemma}

\maybehide{\begin{proof}
    By case analysis on the form of $t \in \val$:
    \begin{itemize}
        \item Case $t = x$. Then we have to consider three cases according to the last rule used in $\Phi$:
        \begin{itemize}
            \item Case $\Phi$ ends with rule (\ruleAx), then $t$ can only be assigned $\sig$. Therefore, this case does not apply.
            \item Case $\Phi$ ends with rule (\ruleMany), then $\tau = \M \neq \tneutral$.
        
            \item Case $\Phi$ ends with rule (\ruleLift). Then $\tau \in \{\tvar, \tabs, \M\}$, which means that $\tau \not= \tneutral$.
        \end{itemize}
        \item Case $t = \lam x.t$. Then we have to consider three cases according to the last rule used in $\Phi$:
        \begin{itemize}
            \item Case $\Phi$ ends with rule (\ruleLam), then $t$ can only be assigned $\sig$. Therefore, this case does not apply.
            \item Case $\Phi$ ends with rule (\ruleMany), then $\tau = \M  \neq \tneutral$.
            \item Case $\Phi$ ends with rule (\ruleLamP), then $\tau = \vl$. Therefore, this case does not apply.
            \item Case $\Phi$ ends with rule (\ruleLift). $\tau \in \{\tabs, \M\}$, which means that $\tau \not= \tneutral$.
        \end{itemize}
    \end{itemize}
\end{proof}} 

\begin{lemma}
    \label{lem:comp-notabs-implies-negabs}
    Let $\Phi \tr \seqi{\Gam}{t}{\tcomptype{\stype}{\tau}{\stype'}}{(b,m,d)}$, such that $\Gam$ is tight. If $\tau \in \nott{\vl}$, then $\neg\isabs{t}$.
\end{lemma}

\maybehide{\begin{proof}
    By induction over $\Phi$:
    \begin{itemize}
        \item Case $\Phi$ ends with rule (\ruleAx), (\ruleLam), (\ruleLamP), or (\ruleMany). These cases do not apply since they do not end with a monadic type.
        \item Case $\Phi$ ends with rule (\ruleApp), (\ruleGet), (\ruleSet), (\ruleAppPOne), or (\ruleAppPTwo), then $\neg\isabs{t}$ holds by definition.
        \item Case $\Phi$ ends with rule (\ruleLift), then we have to consider three cases, depending on the form of $\mu$:
        \begin{itemize}
            \item If $\mu = \M$, then $\tau \in \nott{\tabs}$ does not hold.
            \item If $\mu = \tvar$, then $t = x$. Therefore, $\neg\isabs{x}$ holds by definition.
            \item If $\mu = \tabs$, then $\tau \in \nott{\tabs}$ does not hold.
        \end{itemize}
    \end{itemize}
\end{proof}} 

\subsubsection{Completeness (Auxiliary Lemmas)}

\begin{lemma}[{\bf Merge for Values}]
    \label{lem:comp-merge-values}
    Let $(\Phi^i_v \tr \seqi{\Gam_i}{v}{\M_i}{(b_i,m_i,d_i)})_{\iI}$. Then, there exists $\Phi_v \tr \seqi{\Gam}{v}{\M}{(b,m,d)}$, such that $\Gam = +_{\iI} \Gam_i$, $\M = +_{\iI} \M_i$, $b = +_{\iI} b_i$, $m = +_{\iI} m_i$, and $d = +_{\iI}$.
\end{lemma}

We omit this proof given its similarity with the proof for system $\syscbv$.

\lemcomtightspreading*

\maybehide{\begin{proof}
  We want to show that, if $t \in \neutral$, then $\tau \in \tightt$, for some $\stype'$. We proceed by induction on the predicate  $t \in \neutral$:
    \begin{itemize}
        \item Case $t = xu$, such that $u \in \normal$. Then we have to consider the following two cases depending on the last rule in $\Phi$:
        \begin{itemize}
            \item Case $\Phi$ ends with rule ($\ruleApp$), then it must be of the following form:
            \[ \begin{prooftree}
                \infer0[(\ruleAx)]{\seqi{x : \mul{\M \ta (\comptype{\stype'}{\ctype})}}{x}{\M \ta (\comptype{\stype'}{\ctype})}{(0,0,0)}}
                \hypo{\Phi_u \tr \seqi{\Gam_u}{u}{\tcomptype{\stype}{\M}{\stype'}}{(b_u,m_u,d_u)}}
                \infer2[(\ruleApp)]{\seqi{(x : \mul{\M \ta (\comptype{\stype'}{\ctype})}) + \Gam_u}{xu}{\comptype{\stype}{\ctype}}{(1+b_u,m_u,d_u)}}
            \end{prooftree} \]
            where $\Gam = (x : \mul{\M \ta (\comptype{\stype'}{\ctype})}) + \Gam_p$ is tight, $b = 1+b_u$, $m = m_u$, and $d = d_u$. But $\M \ta (\comptype{\stype'}{\ctype}) \not\in \tightt$, therefore $\Gam$ is not tight and we have a contraction. Thus, this case does not apply.
            \item Case $\Phi$ ends with rule (\ruleAppPOne), then $\tau = \tneutral \in \tightt$, so we can conclude immediately.
        \end{itemize}
        \item Case $t = (\lambda x.p)u$, such that $u \in \neutral$. Then we have to consider the following two cases depending on the last rule in $\Phi$:
        \begin{itemize}
            \item Case $\Phi$ ends with rule ($\ruleApp$), then it must be of the following form:
            \[ \begin{prooftree}
                \hypo{\seqi{\Gam_p}{\lam x.p}{\M \ta (\comptype{\stype'}{\ctype})}{(b_p,m_p,d_p)}}
                \hypo{\Phi_u \tr \seqi{\Gam_u}{u}{\tcomptype{\stype}{\M}{\stype'}}{(b_u,m_u,d_u)}}
                \infer2[(\ruleApp)]{\seqi{\Gam_p + \Gam_u}{(\lam x.p)u}{\comptype{\stype}{\ctype}}{(1+b_p+b_u,m_p+m_u,d_p+d_u)}}
            \end{prooftree} \]
            where $\Gam = \Gam_u + \Gam_p$ is tight, $b = 1 + b_p + b_u$, $m = m_p + m_u$, and $d = d_p + d_u$. By the \ih on $u$, we have that $\M \in \tightt$, which is a contradiction. Therefore, this case does not apply.
            \item Case $\Phi$ ends with rule (\ruleAppPTwo). Then $\tau = \tneutral \in \tightt$, so we can conclude immediately.
        \end{itemize}
    \end{itemize}
\end{proof}
} 

\typstates*

\maybehide{\begin{proof} \mbox{}
    \begin{enumerate}
        \item 
    By induction on $s$: \begin{itemize}
        \item Case $s = \estate$. Then we can build $\Phi_s$ as follows:
        \[ \begin{prooftree}
            \infer0[(\ruleAx)]{\seqi{}{\estate}{\eset}{(0,0,0)}}
        \end{prooftree} \]
        And we can conclude with $\stype = \eset$ tight.
        \item Case $s = \upd{l}{v}{q}$. By the \ih, there exists $\Phi_q \tr \seqi{\Del_q}{q}{\stype_q}{(0,0,0)}$ tight. Therefore, we can build $\Phi_s$ as follows:
        \[ \begin{prooftree}
            \infer0[(\ruleMany)]{\seqi{}{v}{\emul}{(0,0,0)}}
            \hypo{\Phi_q \tr \seqi{}{q}{\stype_q}{(0,0,0)}}
            \infer2[(\ruleUpd)]{\seqi{}{\upd{l}{v}{q}}{\conj{(l : \emul)}; \stype_q}{(0,0,0)}}
        \end{prooftree} \]
        And we can conclude with $\stype = \conj{(l : \emul)}; \stype_q$ tight.
    \end{itemize}
        \item 
    By simultaneous induction on the following claims: \begin{enumerate}
        \item \label{lem:typ-comp-nfs:1} If $t \in \neutral$, then $\Phi \tr \seqi{\Gam}{t}{\stype \ra \tneutral \tim \stype}{(0,0,d)}$ tight, such that $d = \size{t}$, for any tight $\stype$.
        \item \label{lem:typ-comp-nfs:2} If $t \in \normal$, then $\Phi \tr \seqi{\Gam}{t}{\stype \ra \tightt \tim \stype}{(0,0,d)}$ tight, such that $d = \size{t}$, for any tight $\stype$.
    \end{enumerate}
    \begin{enumerate}
        \item By induction on $t \in \neutral$: \begin{itemize}
            \item Case $t = xu$, such that $u \in \normal$. By the \ih (\cref{lem:typestatesnfs}.\ref{lem:typ-comp-nfs:2}), we have $\Phi_u \tr \seqi{\Gam_u}{u}{\tcomptype{\stype}{\tightt}{\stype}}{(0,0,d_u)}$ tight, such that $d_u = \size{u}$, for any tight $\stype$. Therefore, we can build $\Phi$ is as follows:
            \[ \begin{prooftree}
                \hypo{\seqi{\Gam_u}{u}{\tcomptype{\stype}{\tightt}{\stype}}{(0,0,d_u)}}
                \infer1[(\ruleAppPOne)]{\seqi{(x : \mul{\tvar}) + \Gam_u}{xu}{\tcomptype{\stype}{\tneutral}{\stype}}{(0,0,1+d_u)}}
            \end{prooftree} \]
            And we can conclude with $\Gam = (x : \mul{\tvar}) + \Gam_u$ tight and $d = 1+d_u = 1 + \size{u} = 1 + \size{x} + \size{u} = \size{xu}$.
            \item Case $t = (\lam x.p)u$, such that $u \in \neutral$. By the \ih (\cref{lem:typestatesnfs}.\ref{lem:typ-comp-nfs:1}), we have $\Phi_u \tr \seqi{\Gam}{u}{\tcomptype{\stype}{\tneutral}{\stype}}{(0,0,d_u)}$ tight, such that $d_u = \size{u}$, for any tight $\stype$. Therefore, we can build $\Phi$ is as follows:
            \[ \begin{prooftree}
                \hypo{\Phi_u \tr \seqi{\Gam}{u}{\tcomptype{\stype}{\tneutral}{\stype}}{(0,0,d_u)}}
                \infer1[(\ruleAppPTwo)]{\seqi{\Gam_u}{(\lam x.p)u}{\tcomptype{\stype}{\tneutral}{\stype}}{(0,0,1+d_u)}}
            \end{prooftree} \]
            And we can conclude with $d = 1 + d_u = 1 + \size{u} = 1 + \size{\lam x.p} + \size{u} = 1 + \size{(\lam x.p) u}$.
        \end{itemize}
        \item By induction on $t \in \normal$:
        \begin{itemize}
            \item Case $t \in \val$:
            \begin{itemize}
                \item Assume $t = x$. Then we can build $\Phi_t$ as follows:
                \[ \begin{prooftree}
                    \infer0[(\ruleAx)]{\seqi{x : \mul{\nott{\tneutral}}}{x}{\nott{\tneutral}}{(0,0,0)}}
                    \infer1[(\ruleLift)]{\seqi{x : \mul{\nott{\tneutral}}}{x}{\tcomptype{\stype}{\nott{\tneutral}}{\stype}}{(0,0,0)}}
                \end{prooftree} \]
                for any $\stype$ tight. And we can conclude with $\Gam = (x : \mul{\nott{\tneutral}})$ tight, and $b = m = d = 0 = \size{x}$.
                \item Assume $t = \lam x.u$. Then we can build $\Phi_t$ as follows:
                \[ \begin{prooftree}
                    \infer0[(\ruleLamP)]{\seqi{}{\lam x.u}{\vl}{(0,0,0)}}
                    \infer1[(\ruleLift)]{\seqi{}{\lam x.u}{\tcomptype{\stype}{\vl}{\stype}}{(0,0,0)}}
                \end{prooftree} \]
                for any $\stype$ tight. And we can conclude with $\Gam = \eset$ tight, $b = m = d = 0 = \size{\lam x.u}$.
            \end{itemize}
            \item Case $t \not\in \val$. Then $t \in \neutral$, and this case is subsumed by the previous cases.
        \end{itemize}
    \end{enumerate}

    \end{enumerate}
\end{proof}

} 

\subsubsection{Soundness and Completeness (Main Lemmas)}

\lemcompsubsantisubs*

\maybehide{\begin{proof} \mbox{}
    \begin{enumerate}
        \item 
    We are going to generalize the original statement by replacing $\del$ with $\gtype$.
    \\ \\
    The proof now follows by induction over the structure of $\Phi_t$:
        \begin{itemize}
            \item Case $\Phi_t$ ends with rule ($\ruleAx$). Then $t$ must be a variable and we must consider two cases:
            \begin{itemize}
                \item Assume $t = y = x$. Then $\Gam_t = \eset$, $\gtype = \M$, $t \subs{x}{v} = v$, $b_t = m_t = d_t = 0$. So we can take $\Phi_{t \subs{x}{v}} = \Phi_v$ and conclude with $\Gam_t + \Gam_v = \Gam_v$, $b_t + b_v = b_v$, $m_t + m_v = m_v$, and $d_t + d_v = d_v$.
                \item Assume $t = y \not= x$. Then $\M = \emul$, $\Gam_v = \eset$, $t \subs{x}{v} = t$, $b_v = 0$, $m_v = 0$, and $d_v = 0$. So we can take $\Phi_{t \subs{x}{v}} = \Phi_t$ and conclude with $\Gam_t + \Gam_v = \Gam_t$, $b_t + b_v = b_t$, $m_t + m_v = m_t$, and $d_t + d_v = d_t$.
            \end{itemize}
            \item Case $\Phi_t$ ends with (\ruleLam). Then $t$ must be of the form $\lam y.u$ and $\Phi_t$ must be of the following form (by $\alpha$-conversion):
            \[ \begin{prooftree}
                \hypo{\Phi_u \tr \seqi{\Gam; x : \M}{u}{\comptype{\stype}{\ctype}}{(b_t,m_t,d_t)}}
                \infer1[(\ruleLam)]{\seqi{(\Gam \sm y); x : \M}{\lam y.u}{\Gam(y) \ta (\comptype{\stype}{\ctype})}{(b_t,m_t,d_t)}}
            \end{prooftree} \]
            where $\Gam_t = (\Gam \sm y)$, and $\gtype = \Gam(y) \ta (\comptype{\stype}{\ctype})$. By the \ih, we have the following derivation $\Phi_{u \subs{x}{v}} \tr \seqi{\Gam + \Gam_v}{u \subs{x}{v}}{\comptype{\stype}{\ctype}}{(b_t+b_v,m_t+m_v,d_t+d_v)}$. Therefore, we can build $\Phi_{t \subs{x}{v}}$ as follows:
            \[ \begin{prooftree}
                \hypo{\Phi_{u \subs{x}{v}} \tr \seqi{\Gam + \Gam_v}{u \subs{x}{v}}{\comptype{\stype}{\ctype}}{(b_t+b_v,m_t+m_v,d_t+d_v)}}
                \infer1[(\ruleLam)]{\seqi{(\Gam + \Gam_v) \sm y}{\lambda y.u \subs{x}{v}}{\Gam(y) \ta (\comptype{\stype}{\ctype})}{(b_t+b_v,m_t+m_v,d_t+d_v)}}
            \end{prooftree} \]
            And we conclude with $(\Gam + \Gam_v) \sm y = (\Gam \sm y) + \Gam_v = \Gam_t + \Gam_v$, by $\alpha$-conversion.
            \item Case $\Phi_t$ ends with ($\ruleApp$). Then $t$ must be of the form $wu$ and $\Phi_t$ must be of following form:
            \[ \begin{prooftree}
                \hypo{\Phi_w \tr \seqi{\Gam; x : \M_1}{w}{\M' \ta (\comptype{\stype'}{\ctype})}{(b_w,m_w,d_w)}}
                \hypo{\Phi_u \tr \seqi{\Del; x : \M_2}{u}{\tcomptype{\stype}{\M'}{\stype'}}{(b_u,m_u,d_u)}}
                \infer2[(\ruleApp)]{\seqi{\Gam + \Del; x : \M_1 \sqcup \M_2}{wu}{\comptype{\stype}{\ctype}}{(1+b_w+b_u,m_w+m_u,d_w+d_u)}}
            \end{prooftree} \]
            such that $\Gam_t = \Gam + \Del$, $\M = \M_1 \sqcup \M_2$, $\gtype = \comptype{\stype}{\ctype}$, $b_t = 1+b_w+b_u$, $m_t = m_w+m_u$, and $d_t = d_w + d_u$. By~\cref{lem:split-values-stores}.\ref{lem:com-split-values}, we know there exist the following derivations $(\Phi^i_v \tr \seqi{\Gam^i_v}{v}{\M_i}{(b_i,m_i,d_i)})_{i \in \{1,2\}}$, such that $\Gam_v = \Gam^1_v + \Gam^2_v$, $b_v = b_1 + b_2$, $m_v = m_1 + m_2$, and $d_v = d_1 + d_2$. By the \ih, we know there exist $\Phi_{w \subs{x}{v}} \tr \seqi{\Gam + \Gam^1_v}{w \subs{x}{v}}{\M' \ta (\comptype{\stype'}{\ctype})}{(b_w+b_1,m_w+m_1,d_w+d_1)}$ and $\Phi_{u \subs{x}{v}} \tr \seqi{\Del + \Gam^2_v}{u \subs{x}{v}}{\tcomptype{\stype}{\M'}{\stype'}}{(b_u+b_2,m_u+m_2,d_u+d_2)}$. 
            We can build $\Phi_{t \subs{x}{v}}$ as follows:
            \[ \begin{prooftree}
                \hypo{\Phi_{w \subs{x}{v}}}
                \hypo{\Phi_{u \subs{x}{v}}}
                \infer2[(\ruleApp)]{\seqi{(\Gam + \Del) + (\Gam^1_v + \Gam^2_v)}{(wu) \subs{x}{v}}{\comptype{\stype}{\ctype}}{(1+b_w+b_u+b_1+b_2,m_w+m_u+m_1+m_2,d_w+d_u+d_1+d_2)}}
            \end{prooftree} \]
            And we can conclude with $\Gam_t + \Gam_v = (\Gam + \Del) + (\Gam^1_v + \Gam^2_v)$, $b_t + b_v = 1 + b_w+b_u+b_1+b_2$, $m_t + m_v = m_w+m_u+m_1+m_2$, and $d_t + d_v = d_w+d_u+d_1+d_2$.
            \item Case $\Phi_w$ ends with ($\ruleMany$). Then $t$ must be of the form $w$ and $\Phi_t$ must be of the following form:
            \[ \begin{prooftree}
                \hypo{(\Phi^i_w \tr \seqi{\Gam_i; x : \M_i}{w}{\rdel_i}{(b_i,m_i,d_i)})_{\iI}}
                \infer1[(\ruleMany)]{\seqi{+_{\iI} \Gam_i; x : \sqcup_{\iI} \M_i}{w}{\mul{\rdel_i}_{\iI}}{(+_{\iI}b_i, +_{\iI}m_i, +_{\iI}d_i)}}
            \end{prooftree} \]
            such that $\Gam_t = +_{\iI} \Gam_i$, $\gtype = \mul{\rdel_i}_{\iI}$, $b_t = +_{\iI} b_i$, $m_t = +_{\iI} m_i$, and $d_t = +_{\iI} d_i$. By~\cref{lem:split-values-stores}.\ref{lem:com-split-values}, $(\Phi^i_v \tr \seqi{\Gam^i_v}{v}{\M_i}{(b^i_v,m^i_v,d^i_v)})_{\iI}$, such that $\Gam_v = +_{\iI} \Gam^i_v$, $b_v = +_{\iI} b^i_v$, $m_v = +_{\iI} m^i_v$, and $d_v = +_{\iI} d^i_v$. By the \ih over each $\Phi^i_v$, we have $(\Phi^i_{w \subs{x}{v}} \tr \seqi{\Gam_i + \Gam^i_v}{w \subs{x}{v}}{\rdel_i}{(b_i+b^i_v,m_i+m^i_v,d_i+d^i_v)})_{\iI}$. Therefore, we can build $\Phi_{t \subs{x}{v}}$ as follows:
            \[ \begin{prooftree}
                \hypo{(\Phi^i_{w \subs{x}{v}} \tr \seqi{\Gam_i + \Gam^i_v}{w \subs{x}{v}}{\rdel_i}{(b_i+b^i_v,m_i+m^i_v,d_i+d^i_v)})_{\iI}}
                \infer1[(\ruleMany)]{\seqi{+_{\iI} (\Gam^i_v + \Gam^i_w)}{w \subs{x}{v}}{\mul{\tau_i}_{\iI}}{(+_{\iI}(b_i+b^i_v),+_{\iI}(m_i+m^i_v),+_{\iI}(d_i+d^i_v))}}
            \end{prooftree} \]
            And we can conclude with $\Gam_t + \Gam_v = +_{\iI} \Gam_i +_{\iI} \Gam^i_v = +_{\iI} (\Gam_i + \Gam^i_v)$, $b_t + b_v = +_{\iI} b_i +_{\iI} b^i_v = +_{\iI} (b_i + b^i_v)$, $m_t + m_v = +_{\iI} m_i +_{\iI} m^i_v = +_{\iI} (m_i + m^i_v)$, and $d_t + d_v = +_{\iI} d_i +_{\iI} d^i_v = +_{\iI} (d_i + d^i_v)$.
            \item Case $\Phi_t$ ends with (\ruleLift). Then $t$ is a value and $\Phi_t$ must be of the following form:
            \[ \begin{prooftree}
                \hypo{\Phi_w \tr \seqi{\Gam; x : \M}{w}{\mu}{(b_t,m_t,d_t)}}
                \infer1[(\ruleLift)]{\seqi{\Gam; x : \M}{w}{\tcomptype{\stype}{\mu}{\stype}}{(b_t,m_t,d_t)}}
            \end{prooftree} \]
            where $\gtype = \tcomptype{\stype}{\mu}{\stype}$. By the \ih, we have $\Phi_{w \subs{x}{v}} \tr \seqi{\Gam + \Gam_v}{w \subs{x}{v}}{\mu}{(b_t+b_v, m_t+m_v, d_t +d_v)}$. Therefore, we can build $\Phi_{t \subs{x}{v}}$ as follows:
            \[ \begin{prooftree}
                \hypo{\Phi_{w \subs{x}{v}} \tr \seqi{\Gam + \Gam_v}{w \subs{x}{v}}{\mu}{(b_t+b_v, m_t+m_v, d_t +d_v)}}
                \infer1[(\ruleLift)]{\seqi{\Gam + \Gam_v}{w \subs{x}{v}}{\tcomptype{\stype}{\mu}{\stype}}{(b_t+b_v, m_t+m_v, d_t +d_v)}}
            \end{prooftree} \]
            And we can conclude.
            \item Case $\Phi_t$ ends with ($\ruleGet$). Then $t$ must be of the form $\get{l}{y}{u}$ and $\Phi_t$ must be of the following form:
            \[ \begin{prooftree}
                \hypo{\Phi_u \tr \seqi{\Gam_u; x : \M}{u}{\comptype{\stype}{\ctype}}{(b_u,m_u,d_u)}}
                \infer1[(\ruleGet)]{\seqi{(\Gam_u \sm y); x : \M}{\get{l}{y}{u}}{\comptype{\conj{(l : \Gam_{u}(y))} \splus \stype}{\ctype}}{(b_u,1+m_u,d_u)}}
            \end{prooftree} \]
          where $\gtype = \comptype{\conj{(l : \Gam_{u}(y))} \splus \stype}{\ctype}$, $\Gam_t = \Gam_u \sm y$, $b_t = b_u$, $m_t = 1+m_u$, and $d_t = d_u$. By the \ih, we have $\Phi_{u \subs{x}{v}} \tr \seqi{\Gam_u + \Gam_v}{u \subs{x}{v}}{\stype \ra \ctype}{(b_u+b_v,m_u+m_v,d_u+d_v)}$. Therefore, we can build $\Phi_{t \subs{x}{v}}$ as follows:
            \[ \begin{prooftree}
                \hypo{\Phi_{u \subs{x}{v}} \tr \seqi{\Gam_u + \Gam_v}{u \subs{x}{v}}{\comptype{\stype}{\ctype}}{(b_u+b_v,m_u+m_v,d_u+d_v)}}
                \infer1[(\ruleGet)]{\seqi{(\Gam_u  + \Gam_v) \sm y}{\get{l}{y}{u} \subs{x}{v}}{\comptype{\conj{(l : \Gam_u(y))} \splus \stype}{\ctype}}{(b_u+b_v,1+m_u+m_v,d_u+d_v)}}
            \end{prooftree} \]
            And we can conclude with $(\Gam_u + \Gam_v) \sm y = (\Gam \sm y) + \Gam_v = \Gam_t + \Gam_v$ by $\alpha$-conversion, $b_t + b_v = b_u+b_v$, $m_t + m_v = 1+m_u+m_v$, and $d_t + d_v = d_u +d_v$.
            \item Case $\Phi_t$ ends with ($\ruleSet$). Then $t$ must be of the form $\set{l}{w}{u}$ and $\Phi_t$ must be of the following form:
            \[ \begin{prooftree}
                \hypo{\Phi_w \tr \seqi{\Gam_w; x : \M_1}{w}{\M'}{(b_w,m_w,d_w)}}
                \hypo{\Phi_u \tr \seqi{\Gam_u; x : \M_2}{u}{\comptype{\conj{(l : \M')}; \stype}{\ctype}}{(b_u,m_u,d_u)}}
                \infer2[(\ruleSet)]{\seqi{\Gam_w + \Gam_u; x : \M_1 \sqcup \M_2}{\set{l}{w}{u}}{\comptype{\stype}{\ctype}}{(b_w+b_u,1+m_w+m_u,d_w+d_u)}}
            \end{prooftree} \]
            where $\gtype = \comptype{\stype}{\ctype}$, $\Gam_t = \Gam_w + \Gam_u$, $\del = \comptype{\stype}{\ctype}$, $b_t = b_w+b_u$, $m_t = 1+m_w + m_u$, and $d_t = d_w + d_u$. By~\cref{lem:split-values-stores}.\ref{lem:com-split-values}, we have $\Phi^1_v \tr \seqi{\Gam^1_v}{v}{\M_1}{(b^1_v,m^1_v,d^1_v)}$ and $\Phi^2_v \tr \seqi{\Gam^2_v}{v}{\M_2}{(b^2_v,m^2_v,d^2_v)}$, such that $\Gam_v = \Gam^1_v + \Gam^2_v$, $b_v = b^1_v + b^2_v$, $m_v = m^1_v + m^2_v$, and $d_v = d^1_v + d^2_v$. By the \ih, we have $\Phi_{w \subs{x}{v}} \tr \seqi{\Gam_w + \Gam^1_v}{w \subs{x}{v}}{\M'}{(b_w+b^1_v,m_w+m^1_v,d_w+d^1_v)}$ and $\Phi_{u \subs{x}{v}} \tr \seqi{\Gam_u + \Gam^2_v}{u \subs{x}{v}}{\comptype{\conj{(l : \M')}; \stype}{\ctype}}{(b_u+b^2_v,m_u+m^2_u,d_u+d^2_v)}$. Assume $\Phi_{w \subs{x}{v}} \tr \seqi{\Gam_w + \Gam^1_v}{w \subs{x}{v}}{\M'}{(b_w+b^1_v,m_w+m^1_v,d_w+d^1_v)}$ and $\Phi_{u \subs{x}{v}} \tr \seqi{\Gam_u + \Gam^2_v}{u \subs{x}{v}}{\comptype{\conj{(l : \M')}; \stype}{\ctype}}{(b_u+b^2_v,m_u+m^2_u,d_u+d^2_v)}$. We can build $\Phi_{t \subs{x}{v}}$ as follows:
            \[ \begin{prooftree}
                \hypo{\Phi_{w \subs{x}{v}}}
                \hypo{\Phi_{u \subs{x}{v}}}
                \infer2[(\ruleSet)]{\seqi{(\Gam_w + \Gam_u) + (\Gam^1_v + \Gam^2_v)}{(wu) \subs{x}{v}}{\comptype{\stype}{\ctype}}{(b_w+b_u+b^1_v+b^2_v,1+m_w+m_u+m^1_v+m^2_v,d_w+d_u+d^1_v+d^2_v)}}
            \end{prooftree} \]
            And we can conclude with $\Gam_t + \Gam_v = (\Gam_w + \Gam_u) + (\Gam^1_v + \Gam^2_v)$, $b_t + b_v = b_w+b_u+b^1_v+b^2_v$, $m_t + m_v = 1+m_w+m_u+m^1_v+m^2_v$, $d_t + d_v = d_w+d_u+d^1_v+d^2_v$.
            \item Case $\Phi_t$ ends with (\ruleLamP). Then $t$ is of the form $\lam y.u$, $\Gam_t = \eset$, $\gtype = \vl$, $\M = \emul$, $\Gam_v = \eset$, $t \subs{x}{v} = \lam y.(u \subs{x}{v}) = (\lam y.u) \subs{x}{v}$, $b_t = b_v = 0$, $m_t = m_v = 0$, and $d_t = d_v = 0$. So we can build $\Phi_{t \subs{x}{v}}$ as follows:
            \[ \begin{prooftree}
                \infer0[(\ruleLamP)]{\seqi{}{(\lam y.u) \subs{x}{v}}{\tabs}{(0,0,0)}}
            \end{prooftree} \]
            And conclude with $\Gam_t + \Gam_v = \eset$, $b_t = b_v = 0$, $m_t = m_v = 0$, and $d_t = d_v = 0$.
            \item Case $\Phi_t$ ends with (\ruleAppPOne). Then $t$ is of the form $yu$ and we have to consider two cases:
            \begin{itemize}
                \item Case $y = x$. Then $\Phi_t$ must be of the following form:
                \[ \begin{prooftree}
                    \hypo{\seqi{\Gam_u}{u}{\tcomptype{\stype}{\tightt}{\stype'}}{(b_u,m_u,d_u)}}
                    \infer1[(\ruleAppPOne)]{\seqi{(x : \mul{\tvar} \sqcup \Gam_u(x)); (\Gam_u \sm x)}{x u}{\tcomptype{\stype}{\tneutral}{\stype'}}{(b_u,m_u,1+d_u)}}
                \end{prooftree} \]
                such that $\Gam_t = (\Gam_u \sm x)$, $b = b_u$, $m = m_u$, and $d = 1+d_u$. Then $\M = \mul{\tvar} \sqcup \Gam_u(x)$ and, by~\cref{lem:split-values-stores}.\ref{lem:com-split-values}, we have $\Phi^1_v \tr \seqi{\Gam^1_v}{v}{\mul{\tvar}}{(b^1_v,m^1_v,d^1_v)}$ and $\Phi^2_v \tr \seqi{\Gam^2_v}{v}{\Gam_u(x)}{(b^2_v,m^2_v,d^2_v)}$, such that $\Gam_v = \Gam^1_v + \Gam^2_v$, $b_v = b^1_v + b^2_v$, $m_v = m^1_v + m^2_v$, and $d_v = d^1_v + d^2_v$. By the \ih, we know there exists $\Phi_{u \subs{x}{v}} \tr \seqi{(\Gam_u \sm x) + \Gam^2_u}{u \subs{x}{v}}{\tcomptype{\stype}{\tightt}{\stype'}}{(b_u+b^2_v, m_u+m^2_v, d_u+d^2_v)}$. Now, we need to consider two cases:
                \begin{itemize}
                    \item Case $v = z$. Then $\Phi^1_v \tr \seqi{z : \mul{\tvar}}{z}{\mul{\tvar}}{(0,0,0)}$ and $\Phi^2_v \tr \seqi{z : \Gam_u(x)}{z}{\Gam_u(x)}{(0,0)}$. Therefore, we can build $\Phi_{t \subs{x}{v}} = \Phi_{v \subs{x}{v}}$ as follows:
                    \[ \begin{prooftree}
                        \hypo{\Phi_{u \subs{x}{v}} \tr \seqi{(\Gam_u \sm x) + \Gam_u(x)}{u \subs{x}{v}}{\tcomptype{\stype}{\tightt}{\stype'}}{(b_u+b^2_v, m_u+m^2_v, d_u+d^2_v)}}
                        \infer1[(\ruleAppPOne)]{\seqi{(z : \mul{\tvar}) + (\Gam_u \sm x + (z : \Gam_u(x)))}{z (u \subs{x}{v})}{\tcomptype{\stype}{\tneutral}{\stype'}}{(b_u+b^2_v,m_u+m^2_v,1+d_u+d^2_v)}}
                    \end{prooftree} \]
                    where $(z : \mul{\tvar}) + (\Gam_u \sm x + (z : \Gam_u(x))) = (\Gam_u \sm x) + (z : \mul{\tvar} \cup \Gam_u(x)) = \Gam_u + \Gam_v$, $b_u + b^2_v = b + b^1_v + b^2_v = b + b_v$, $m_u + m^2_v = m + m^1_v + m^2_v = m + m_v$, and $d_u + d^2_v = d + d^1_v + d^2_v = d + d_v$.
                    \item Case $v = \lam z.p$. This case does not apply, since it is clearly not possible to assign $\tvar$ to $\lam z.p$.
                \end{itemize}
                \item Case $y \neq x$. Then, the proof is very similar to when $\Phi_t$ ends with rule ($\ruleApp$).
            \end{itemize}
            \item Case $\Phi_t$ ends with (\ruleAppPTwo), the proof is very similar to when $\Phi_t$ ends with rule (\ruleAppPOne).
        \end{itemize}

        \item 
    We are going to generalize the original statement by replacing $\del$ with $\gtype$.
    \\ \\
    The proof follows by induction over $t$:
    \begin{itemize}
        \item Case $t = y$. Then we have to consider two cases:
        \begin{itemize}
            \item Let $t = y \not= x$. Then $t \subs{x}{v} = y$. Let $\Gam_v = \eset$, $\M = \emul$, $b_v = m_v = d_v = 0$. Then, $\Phi_v$ is derivable using rule ($\ruleMany$) with no premise. We also take $\Phi_t = \Phi_{t \subs{x}{v}}$, so that, in particular $\Gam_t = \Gam_{t \subs{x}{v}}$. Then, we can conclude with $\Gam_{t \subs{x}{v}} = \Gam_t + \Gam_v = \Gam_t$, $b = b_t + b_v = b_t$, $m = m_t + m_v = m_t$, and $d = d_t + d_v = d_t$.
            \item Let $t = y = x$. Then $t \subs{x}{v} = v$. Let $\Gam_t = \eset$, and $b_t = m_t = s_t = 0$. Now we will consider two cases depending on the form of $v$:
            \begin{itemize}
                \item Case $v = z$. Then $t \subs{x}{v} = z$ and we can proceed by case analysis of the last rule in $\Phi_{t\subs{x}{v}}$. In all of them we can build $\Phi_t$ from $\Phi_{t \subs{x}{v}}$ by simply replacing $x$ with $z$, and we can build $\Phi_v$ as follows:
                \[ \begin{prooftree}
                    \infer0[(\ruleAx)]{\seqi{z : \mul{\sig}}{z}{\sig}{(0,0,0)}}
                    \infer1[(\ruleMany)]{\seqi{z : \mul{\sig}}{z}{\mul{\sig}}{(0,0,0)}}
                \end{prooftree} \]
                And we can conclude since all the counters are zero.
                \item Case $v = \lam z.p$. Then $t \subs{x}{v} = \lam z.p$ and we can proceed by case analysis of the last rule in $\Phi_{t \subs{x}{v}}$:
                \begin{itemize}
                    \item Case (\ruleLam). Then $\del = \M' \ta (\comptype{\stype}{\ctype})$. So we can build $\Phi_t \tr \seqi{x : \mul{\M' \ta (\comptype{\stype}{\ctype})}}{x}{\M' \ta (\comptype{\stype}{\ctype})}{(0,0,0)}$ using rule (\ruleAx). And $\Phi_v$ as follows:
                    \[ \begin{prooftree}
                        \hypo{\Phi_{t \subs{x}{v}}}
                        \infer1[(\ruleMany)]{\seqi{\Gam_{t \subs{x}{v}}}{\lam z.p}{\mul{\M' \ta (\comptype{\stype}{\ctype})}}{(b,m,d)}}
                    \end{prooftree} \]
                    \item Case (\ruleMany). Then $\del = \mul{\sig_i}_{\iI}$. So we can build $\Phi_t \tr \seqi{x : \mul{\sig_i}_{\iI}}{x}{\mul{\sig_i}_{\iI}}{(0,0,0)}$, by applying rule (\ruleMany) to the premises $(\seqi{x : \mul{\sig_i}}{x}{\sig_i}{(0,0,0)})_{\iI}$. And we can take $\Phi_v = \Phi_{t \subs{x}{v}}$.
                    \item Case (\ruleLift). Then $\del = \tcomptype{\stype}{\mu}{\stype}$ and we have to consider three cases, depending on the form of $\mu$. Clearly, it cannot be the case that $\mu = \tvar$. Therefore, we have to consider the cases where $\mu = \M'$ and $\mu = \tabs$. Let $\mu = \M' = \mul{\sig_i}_{\iI}$. Then we proceed similarly to the case where $\Phi_{t \subs{x}{v}}$ ends with rule (\ruleMany) and obtain $\Phi_t$ by applying rule (\ruleLift). Let $\mu = \tabs$, then $\Gam_{t \subs{x}{v}} = \eset$ and $b = m = d = 0$. Then we can build $\Phi_t$ as follows:
                    \[ \begin{prooftree}
                        \infer0[(\ruleAx)]{\seqi{x : \mul{\tabs}}{x}{\tabs}{(0,0,0)}}
                        \infer1[(\ruleLift)]{\seqi{x : \mul{\tabs}}{x}{\tcomptype{\stype}{\tabs}{\stype}}{(0,0,0)}}
                    \end{prooftree} \]
                    And $\Phi_v$ as follows:
                    \[ \begin{prooftree}
                        \infer0[(\ruleLamP)]{\seqi{}{\lam z.p}{\tabs}{(0,0,0)}}
                        \infer1[(\ruleMany)]{\seqi{}{\lam z.p}{\mul{\tabs}}{(0,0,0)}}
                    \end{prooftree} \]
                    \item Case (\ruleLamP). Then $\del = \tabs$, $\Gam_{t \subs{x}{v}} = \eset$, and $b = m = d = 0$. So we can build $\Phi_t \tr \seqi{x : \mul{\tabs}}{x}{\tabs}{(0,0,0)}$ using rule (\ruleAx) and $\Phi_v$ as follows:
                    \[ \begin{prooftree}
                        \infer0[(\ruleLamP)]{\seqi{}{\lam z.p}{\tabs}{(0,0,0)}}
                        \infer1[(\ruleMany)]{\seqi{}{\lam z.p}{\mul{\tabs}}{(0,0,0)}}
                    \end{prooftree} \]
                \end{itemize}
            \end{itemize}
        \end{itemize}
        \item Case $t = \lam y.u$. Then $t \subs{x}{v} = (\lam y.u)\subs{x}{v} = \lam y.(u \subs{x}{v})$ and we must consider three cases:
        \begin{itemize}
            \item Case $\Phi_{t \subs{x}{v}}$ ends with rule (\ruleLam), then it must be of the following form: 
            \[ \begin{prooftree}
                \hypo{\Phi_{u \subs{x}{v}} \tr \seqi{\Gam_{u \subs{x}{v}}; y : \M'}{u \subs{x}{v}}{\comptype{\stype}{\ctype}}{(b,m,d)}}
                \infer1[(\ruleLam)]{\seqi{\Gam_{u \subs{x}{v}}}{\lam y.(u \subs{x}{v})}{\M' \ta (\comptype{\stype}{\ctype})}{(b,m,d)}}
            \end{prooftree} \]
            where $\gtype = \M' \ta (\comptype{\stype}{\ctype})$ and $\Gam_{t \subs{x}{v}} = \Gam_{u \subs{x}{v}}$. By the \ih, we have $\Phi_u \tr \seqi{\Gam_u; y : \M'; x : \M}{u}{\comptype{\stype}{\ctype}}{(b_u,m_u,d_u)}$ and $\Phi_v \tr \seqi{\Gam_v}{v}{\M}{(b_v,m_v,d_v)}$, such that $\Gam_{u \subs{x}{v}} = \Gam_u + \Gam_v$, $b = b_u + b_v$, $m = m_u + m_v$, and $d = d_u + d_v$. So we can build $\Phi_{\lam y.u}$ as follows:
            \[ \begin{prooftree}
                \hypo{\Phi_u \tr \seqi{\Gam_u; y: \M'; x : \M}{u}{\comptype{\stype}{\ctype}}{(b_u,m_u,d_u)}}
                \infer1[(\ruleLam)]{\seqi{\Gam_u; x : \M}{\lam y.u}{\M' \ta (\comptype{\stype}{\ctype})}{(b_u,m_u,d_u)}}
            \end{prooftree} \]
            And we can pick $\Phi_t = \Phi_{\lam y.u}$, and conclude with $\Gam_{t \subs{x}{v}} = \Gam_{u \subs{x}{v}} = \Gam_u + \Gam_v$, $b = b_u + b_v$, $m = m_u + m_v$, and $d = d_u + d_v$.
            \item Case $\Phi_{t \subs{x}{v}}$ ends with rule ($\ruleMany$). Then $t \subs{x}{v}$ is a value, and $\Phi_{t \subs{x}{v}}$ must be of the following form:
            \[ \begin{prooftree}
                \hypo{(\Phi_i \tr \seqi{\Gam_i}{t \subs{x}{v}}{\rdel_i}{(b_i,m_i,d_i)})_{\iI}}
                \infer1[(\ruleMany)]{\seqi{+_{\iI} \Gam_i}{t \subs{x}{v}}{\mul{\rdel_i}_{\iI}}{(+_{\iI} b_i, +_{\iI} m_i, +_{\iI} d_i)}}
            \end{prooftree} \]
            where $\gtype = \mul{\rdel_i}_{\iI}$, $\Gam_{t \subs{x}{v}} = +_{\iI} \Gam_i$, $b = +_{\iI} b_i$, $m = +_{\iI} m_i$, and $d = +_{\iI} d_i$. By the \ih over each $\Phi_i$, we have the following derivations $\Phi^i_t \tr \seqi{\Gam^i_t; x : \M_i}{t}{\rdel_i}{(b^i_t,m^i_t,d^i_t)}$ and $\Phi^i_v \tr \seqi{\Gam^i_v}{v}{\M_i}{(b^i_v, m^i_v, d^i_v)}$, such that $\Gam_i = \Gam^i_t + \Gam^i_v$, $b = b^i_t + b^i_v$, $m = m^i_t + m^i_v$, and $d = d^i_t + d^i_v$, for each $\iI$. So we can construct $\Phi_t$ as follows:
            \[ \begin{prooftree}
                \hypo{(\Phi^i_t \tr \seqi{\Gam^i_t; x : \M_i}{t}{\rdel_i}{(b^i_t,m^i_t,d^i_t)})_{\iI}}
                \infer1[(\ruleMany)]{\seqi{+_{\iI} \Gam^i_t; x : \sqcup_{\iI} \M_i}{t}{\mul{\rdel_i}_{\iI}}{(+_{\iI} b^i_t, +_{\iI} m^i_t, +_{\iI} d^i_t)}}
            \end{prooftree} \]
            such that $\Gam_t = +_{\iI} \Gam^i_t$, $\M = \sqcup_{\iI} \M_i$, $b_t = +_{\iI} b^i_t$, $m_t = +_{\iI} m^i_t$, and $d_t = +_{\iI} d^i_t$. By~\cref{lem:comp-merge-values}, we can take the following derivation $\Phi_v \tr \seqi{+_{\iI} \Gam^i_v}{v}{\M}{(+_{\iI} b^i_v, +_{\iI} m^i_v, +_{\iI} d^i_v)}$. And we can conclude with $\Gam_{t \subs{x}{v}} = +_{\iI} \Gam_i = +_{\iI} (\Gam^i_t + \Gam^i_v) = +_{\iI} \Gam^i_t +_{\iI} \Gam^i_v = \Gam_t + \Gam_v$, $b = +_{\iI} b_i = +_{\iI} (b^i_t + b^i_v) = +_{\iI} b^i_t +_{\iI} b^i_v = b_t + b_v$, $m = +_{\iI} m_i = +_{\iI} (m^i_t + m^i_v) = +_{\iI} m^i_t +_{\iI} m^i_v = m_t + m_v$, and $d = +_{\iI} d_i = +_{\iI} (d^i_t + d^i_v) = +_{\iI} d^i_t +_{\iI} d^i_v = d_t + d_v$.
            \item Case $\Phi_{t \subs{x}{v}}$ ends with rule (\ruleLift), then it must be of the following form:
            \[ \begin{prooftree}
                \hypo{\Phi'_{t \subs{x}{v}} \tr \seqi{\Gam_{t \subs{x}{v}}}{(\lam y.u) \subs{x}{v}}{\mu}{(b,m,d)}}
                \infer1[(\ruleLift)]{\seqi{\Gam_{t \subs{x}{v}}}{(\lam y.u) \subs{x}{v}}{\tcomptype{\stype}{\mu}{\stype}}{(b,m,d)}}
            \end{prooftree} \]
            where $\gtype = \tcomptype{\stype}{\mu}{\stype}$. By the \ih over $\Phi'_{t \subs{x}{v}}$, we have $\Phi'_t \tr \seqi{\Gam'_t; x : \M}{\lam y.u}{\mu}{(b',m',d')}$ and $\Phi_v \tr \seqi{\Gam_v}{v}{\M}{(b_v,m_v,d_v)}$, such that $\Gam_{t \subs{x}{v}} = \Gam'_t + \Gam_v$, $b = b' + b_v$, $m = m' + m_v$, and $d = d' + d_v$. So we can build $\Phi_{\lam y.u}$ as follows:
            \[ \begin{prooftree}
                \hypo{\Phi'_t \tr \seqi{\Gam'_t; x : \M}{\lam y.u}{\mu}{(b',m',d')}}
                \infer1[(\ruleLift)]{\seqi{\Gam'_t; x : \M}{\lam y.u}{\tcomptype{\stype}{\mu}{\stype}}{(b',m',d')}}
            \end{prooftree} \]
            And we can pick $\Phi_t = \Phi'_{t \subs{x}{v}}$ to conclude.
            \item Case $\Phi_{t \subs{x}{v}}$ ends with rule (\ruleLamP). Then it must be of the following form:
            \[ \begin{prooftree}
                \infer0[(\ruleLamP)]{\seqi{}{\lam y.(u \subs{x}{y})}{\tabs}{(0,0,0)}}
            \end{prooftree} \]
            where $\Gam_{t \subs{x}{v}} = \eset$, $\gtype =\tabs$, and $b = m = d = 0$. Let $\Gam_t = \eset$, $\M = \emul$, and $b_t = m_t = d_t = 0$. Then, we can construct $\Phi_t$ as follows:
            \[ \begin{prooftree}
                \infer0[(\ruleLamP)]{\seqi{}{\lam y.u}{\tabs}{(0,0,0)}}
            \end{prooftree} \]
            Let $\Gam_v = \eset$, and $b_v = m_v = d_v = 0$. Then $\Phi_v$ can be constructed by using rule ($\ruleMany$) with no premises. So we can conclude with $\Gam_{t \subs{x}{v}} = \eset = \Gam_t + \Gam_v$, and $b = 0 = b_t + b_v$, $m = 0 = m_t + m_v$, and $d = 0 = d_t + d_v$.
        \end{itemize}
        \item Let $t = wu$. Then $t \subs{x}{v} = (wu) \subs{x}{v} = (w \subs{x}{v})(u \subs{x}{v})$, and we have to consider three cases:
        \begin{itemize}
            \item Case $\Phi_{t \subs{x}{v}}$ ends with ($\ruleApp$). Assume $\Phi_{w \subs{x}{v}} \tr \seqi{\Gam_{w \subs{x}{v}}}{w \subs{x}{v}}{\M' \ta (\comptype{\stype'}{\ctype})}{(b',m',d')}$ and $\Phi_{u \subs{x}{v}} \tr \seqi{\Gam_{u \subs{x}{v}}}{u \subs{x}{v}}{\tcomptype{\stype}{\M'}{\stype'}}{(b'',m'',d'')}$. $\Phi_{t \subs{x}{v}}$ must be of the following form:
            \[ \begin{prooftree}
                \hypo{\Phi_{w \subs{x}{v}}}
                \hypo{\Phi_{u \subs{x}{v}}}
                \infer2[(\ruleApp)]{\seqi{\Gam_{w \subs{x}{v}} + \Gam_{u \subs{x}{v}}}{(w \subs{x}{v})(u \subs{x}{v})}{\comptype{\stype}{\ctype}}{(1+b'+b'',m'+m'',d'+d'')}}
            \end{prooftree} \]
            where $\gtype = \comptype{\stype}{\ctype}$, $\Gam_{t \subs{x}{v}} = \Gam_{w \subs{x}{v}} + \Gam_{u \subs{x}{v}}$, $b = 1+b'+b''$, $m = m'+m''$, and $d = d'+d''$. By the \ih over $\Phi_{w \subs{x}{v}}$, we have $\Phi_w \tr \seqi{\Gam_w; x : \M_1}{w}{\M' \ta (\comptype{\stype'}{\ctype})}{(b_w,m_w,d_w)}$ and $\Phi^1_v \tr \seqi{\Gam^1_v}{v}{\M_1}{(b^1_v,m^1_v,d^1_v)}$, such that $\Gam_{w \subs{x}{v}} = \Gam_w + \Gam^1_v$, $b' = b_w + b^1_v$, $m' = m_w + m^1_v$, and $d' = d_w + d^1_v$. And by the \ih over $\Phi_{u \subs{x}{v}}$, we have $\Phi_u \tr \seqi{\Gam_u; x : \M_2}{u}{\tcomptype{\stype}{\M'}{\stype'}}{(b_u, m_u,d_u)}$ and $\Phi^2_v \tr \seqi{\Gam^2_v}{v}{\M_2}{(b^2_v,m^2_v,d^2_v)}$, such that $\Gam_{u \subs{x}{v}} = \Gam_u + \Gam^2_v$, $b'' = b_u + b^2_v$, $m'' = m_u + m^2_v$, and $d'' = d_u + d^2_v$. By~\cref{lem:comp-merge-values}, we can take $\Phi_v \tr \seqi{\Gam^1_v + \Gam^2_v}{v}{\M_1 \sqcup \M_2}{(b^1_v+b^2_v, m^1_v+m^2_v, d^1_v+d^2_v)}$, such that $\Gam_v = \Gam^1_v + \Gam^2_v$, $b_v = b^1_v + b^2_v$, $m_v = m^1_v + m^2_v$, and $d_v = d^1_v + d^2_v$. And we can build $\Phi_{wu}$ as follows:
            \[ \begin{prooftree}
                \hypo{\Phi_w}
                \hypo{\Phi_u}
                \infer2[(\ruleApp)]{\seqi{(\Gam_w + \Gam_u); x : \M_1 \sqcup \M_2}{wu}{\comptype{\stype}{\kappa}}{(1+b_w+b_u,m_w+m_u,d_w+d_u)}}
            \end{prooftree} \]
            such that $\Gam_t = \Gam_w + \Gam_u$, $b_t = 1 + b_w + b_u$, $m_t = b_w + b_u$, and $d_t = d_w + d_u$. So we can pick $\Phi_t = \Phi_{wu}$, and conclude with $\Gam_{t \subs{x}{v}} = \Gam_{w \subs{x}{v}} + \Gam_{u \subs{x}{v}} = (\Gam_w + \Gam^1_v) + (\Gam_u + \Gam^2_v) = (\Gam_w + \Gam_u) + (\Gam^1_v + \Gam^2_v) = \Gam_t + \Gam_v$, $b = 1 + b' + b'' = 1 + b_w + b^1_v + b_u + b^2_v = (1 + b_w + b_u) + (b^1_v + b^2_v) = b_t + b_v$, $m = m' + m'' = m_w + m^1_v + m_u + m^2_v = (m_w + m_u) + (m^1_v + m^2_v) = m_t + m_v$, and $d = d' + d'' = d_w + d^1_v + d_u + d^2_v = (d_w + d_u) + (d^1_v + d^2_v) = d_t + d_v$.
            \item Case $\Phi_{t \subs{x}{v}}$ ends with (\ruleAppPOne) or (\ruleAppPTwo). These cases are very similar to the case where $\Phi_{t \subs{x}{v}}$ ends with (\ruleApp).
        \end{itemize}
        \item Let $t = \get{l}{y}{u}$. Then $t \subs{x}{v} = \get{l}{y}{u \subs{x}{v}}$ and $\Phi_{t \subs{x}{v}}$ must be of the following form:
        \[ \begin{prooftree}
            \hypo{\Phi_{u \subs{x}{v}} \tr \seqi{\Gam_{u \subs{x}{v}}; y : \M'}{u \subs{x}{v}}{\comptype{\stype}{\ctype}}{(b,m',d)}}
            \infer1[(\ruleGet)]{\seqi{\Gam_{u \subs{x}{v}}}{\get{l}{y}{u \subs{x}{v}}}{\comptype{\conj{(l : \M')} \splus \stype}{\ctype}}{(b,1+m',d)}}
        \end{prooftree} \]
        where $\Gam_{t \subs{x}{v}} = \Gam_{u \subs{x}{v}}$ and $m = 1+m'$. By the \ih, we have $\Phi_u \tr \seqi{\Gam_u; y : \M'; x : \M}{u}{\comptype{\stype}{\ctype}}{(b_u,m_u,d_u)}$ and $\Phi_v \tr \seqi{\Gam_v}{v}{\M}{(b_v,m_v,d_v)}$, such that $\Gam_{u \subs{x}{v}} = \Gam_u + \Gam_v$, $b = b_u + b_v$, $m' = m_u + m_v$, and $d = d_u + d_v$. So we can build $\Phi_{\get{l}{y}{u}}$ as follows:
        \[ \begin{prooftree}
            \hypo{\Phi_u \tr \seqi{\Gam_u; y : \M'; x : \M}{u}{\comptype{\stype}{\ctype}}{(b_u,m_u,d_u)}}
            \infer1[(\ruleGet)]{\seqi{\Gam_{u}; x : \M}{\get{l}{y}{u}}{\comptype{\conj{(l : \M')} \splus \stype}{\ctype}}{(b_u,1+m_u,d_u)}}
        \end{prooftree} \]
        And we can pick $\Phi_t = \Phi_{\get{l}{y}{u}}$, and conclude with $\Gam_{t \subs{x}{v}} = \Gam_{u \subs{x}{v}} = \Gam_u + \Gam_v$, $b = b_u + b_v$, $m = 1 + m' = 1 + m_u + m_v = (1 + m_u) + m_v$, and $d = d_u + d_v$.
        \item Let $t = \set{l}{w}{u}$. Then $t \subs{x}{v} = (\set{l}{w}{u}) \subs{x}{v} = \set{l}{w \subs{x}{v}}{u \subs{x}{v}}$. Assume $\Phi_{w \subs{x}{v}} \tr \seqi{\Gam_{w \subs{x}{v}}}{w \subs{x}{v}}{\M}{(b',m',d')}$ and $\Phi_{u \subs{x}{v}} \tr \seqi{\Gam_{u \subs{x}{v}}}{u \subs{x}{v}}{\comptype{\conj{(l : \M)}; \stype}{\ctype}}{(b'',m'',d'')}$. $\Phi_{t \subs{x}{v}}$ must be of the following form:
        \[ \begin{prooftree}
            \hypo{\Phi_{w \subs{x}{v}}}
            \hypo{\Phi_{t \subs{x}{v}}}
            \infer2[(\ruleSet)]{\seqi{\Gam_{w \subs{x}{v}} + \Gam_{u \subs{x}{v}}}{\set{l}{w \subs{x}{v}}{u \subs{x}{v}}}{\comptype{\stype}{\ctype}}{(b'+b'',1+m'+m'',d'+d'')}}
        \end{prooftree} \]
        where $\Gam_{t \subs{x}{v}} = \Gam_{w \subs{x}{v}} + \Gam_{u \subs{x}{v}}$, $b = b'+ b''$, $m = 1 + m'+m''$, and $d = d' + d''$. By the \ih over $\Phi_{w \subs{x}{v}}$, we have $\Phi_w \tr \seqi{\Gam_w; x : \M_1}{w}{\M}{(b_w,m_w,d_w)}$ and $\Phi^1_v \tr \seqi{\Gam^1_v}{v}{\M_1}{(b^1_v,m^1_v,d^1_v)}$, such that $\Gam_{w \subs{x}{v}} = \Gam_w + \Gam^1_v$, $b' = b_w + b^1_v$, $m'= m_w + m^1_v$, and $d' = d_w+d^1_v$. And by the \ih over $\Phi_{u \subs{x}{v}}$, we have $\Phi_u \tr \seqi{\Gam_u; x : \M_2}{u}{\comptype{\conj{(l : \M)}; \stype}{\ctype}}{(b_u,m_u,d_u)}$ and $\Phi^2_v \tr \seqi{\Gam^2_v}{v}{\M_2}{(b^2_v,m^2_v,d^2_v)}$, such that $\Gam_{u \subs{x}{v}} = \Gam_u + \Gam^2_v$, $b'' = b_u + b^2_v$, $m'' = m_u + m^2_v$, and $d'' = d_u + d^2_v$. By~\cref{lem:comp-merge-values}, we can take $\Phi_v \tr \seqi{\Gam^1_v + \Gam^2_v}{v}{\M_1 \sqcup \M_2}{(b^1_v + b^2_v, m^1_v+m^2_v, d^1_v + d^2_v)}$, such that $\Gam_v = \Gam^1_v + \Gam^2_v$, $b_v = b^1_v + b^2_v$, $m_v = m^1_v + m^2_v$, and $d_v = d^1_v + d^2_v$. And we can build $\Phi_{\set{l}{w}{u}}$ as follows:
        \[ \begin{prooftree}
            \hypo{\Phi_w \tr \seqi{\Gam_w; x : \M_1}{w}{\M}{(b_w,m_w,d_w)}}
            \hypo{\Phi_u \tr \seqi{\Gam_u; x : \M_2}{u}{\comptype{\conj{(l : \M)}; \stype}{\ctype}}{(b_u,m_u,d_u)}}
            \infer2[(\ruleSet)]{\seqi{(\Gam_w + \Gam_u); x : \M_1 \sqcup \M_2}{\set{l}{w}{u}}{\comptype{\stype}{\ctype}}{(b_w+b_u, 1+m_w+m_u,d_w+d_u)}}
        \end{prooftree} \]
        such that $\Gam_t = \Gam_w + \Gam_u$, $b_t = b_w + b_u$, $m_t = 1 + m_w + m_u$, and $d_t = d_w + d_u$. So we can pick $\Phi_t = \Gam_{\set{l}{w}{u}}$, and conclude with $\Gam_{t \subs{x}{v}} = \Gam_{w \subs{x}{v}} + \Gam_{u \subs{x}{v}} = (\Gam_w + \Gam^1_u) + (\Gam_u + \Gam^2_v) = (\Gam_w + \Gam_u) + (\Gam^1_v + \Gam^2_v) = \Gam_t + \Gam_v$, $b = b' + b'' = (b_w + b^1_v) + (b_u + b^2_v) = (b_w + b_u) + (b^1_v + b^2_v) = b_t + b_v$, $m = 1+ m' + m'' = 1+ (m_w + m^1_v) + (m_u + m^2_v) = (1 + m_w + m_u) + (m^1_v + m^2_v) = m_t + m_v$, and $d = d' + d'' = (d_w + d^1_v) + (d_u + d^2_v) = (d_w + d_u) + (d^1_v + d^2_v) = d_t + d_v$.
    \end{itemize}

    \end{enumerate}
\end{proof}}

\lemexactredexp*

\maybehide{\begin{proof} \mbox{}
    \begin{enumerate} 
        \item 
  We show a stronger statement of the form:

  Let $(t,s) \red[\gname] (u,q)$. If $\Phi \tr \seqi{\Gam}{(t,s)}{\ctype}{(b,m,d)}$, $\Gam$ is tight,  and ($\ctype$ is tight or $\neg \isvalue{t}$), then $\Phi' \tr \seqi{\Gam}{(u,q)}{\ctype}{(b',m',d)}$, where $\gname =\beta$ implies $b' = b - 1$ and $m' = m$, while  $\gname \in \{\getname, \setname\}$ implies $b'=b$ and $m' = m - 1$.

  We proceed by induction on $(t,s) \ra (u,q)$: 
  \begin{itemize}
    \item Case $(t,s) = ((\lam x.p) v,s) \redbeta (p \subs{x}{v}, s) = (u,q)$. Let $\Phi_{(\lam x.p) v}$ be the sub-derivation for $(\lam x.p) v$ in $\Phi$. Assume that $\Phi_{(\lam x.p)v}$ ends with rule (\ruleAppPTwo). Then $v$ must be assigned type $\comptype{\stype}{\tneutral \tim \stype}$, which is not possible by~\cref{lem:comp-values-not-neutral}. Let $\Phi_0$ be the following derivation:
    \[ \begin{prooftree}
      \hypo{\Phi_p \tr \seqi{\Gam_{\lam x.p}; x : \M}{p}{\comptype{\stype}{\ctype}}{(b_p,m_p,d_p)}}
        \infer1[(\ruleLam)]{\seqi{\Gam_{\lam x.p}}{\lam x.p}{\M \ta (\comptype{\stype}{\ctype})}{(b_p,m_p,d_p)}}
        \hypo{\Phi_v \tr \seqi{\Gam_v}{v}{\M}{(b_v,m_v,d_v)}}
        \infer1[(\ruleLift)]{\seqi{\Gam_v}{v}{\tcomptype{\stype}{\M}{\stype}}{(b_v,m_v,d_v)}}
        \infer2[(\ruleApp)]{\seqi{\Gam_{\lam x.p}+\Gam_v}{(\lam x.p)v}{\comptype{\stype}{\ctype}}{(1+b_v+b_p,m_v+m_p,d_v+d_p)}}
    \end{prooftree} \]
    $\Phi_{(\lam x.p) v}$ must end with rule ($\ruleApp$) and $\Phi$ must be of the following form:
    \[ \begin{prooftree}
        \hypo{\Phi_0}
        \hypo{\Phi_s \tr \seqi{\Del}{s}{\stype}{(b_s,m_s,d_s)}}
        \infer2[(\ruleConf)]{\seqi{\Gam_{\lam x.p}+ \Gam_v + \Del}{((\lam x.p)v, s)}{\ctype}{(1+b_v+b_p+b_s,m_v+m_p+m_s,d_v+d_p+d_s)}}
    \end{prooftree} \]
    where  $\Gam = \Gam_{\lam x.p}+ \Gam_v + \Del$,  $b = 1+ b_v + b_p + b_s$, $m = m_v + m_p + m_s$, and $d = d_v + d_p + d_s$. By~\cref{lem:comp-subs-antisubs}.\ref{lem:comp-subs}, there exists $\Phi_{p \subs{x}{v}} \tr \seqi{\Gam_{\lam x.p} +\Gam_v}{p \subs{x}{v}}{\comptype{\stype}{\ctype}}{(b_v+b_p,m_v+m_p,d_v+d_p)}$, therefore we can build $\Phi_{(p\subs{x}{v},s)}$ as follows:
    \[ \begin{prooftree}
        \hypo{\Phi_{p \subs{x}{v}} \tr \seqi{\Gam_{\lam x.p}+\Gam_v}{p \subs{x}{v}}{\comptype{\stype}{\ctype}}{(b_v+b_p,m_v+m_p,d_v+d_p)}}
        \hypo{\Phi_s \tr \seqi{\Del}{s}{\stype}{(b_s,m_s,d_s)}}
        \infer2[(\ruleConf)]{\seqi{\Gam_{\lam x.p} + \Gam_v + \Del}{(p \subs{x}{v}, s)}{\ctype}{(b_v+b_p+b_s,m_v+m_p+m_s,d_v+d_p+d_s)}}
    \end{prooftree} \]
    We can finally conclude since the first counter is equal to $b-1$, while the second and third remain the same.
  \item Case $(t,s) = (vp,s) \ra (vp',q) = (u,q)$, such that $(p,s) \ra (p',q)$. Then we have three cases for the type derivation $\Phi_p$ of $p$ inside $\Phi$: 
    \begin{itemize}
      \item Case $\Phi_p$ ends with ($\ruleApp$). Let $\Phi_0$ be the following derivation:
      \[ \begin{prooftree}
        \hypo{\Phi_v \tr \seqi{\Gam_v}{v}{\M \ta (\comptype{\stype'}{\ctype})}{(b_v,m_v,d_v)}}
            \hypo{\Phi_p \tr \seqi{\Gam_p}{p}{\tcomptype{\stype}{\M}{\stype'}}{(b_p,m_p,d_p)}}
            \infer2[(\ruleApp)]{\seqi{\Gam_v + \Gam_p}{vp}{\comptype{\stype}{\ctype}}{(1+b_v+b_p,m_v+m_p,d_v+d_p)}}
      \end{prooftree} \]
      $\Phi$ must be of the following form:
        \[ \begin{prooftree}
            \hypo{\Phi_0}
            \hypo{\Phi_s \tr \seqi{\Del}{s}{\stype}{(b_s,m_s,d_s)}}
            \infer2[(\ruleConf)]{\seqi{\Gam_v + \Gam_p + \Del}{(vp, s)}{\kappa}{(1+b_v+b_p+b_s,m_v+m_p+m_s,d_v+d_p+d_s)}}
        \end{prooftree} \]
        where $\Gam = \Gam_v + \Gam_p + \Del$ is tight, $b = 1+b_v+b_p+b_s$, $m = m_v+m_p+m_s$, and $s = d_v+d_p+d_s$. Therefore, we can build the following derivation for $(p,s)$:
        \[ \begin{prooftree}
            \hypo{\Phi_p \tr \seqi{\Gam_p}{p}{\tcomptype{\stype}{\M}{\stype'}}{(b_p,m_p,d_p)}}
            \hypo{\Phi_s \tr \seqi{\Del}{s}{\stype}{(b_s,m_s,d_s)}}
            \infer2[(\ruleConf)]{\seqi{\Gam_p + \Del}{(p,s)}{\conftype{\M}{\stype'}}{(b_p+b_s,m_p+m_s,d_p+d_s)}}
        \end{prooftree} \]
        Since $\Gam$ is tight, then $\Gam_p + \Del$ is tight. Moreover, $(p,s) \ra (p',q)$ implies that $\neg \isvalue{p}$. Then we can apply the \ih, and thus there exists a derivation for $(p',q)$ that must be of the following form:
        \[ \begin{prooftree}
            \hypo{\Phi_{p'} \tr \seqi{\Gam_{p'}}{p'}{\tcomptype{\stype''}{\M}{\stype'}}{(b_{p'},m_{p'},d_{p'})}}
            \hypo{\Phi_q \tr \seqi{\Del_q}{q}{\stype''}{(b_q,m_q,d_q)}}
            \infer2[(\ruleConf)]{\seqi{\Gam_{p'} + \Del_q}{(p',q)}{\conftype{\M}{\stype'}}{(b_{p'}+b_q,m_{p'}+m_q,d_{p'}+d_q)}}
        \end{prooftree} \]
        where $\Gam_{p'} + \Del_q = \Gam_p + \Del$ is tight,  and the counters are related properly. Let $\Phi_0$ be the following derivation:
        \[ \begin{prooftree}
          \hypo{\Phi_v \tr \seqi{\Gam_v}{v}{\M \ta \comptype{\stype'}{\ctype'}}{(b_v,m_v,d_v)}}
            \hypo{\Phi_{p'} \tr \seqi{\Gam_{p'}}{p'}{\tcomptype{\stype''}{\M}{\stype'}}{(b_{p'},m_{p'},d_{p'})}}
            \infer2[(\ruleApp)]{\seqi{\Gam_v+\Gam_{p'}}{vp'}{\comptype{\stype''}{\ctype'}}{(1+b_v+b_{p'},m_v+m_{p'},d_v+d_{p'})}}
        \end{prooftree} \]
        We can build $\Phi_{(u,q)}$ as follows:
        \[ \begin{prooftree}
            \hypo{\Phi_0}
            \hypo{\Phi_q \tr \seqi{\Del_q}{q}{\stype''}{(b_q,m_q,d_q)}}
            \infer2[(\ruleConf)]{\seqi{\Gam_v + \Gam_{p'} + \Del_q}{(vp', q)}{\ctype'}{(1+b_v+b_{p'}+b_q,m_v+m_{p'}+m_q,d_v+d_{p'}+d_q)}}
        \end{prooftree} \]
        where $\Gam_{p'} + \Gam_v + \Del_q = \Gam_v + \Gam_p + \Del = \Gam$, $b' = 1+b_v+b_{p'}+b_q$, $m' = m_v+m_{p'}+m_q$, and $d' = d_v+d_{p'}+d_q$. We can conclude since the counters are related properly according to the \ih.  
        \item Case $\Phi_p$ ends with (\ruleAppPOne) or (\ruleAppPTwo). These two cases are very similar to the previous case.
      \end{itemize}
      \item Case $(t,s) = (\get{l}{x}{p},s) \ra (p \subs{x}{v},s) = (u,q)$, where $s \equivstate \upd{l}{v}{s'}$. Let $\Phi_0$ be the following derivation:
      \[ \begin{prooftree}
        \hypo{\Phi_{p} \tr \seqi{\Gam_{p}}{p}{\comptype{\stype}{\ctype}}{(b_p,m_p,d_p)}}
        \infer1[(\ruleGet)]{\seqi{\Gam_{p} \sm x}{\get{l}{x}{p}}{\comptype{\conj{(l : \Gam_{p}(x))} \splus \stype}{\ctype}}{(b_p,1+m_p,d_p)}}
      \end{prooftree} \]
      $\Phi$ must be of the following form:
        \[ \begin{prooftree}
        \hypo{\Phi_0}
          \hypo{\Phi_{s} \tr \seqi{\Del}{s}{\conj{(l : \Gam_{p}(x))} \splus  \stype}{(b_s,m_s,d_s)}}
          \infer2[(\ruleConf)]{\seqi{(\Gam_{p} \sm x) + \Del}{(\get{l}{x}{p}, s)}{\kap}{(b_p+b_s,1+m_p+m_s,d_p+d_s)}}
        \end{prooftree} \] 
        where $\Gam = (\Gam_{p} \sm x) + \Del$ is tight, $b = b_p + b_s$, $m = 1+ m_p + m_s$, and  $d = d_p + d_s$. Since $\Phi_{s} \tr \seqi{\Del}{s}{\conj{(l : \Gam_{p}(x))} \splus \stype}{(b_s,m_s,d_s)}$, then~\cref{lem:split-values-stores}.\ref{lem:split-state} gives $s \equivstate \upd{l}{v_0}{s'_0}$, but we necessarily have $v_0 = v$ and $s'_0 = s'$. Moreover, the lemma also gives $\Phi_v \tr \seqi{\Del_v}{v}{\Gam_{p}(x) \sqcup \stype(l)}{(b_v,m_v,d_v)}$ and $\Phi_{s'} \tr \seqi{\Del_{s'}}{s'}{\stype'}{(b_{s'},m_{s'},d_{s'})}$, where $\conj{(l : \Gam_{p}(x))} \splus \stype = \conj{(l : \Gam_{p}(x) \sqcup \stype(l))};\stype'$, $\Del = \Del_v + \Del_{s'}$, $b_s=b_v + b_{s'}$, $m_s=m_v + m_{s'}$, and $d_s=d_v + d_{s'}$. Thus, by~\cref{lem:split-values-stores}.\ref{lem:com-split-values} there exist $\Phi^1_v \tr \seqi{\Del^1_v}{v}{\Gam_{p}(x)}{(b^1_v,m^1_v,d^1_v)}$ and $\Phi^2_v \tr \seqi{\Del^2_v}{v}{\stype(l)}{(b^2_v,m^2_v,d^2_v)}$, such that $\Del_v = \Del^1_v + \Del^2_v$, $b_v = b^1_v+b^2_v$, $m_v = m^1_v+m^2_v$, and $d_v = d^1_v+d^2_v$. From $\Phi_{p} \tr \seqi{\Gam_{p}}{p}{\comptype{\stype}{\ctype}}{(b_p,m_p,d_p)}$ and $\Phi^1_v \tr \seqi{\Del^1_v}{v}{\Gam_{p}(x)}{(b^1_v,m^1_v,d^1_v)}$, we obtain $\Phi_{p \subs{x}{v}} \tr \seqi{(\Gam_{p} \sm x) +\Del^1_v}{p\subs{x}{v}}{\comptype{\stype}{\ctype}}{(b_p+b^1_v,m_p+m^1_v,d_p+d^1_v)}$, by~\cref{lem:comp-subs-antisubs}.\ref{lem:comp-subs}. We now construct an alternative type derivation for $s$ of the form:
        \[ \begin{prooftree}
            \hypo{\Phi^2_v \tr  \seqi{\Del^2_v}{v}{\stype(l)}{(b^2_v,m^2_v,d^2_v)}}
            \hypo{\Phi_{s'} \tr \seqi{\Del_{s'}}{s'}{\stype'}{(b_{s'},m_{s'},d_{s'})}}
            \infer2[(\ruleUpd)]{\seqi{\Del^2_v+ \Del_{s'}}{\upd{l}{v}{s'}}{\conj{(l:\stype(l))};\stype'}{(b^2_v+b_{s'},m^2_v+m_{s'},d^2_v+d_{s'})}}
        \end{prooftree} \]
        Let $q = s= \upd{l}{v}{s'}$ and let $\Phi_q$ be this new derivation above. Notice also that $\stype = \conj{(l:\stype(l))}; \stype'$. Then we can construct $\Phi'$ as follows:
        \[ \begin{prooftree}
            \hypo{\Phi_{p\subs{x}{v}}}
            \hypo{\Phi_q}
            \infer2[(\ruleConf)]{\seqi{(\Gam_{p} \sm x) + \Del^1_v + \Del^2_v + \Del_{s'}}{(p \subs{x}{v}, s)}{\kap}{(b,m,d)}}
        \end{prooftree} \]
        Notice that the type environment of the conclusion is $(\Gam_{p} \sm x) + \Del^1_v + \Del^2_v + \Del_{s'} = (\Gam_{p} \sm x) + \Del_v + \Del_{s'} = (\Gam_{p} \sm x) + \Del = \Gam $, and the counters are as expected.
        \item Case $(t,s) = (\set{l}{v}{p},s) \ra (p, \upd{l}{v}{s}) = (u,q)$. Let $\Phi_0$ be the following derivation:
        \[ \begin{prooftree}
          \hypo{\Phi_{v} \tr \seqi{\Gam_v}{v}{\M}{(b_v,m_v,d_v)}}
            \hypo{\Phi_{p} \tr \seqi{\Gam_{p}}{p}{\comptype{\conj{(l : \M)}; \stype}{\ctype}}{(b_p,m_p,d_p)}}
            \infer2[(\ruleSet)]{\seqi{\Gam_v + \Gam_{p}}{\set{l}{v}{p}}{\comptype{\stype}{\ctype}}{(b_v+b_p,1+m_v+m_p,s_v+s_p)}}
        \end{prooftree} \]
        $\Phi$ must be of the following form:
        \[ \begin{prooftree}
          \hypo{\Phi_0}    
            \hypo{\Phi_{s} \tr \seqi{\Gam_{s}}{s}{\stype}{(b_s,m_s,d_s)}}
            \infer2[(\ruleConf)]{\seqi{(\Gam_v + \Gam_{p}) + \Gam_{s}}{(\set{l}{v}{p}, s)}{\kap}{(b_v+b_p+b_s,1+m_v+m_p+m_s,d_v+d_p+d_s)}}
        \end{prooftree} \]
        where  $\Gam = (\Gam_v + \Gam_{p}) + \Gam_{s}$ is tight, $b = b_v+b_p+b_s$, $m=1+m_v+m_p+m_s$ and $d=d_v+d_p+d_s$. Therefore, we can build $\Phi_{\upd{l}{v}{s}}$ as follows:
        \[ \begin{prooftree}
          \hypo{\Phi_{v} \tr \seqi{\Gam_v}{v}{\M}{(b_v,m_v,d_v)}}
          \hypo{\Phi_{s} \tr \seqi{\Gam_{s}}{s}{\stype}{(b_s,m_s,d_s)}}
          \infer2[(\ruleUpd)]{\seqi{\Gam_v + \Gam_{s}}{\upd{l}{v}{s}}{\conj{(l : \M)}; \stype}{(b_v+b_s,m_v+m_s,d_v+d_s)}}
        \end{prooftree} \]
        Assume And we can build $\Phi'$ as follows:
        \[ \begin{prooftree}
            \hypo{\Phi_{p} \tr \seqi{\Gam_{p}}{p}{\comptype{\conj{(l : \M)}; \stype}{\ctype}}{(b_p,m_p,d_p)}}
            \hypo{\Phi_{\upd{l}{v}{s}}}
            \infer2[(\ruleConf)]{\seqi{\Gam_{p} + (\Gam_v + \Gam_{s})}{(p, \upd{l}{v}{s})}{\kap}{(b_v+b_v+b_s,m_v+m_v+m_s,d_v+d_v+d_s)}}
        \end{prooftree} \]
        Notice that the type environment of the conclusion is $\Gam_{p} + (\Gam_v + \Gam_{s}) = \Gam$, and the counters are as expected.
    \end{itemize}

        \item 
    We show a stronger statement of the form:

    Let $(t,s) \red[\gname] (u,q)$. If  $\Phi' \tr \seqi{\Gam}{(u,q)}{\ctype}{(b',m',d)}$, $\Gam$ is tight, and ($\ctype$ is tight or $\neg \isvalue{t}$), then $\Phi \tr \seqi{\Gam}{(t,s)}{\ctype}{(b,m,d)}$, where $\gname =\beta$ implies $b' = b - 1$ and $m' = m$, while $\gname \in \{\getname, \setname\}$ implies $b'=b$ and $m' = m - 1$.

    We proceed by induction on $(t, s) \red (u,q)$: 
    \begin{itemize}
        \item Case $(t,s) = ((\lam x.p) v,s) \redbeta (p \subs{x}{v}, s) = (u,q)$. Then $\Phi'$ must be of the following form:
        \[ \begin{prooftree}
            \hypo{\Phi_{p \subs{x}{v}} \tr \seqi{\Gam_{p \subs{x}{v}}}{p \subs{x}{v}}{\comptype{\stype}{\ctype}}{(b'',m'',d'')}}
            \hypo{\Phi_s \tr \seqi{\Gam_s}{s}{\stype}{(b_s,m_s,d_s)}}
            \infer2[(\ruleConf)]{\seqi{\Gam_{p \subs{x}{v}} + \Gam_s}{(p \subs{x}{v}, s)}{\ctype}{(b''+b_s,m''+m_s,d''+d_s)}}
        \end{prooftree} \]
        such that $\Gam = \Gam_{p \subs{x}{v}} + \Gam_s$, $b' = b''+b_s$, $m' = m''+m_s$, and $d' = d''+d_s$. By~\cref{lem:comp-subs-antisubs}.\ref{lem:comp-antisubs}, there exist $\Phi_p \tr \seqi{\Gam_p; x : \M}{p}{\comptype{\stype}{\ctype}}{(b_p,m_p,d_p)}$ and $\Phi_{v} \tr \seqi{\Gam_v}{v}{\M}{(b_v,m_v,d_v)}$, such that $\Gam_{p \subs{x}{v}} = \Gam_p + \Gam_v$, $b'' = b_p+b_v$, $m'' = m_p+m_v$, and $d'' = d_p + d_v$. We can build $\Phi$ as follows:
        \[ \begin{prooftree}
            \hypo{\Phi_p \tr \seqi{\Gam_p; x : \M}{p}{\comptype{\stype}{\ctype}}{(b_p,m_p,d_p)}}
            \infer1[(\ruleLam)]{\seqi{\Gam_p}{\lam x.p}{\M \ta (\comptype{\stype}{\ctype})}{(b_p,m_p,d_p)}}
            \hypo{\Phi_{v} \tr \seqi{\Gam_v}{v}{\M}{(b_v,m_v,d_v)}}
            \infer1[(\ruleLift)]{\seqi{\Gam_v}{v}{\tcomptype{\stype}{\M}{\stype}}{(b_v,m_v,d_v)}}
            \infer2[(\ruleApp)]{\seqi{\Gam_p + \Gam_v}{(\lam x.p)v}{\comptype{\stype}{\ctype}}{(1+b_p+b_v,m_p+m_v,d_p+d_v)}}
            \hypo{\Phi_s}
            \infer2[(\ruleConf)]{\seqi{(\Gam_p + \Gam_v) + \Gam_s}{((\lam x.t')v, s)}{\ctype}{(1+b_p+b_v+b_s,m_p+m_v+m_s,d_p+d_v+d_s)}}
        \end{prooftree} \]
        such that $b = 1+b_p+b_v+b_s$, $m = m_p+m_v+m_s$, and $d = d_p+d_v+d_s$. And we can conclude with $\Gam = \Gam_{p \subs{x}{v}} + \Gam_s = (\Gam_p + \Gam_v) + \Gam_s$, $b' = b'' + b_s = b_p + b_v + b_s = (1 + b_p + b_v + b_s) - 1 = b - 1$, $m' = m'' + m_s = (m_p + m_v) + m_s = m$, and $d' = d'' + d_s = (d_p + d_v) + d_s = d$.
        \item Case $(t,s) = (vp,s) \ra (vp',q) = (u,q)$, such that $(p,s) \ra (p',q)$. Then we have three cases for the type derivation $\Phi_{p'}$ of $p'$ inside $\Phi'$:
        \begin{itemize}
            \item Case $\Phi_{vp'}$ ends with ($\ruleApp$). Let $\Phi_0$ be the following derivation:
            \[ \begin{prooftree}
                \hypo{\Phi_{v} \tr \seqi{\Gam_v}{v}{\M \ta \comptype{\stype'}{\ctype}}{(b_v,m_v,d_v)}}
                    \hypo{\Phi_{p'} \tr \seqi{\Gam_{p'}}{p'}{\tcomptype{\stype}{\M}{\stype'}}{(b'',m'',d'')}}
                    \infer2[(\ruleApp)]{\seqi{\Gam_v + \Gam_{p'}}{v p'}{\comptype{\stype}{\ctype}}{(1+b_v+b'', m_v+m'', d_v+d'')}}
            \end{prooftree} \]
            $\Phi'$ must be of the following form: 
            \[ \begin{prooftree}
                \hypo{\Phi_0}
                \hypo{\Phi_q \tr \seqi{\Gam_q}{q}{\stype}{(b_q,m_q,d_q)}}
                \infer2[(\ruleConf)]{\seqi{(\Gam_v + \Gam_{p'}) + \Gam_q}{(v p', q)}{\ctype}{(1+b_v+b''+b_q,m_v+m''+m_q,d_v+d''+d_q)}}
            \end{prooftree} \]
            such that $\Gam = (\Gam_v + \Gam_{p'}) + \Gam_q$ tight, $b' = 1+b_v+b''+b_q$, $m' = m_v+m''+m_q$, and $d' = d_v+d''+d_q$. So we can build $\Phi_{(p',q)}$ as follows:
            \[ \begin{prooftree}
                \hypo{\Phi_{p'} \tr \seqi{\Gam_{p'}}{p'}{\tcomptype{\stype}{\M}{\stype'}}{(b'',m'',d'')}}
                \hypo{\Phi_q \tr \seqi{\Gam_q}{q}{\stype}{(b_q,m_q,d_q)}}
                \infer2[(\ruleConf)]{\seqi{\Gam_{p'} + \Gam_q}{(p', q)}{\conftype{\M}{\stype'}}{(b''+b_q,m''+m_q,d''+d_q)}}
            \end{prooftree} \]
            Since $\Gam$ is tight, then $\Gam_{p'} + \Gam_q$ is tight. Moreover, $(p, s) \red (p',q)$ implies $\neg\isvalue{p}$. Then we can apply the \ih, and thus there exists a derivation for $(p,s)$ that must be of the following form:
            \[ \begin{prooftree}
                \hypo{\Phi_p \tr \seqi{\Gam_p}{p}{\tcomptype{\stype''}{\M}{\stype'}}{(b_p,m_p,d_p)}}
                \hypo{\Phi_s \tr \seqi{\Gam_s}{s}{\stype''}{(b_s,m_s,d_s)}}
                \infer2[(\ruleConf)]{\seqi{\Gam_p + \Gam_s}{(p, s)}{\conftype{\M}{\stype'}}{(b_p+b_s,m_p+m_s,d_p+d_s)}}
            \end{prooftree} \]
            where $\Gam_p + \Gam_s = \Gam_{p'} + \Gam_q$ is tight, and either (1) $b''+b_q = b_p+b_s-1$, $m''+m_q=m_p+m_s$, and $d''+d_q = d_p+d_s$, or (2) $b''+b_q = b_p+b_s$, $m''+m_q=m_p+m_s-1$, and $d''+d_q = d_p+d_s$. So, we can build $\Phi$ as follows:
            \[ \begin{prooftree}
                \hypo{\Phi_{v} \tr \seqi{\Gam_v}{v}{\M \ta (\comptype{\stype'}{\ctype})}{(b_v,m_v,d_v)}}
                \hypo{\Phi_p \tr \seqi{\Gam_p}{p}{\tcomptype{\stype''}{\M}{\stype'}}{(b_p,m_p,d_p)}}
                \infer2[(\ruleApp)]{\seqi{\Gam_v + \Gam_p}{vp}{\comptype{\stype''}{\ctype}}{(1+b_v+b_p,m_v+m_p,d_v+d_p)}}
                \hypo{\Phi_s}
                \infer2[(\ruleConf)]{\seqi{(\Gam_v + \Gam_p) + \Gam_s}{(vp, s)}{\kappa}{(1 + b_v + b_p+b_s,m_v+m_p+m_s,d_v+d_p+d_s)}}
            \end{prooftree} \]
            where $\Gam_v + \Gam_p + \Gam_s = \Gam_v + \Gam_{p'} + \Gam_q = \Gam$, $b = 1+b_v+b_p+b_s$, $m = m_v+m_p+m_s$, and $d = d_v+d_p+d_s$. We can conclude since:
            \begin{itemize}
                \item Case (1): $b' = 1 + b_v + b'' + b_q = 1 + b_v + b_p + b_s - 1 = b -1$, and the other counters are easy to check;
                \item Case (2): $m' = m_v + m'' + m_q = m_v + m_p + m_s - 1 = m - 1$, and the other counters are easy to check.
            \end{itemize}
            \item Case $\Phi_{vp'}$ ends with (\ruleAppPOne) or (\ruleAppPTwo). These two cases are very similar to the previous case.
        \end{itemize}
        \item Case $(t,s) = (\get{l}{x}{p},s) \ra (p \subs{x}{v},s) = (u,q)$, such that $s \equivstate \upd{l}{v}{s'}$. Let $\Phi_0$ be the following derivation:
        \[ \begin{prooftree}
            \hypo{\Phi^2_v \tr \seqi{\Gam^2_v}{v}{\M_2}{(b^2_v,m^2_v,d^2_v)}}
            \hypo{\Phi_{s'} \tr \seqi{\Gam_{s'}}{s'}{\stype}{(b_{s'},m_{s'},d_{s'})}}
            \infer2[(\ruleUpd)]{\seqi{\Gam^2_v + \Gam_{s'}}{\upd{l}{v}{s'}}{\conj{(l : \M_2)}; \stype}{(b^2_v+b_{s'},m^2_v+m_{s'},d^2_v+d_{s'})}}
        \end{prooftree} \]
        Then $\Phi'$ must be of the following form:
        \[ \begin{prooftree}
            \hypo{\Phi_{p \subs{x}{v}} \tr \seqi{\Gam_{p \subs{x}{v}}}{p \subs{x}{v}}{\comptype{\conj{(l : \M)}; \stype}{\ctype}}{(b'',m'',d'')}}
            \hypo{\Phi_0}
            \infer2[(\ruleConf)]{\seqi{\Gam_{p \subs{x}{v}} + (\Gam^2_v + \Gam_{s'})}{(p \subs{x}{v}, \upd{l}{v}{s'})}{\ctype}{(b''+b^2_v+b_{s'},m''+m^2_v+m_{s'},d''+d^2_v+d_{s'})}}
        \end{prooftree} \]
        such that $\Gam = \Gam_{p \subs{x}{v}} + (\Gam^2_v + \Gam_{s'})$, $b' = b'' + b^2_v + b_{s'}$, $m' = m'' + b^2_v + b_{s'}$, and $d' = d'' +d^2_v+d_{s'}$. By~\cref{lem:comp-subs-antisubs}.\ref{lem:comp-antisubs}, there exist $\Phi_p \tr \seqi{\Gam_p; x : \M_1}{p}{\comptype{\conj{(l : \M_2)}; \stype}{\ctype}}{(b_p,m_p,d_p)}$ and $\Phi^1_v \tr \seqi{\Gam^1_v}{v}{\M_1}{(b^1_v,m^1_v,d^1_v)}$, such that $\Gam_{p \subs{x}{v}} = \Gam_p + \Gam^1_v$, $b'' = b_p + b^1_v$, $m'' = m_p + m^1_v$, and $d'' = d_p + d^1_v$. Therefore, we can build $\Phi_{\get{l}{x}{p}}$ as follows:
        \[ \begin{prooftree}
            \hypo{\Phi_p \tr \seqi{\Gam_p; x : \M_1}{p}{\comptype{\conj{(l : \M_2)}; \stype}{\ctype}}{(b_p,m_p,d_p)}}
            \infer1[(\ruleGet)]{\seqi{\Gam_p}{\get{l}{x}{p}}{\comptype{\conj{(l : \M_1 \sqcup \M_2)}; \stype}{\ctype}}{(b_p,1+m_p,d_p)}}
        \end{prooftree} \]
        By~\cref{lem:comp-merge-values}, we have $\Phi_v \tr \seqi{\Gam^1_v + \Gam^2_v}{v}{\M_1 \sqcup \M_2}{(b^1_v+b^2_v,m^1_v+m^2_v,d^1_v+d^2_v)}$. Thus, we can build $\Phi_{\upd{l}{v}{s'}}$ as follows:
        \[ \begin{prooftree}
            \hypo{\Phi_v \tr \seqi{\Gam^1_v + \Gam^2_v}{v}{\M_1 \sqcup \M_2}{(b^1_v+b^2_v,m^1_v+m^2_v,d^1_v+d^2_v)}}
            \hypo{\Phi_{s'} \tr \seqi{\Gam_{s'}}{s'}{\stype}{(b_{s'},m_{s'},d_{s'})}}
            \infer2[(\ruleUpd)]{\seqi{(\Gam^1_v + \Gam^2_v) + \Gam_{s'}}{\upd{l}{v}{s'}}{\conj{(l : \M_1 \sqcup \M_2)}; \stype}{(b^1_v+b^2_v+b_{s'},m^1_v+m^2_v+m_{s'},d^1_v+d^2_v+d_{s'})}}
        \end{prooftree} \]
        Finally, we can build $\Phi$ as follows:
        \[ \begin{prooftree}
            \hypo{\Phi_{\get{l}{x}{p}}}
            \hypo{\Phi_{\upd{l}{v}{s'}}}
            \infer2[(\ruleConf)]{\seqi{\Gam_p + (\Gam^1_v + \Gam^2_v) + \Gam_{s'}}{(\get{l}{x}{p}, \upd{l}{v}{s'})}{\ctype}{(b_p+b^1_v+b^2_v+b_{s'},1+m_p+m^1_v+m^2_v+m_{s'},d_p+d^1_v+d^2_v+d_{s'})}}
        \end{prooftree} \]
        such that $b = b_p+b^1_v+b^2_v+b_{s'}$, $m = 1+m_p+m^1_v+m^2_v+m_{s'}$, and $d = d_p+d^1_v+d^2_v+d_{s'}$. And we can conclude with $\Gam = \Gam_{p \subs{x}{v}} + (\Gam^2_v + \Gam_{s'}) = \Gam_p + \Gam^1_v + \Gam^2_v + \Gam_{s'}$, $b' = b'' + b^2_v + b_{s'} = b_p + b^1_v + b^2_v + b_{s'} = b$, and $m' = m'' + m^2_v + m_{s'} = m_p + m^1_v + m^2_v + m_{s'} = (1 + m_p + m^1_v + m^2_v + m_{s'}) - 1 = m - 1$, $d' = d'' + d^2_v + d_{s'} = d_p + d^1_v + d^2_v + d_{s'} = d$.
        \item Case $(t,s) = (\set{l}{v}{p},s) \ra (p, \upd{l}{v}{s}) = (u,q)$. Let $\Phi_0$ be the following derivation:
        \[ \begin{prooftree}
            \hypo{\Phi_{v} \tr \seqi{\Gam_v}{v}{\M}{(b_v,m_v,d_v)}}
            \hypo{\Phi_{s} \tr \seqi{\Gam_s}{s}{\stype}{(b_s,m_s,d_s)}}
            \infer2[(\ruleUpd)]{\seqi{\Gam_v + \Gam_s}{\upd{l}{v}{s}}{\conj{(l : \M)}; \stype}{(b_v+b_s,m_v+m_s,d_v+d_s)}}
        \end{prooftree} \]
        $\Phi'$ must be of the following form:
        \[ \begin{prooftree}
            \hypo{\Phi_p \tr \seqi{\Gam_p}{p}{\comptype{\conj{(l : \M)}; \stype}{\ctype}}{(b_p,m_p,d_p)}}
            \hypo{\Phi_0}
            \infer2[(\ruleConf)]{\seqi{\Gam_p + (\Gam_v + \Gam_s)}{(p, \upd{l}{v}{s})}{\kap}{(b_p+b_v+b_s,m_p+m_v+m_s,d_p+d_v+d_s)}}
        \end{prooftree} \]
        such that $\Gam = \Gam_p + (\Gam_v + \Gam_s)$, $b' = b_p + b_v + b_s$, $m' = m_p + m_v + m_s$, and $d' = d_p + d_v + d_s$. Therefore, we can build $\Phi$ as follows:
        \[ \begin{prooftree}
            \hypo{\Phi_{v} \tr \seqi{\Gam_v}{v}{\M}{(b_v,m_v,d_v,)}}
            \hypo{\Phi_p \tr \seqi{\Gam_p}{p}{\comptype{\conj{(l : \M)}; \stype}{\ctype}}{(b_p,m_p,d_p)}}
            \infer2[(\ruleSet)]{\seqi{\Gam_v + \Gam_p}{\set{l}{v}{p}}{\comptype{\stype}{\ctype}}{(b_v+b_p,1+m_v+m_p,d_v+d_p)}}
            \hypo{\Phi_s}
            \infer2[(\ruleConf)]{\seqi{(\Gam_v + \Gam_p) + \Gam_s}{(\set{l}{v}{p}, s)}{\ctype}{(b_v+b_p+b_s,1+m_v+m_p+m_s,d_v+d_p+d_s)}}
        \end{prooftree} \]
        Notice that the type environment of the conclusion is $(\Gam_v + \Gam_p) + \Gam_s = \Gam$, and the counters are as expected.
    \end{itemize}
    \end{enumerate}
\end{proof}}

\compsoundness*

\maybehide{\begin{proof} \mbox{}
    \begin{enumerate}
        \item 
    The proof follows by induction over $b+m$:
    \begin{itemize}
        \item Case $b+m = 0$. Then $b=m=0$, therefore $t \in \normal$, by \cref{lem:zero-counters-size-store}.\ref{lem:zero-counters}, and  $d = \size{t}$, by~\cref{lem:zero-counters-size-store}.\ref{lem:zero-size-store}. Let $u = t$ and $q=s$, then  we can conclude since $\size{(u,q)} = \size{u} =\size{t} = d$.
        \item Case $b+m > 0$. Then $b>0$ or $m>0$, and in either case $t \not\in \normal$,  by~\cref{lem:zero-nfs}. Note that $(t,s)$ is not final because $t$ is unblocked by~\cref{prop:typed-unblock}. Therefore, by~\cref{prop:normal-iff-final} there exists $(t',s')$ such that $(t,s) \gsred (t',s')$. By~\cref{lem-exact-red-exp}.\ref{lem:subj-comp-red}, there exists $\Phi' \tr \seqi{\Gam}{(t',s')}{\ctype}{(b',m',d)}$, such that $b'+m'=b+m-1$. By the \ih, there exists $(u,q)$, such that $u\in \normal$, $(t',s') \gsrred^{(b',m')} (u,q)$ and $d = \size{(u,q)}$. So we can conclude with $(t,s) \gsred (t',s') \gsrred^{(b',m')} (u,q)$, which means that $(t,s) \drred^{(b,m)} (u,q)$, as expected.
    \end{itemize}

        \item 
    By induction over $b + m$: \begin{itemize}
        \item Case $b + m = 0$. Then $b = m = 0$ and $(t,s) = (u,q)$. We can conclude by~\cref{lem:typestatesnfs}.\ref{lem:typ-states} and~\cref{lem:typestatesnfs}.\ref{lem:comp-typ-nfs}.
        \item Case $b + m > 0$. Then there exists $(t',s')$, such that $(t,s) \ra^{(1,0)} (t',s') \rra^{(b-1,m)} (u,q)$ or $(t,s) \ra^{(0,1)} (t',s') \rra^{(b,m-1)} (u,q)$. By the \ih, there exists $\Phi' \tr \seqi{\Gam}{(t',s')}{\kap}{(b',m',\size{(u,q)})}$ tight, such that $b' + m' = b + m - 1$. By~\cref{lem-exact-red-exp}.\ref{lem:comp-subj-exp}, we have $\Phi \tr \seqi{\Gam}{(t,s)}{\kap}{(b'',m'',\size{(u,q)})}$ tight, such that $b'' + m'' = 1+ b' + m'$. Therefore, $b'' + m'' = b + m$, since the fact that $b'' = b$, and $m'' = m$ can be easily checked by a simple case analysis.
    \end{itemize}
    \end{enumerate}
\end{proof}}

\end{document}